\documentclass[twocolumn,tighten]{aastex631}

\usepackage{amsmath,amstext}
\usepackage[T1]{fontenc}
\usepackage[figure,figure*]{hypcap}

\usepackage{graphicx} % Including Figure~files
\usepackage{amssymb}  % Extra maths symbols
\usepackage{multirow}
\usepackage{booktabs}
\usepackage{xspace}
\usepackage{makecell}
\usepackage{tabularx}
\usepackage{enumitem}

\usepackage{xcolor}
\usepackage{totcount}

\newtotcounter{citnum} %From the package documentation
\def\oldbibitem{} \let\oldbibitem=\bibitem
\def\bibitem{\stepcounter{citnum}\oldbibitem}

\newcommand {\chandra} {\textit{Chandra}\xspace}
\newcommand {\xmm} {\textit{XMM}-Newton\xspace}

\newcommand {\swift} {\textit{Swift}\xspace}
\newcommand {\nustar} {\textit{NuSTAR}\xspace}

\newcommand {\suzaku} {\textit{Suzaku}\xspace}

\newcommand {\xis} {\textit{Suzaku}/XIS\xspace}

\newcommand {\bat} {\textit{Swift}/BAT\xspace}
\newcommand {\xrt} {\textit{Swift}/XRT\xspace}
\newcommand {\integral} {\textit{INTEGRAL}\xspace}
\newcommand {\wise} {\textit{WISE}\xspace}
\newcommand {\iras} {\textit{IRAS}\xspace}

\newcommand {\ciao}{\textsc{CIAO}\xspace}
\newcommand {\caldb}{\textsc{CALDB}\xspace}

\newcommand {\mytorus}{\textsc{MYTorus}\xspace}

\newcommand {\nh} {$N_{\mathrm{H}}$\xspace}

\newcommand {\simbad} {\textsc{simbad}\xspace}
\newcommand {\ned} {\textsc{ned}\xspace}

\newcommand {\oiii}{[O\,{\sc iii}]\xspace}

\newcommand {\fluxunit} {erg\,s$^{-1}$~cm$^{-2}$\xspace}

\newcommand {\nhunit} {cm$^{-2}$\xspace}

\defcitealias{deGrijp85}{dG85}
\defcitealias{deGrijp87}{dG87}
\defcitealias{deGrijp92}{dG92}
\defcitealias{Keel94}{K94}
\defcitealias{Ricci17c}{R17}
\defcitealias{Gandhi17}{G17}
\defcitealias{Balokovic14}{B14}
\defcitealias{Boorman16}{B16}
\defcitealias{Kara17}{K17}
\defcitealias{Burlon11}{B11}
\defcitealias{BalokovicBAT}{BIP}
\defcitealias{Asmus15}{A15}

\shorttitle{NuLANDS Paper I}
\shortauthors{Boorman et al.}
\graphicspath{{./}{figures/}}
\begin{document}

\title{The \textit{NuSTAR} Local AGN $N_{\rm H}$ Distribution Survey (NuLANDS) I: Towards a Truly Representative Column Density Distribution in the Local Universe}

\correspondingauthor{Peter~G.~Boorman}
\email{boorman@caltech.edu}

\author[0000-0001-9379-4716]{Peter~G.~Boorman}
\affiliation{Cahill Center for Astrophysics, California Institute of Technology, 1216 East California Boulevard, Pasadena, CA 91125, USA}
\affiliation{Department of Physics \& Astronomy, Faculty of Physical Sciences and Engineering, University of Southampton, Southampton, SO17 1BJ, UK}
\affiliation{Astronomical Institute, Academy of Sciences, Bo\v{c}n\'{i} II 1401, CZ-14131 Prague, Czech Republic}

\author[0000-0003-3105-2615]{Poshak~Gandhi}
\affiliation{Department of Physics \& Astronomy, Faculty of Physical Sciences and Engineering, University of Southampton, Southampton, SO17 1BJ, UK}

\author[0000-0003-0426-6634]{Johannes~Buchner}
\affiliation{Max Planck Institute for Extraterrestrial Physics, Giessenbachstrasse, 85741 Garching, Germany}
\affiliation{Instituto de Astrof{\'{\i}}sica and Centro de Astroingenier{\'{\i}}a, Facultad de F{\'{i}}sica, Pontificia Universidad Cat{\'{o}}lica de Chile, Casilla 306, Santiago 22, Chile}
\affiliation{Excellence Cluster Universe, Boltzmannstr. 2, D-85748, Garching, Germany}

\author[0000-0003-2686-9241]{Daniel~Stern}
\affiliation{Jet Propulsion Laboratory, California Institute of Technology, Pasadena, CA 91109, USA}

\author[0000-0001-5231-2645]{Claudio~Ricci}
\affiliation{Instituto de Estudios Astrof\'isicos, Facultad de Ingenier\'ia y Ciencias, Universidad Diego Portales, Av. Ej\'ercito Libertador 441, Santiago, Chile}
\affiliation{Kavli Institute for Astronomy and Astrophysics, Peking University, Beijing 100871, China}

\author[0000-0003-0476-6647]{Mislav~Balokovi{\'{c}}}
\affiliation{Yale Center for Astronomy \& Astrophysics, 219 Prospect Street, New Haven, CT 06511, USA}
\affiliation{Department of Physics, Yale University, P.O. Box 208120, New Haven, CT 06520, USA}

\author[0000-0003-0220-2063]{Daniel~Asmus}
\affiliation{Department of Physics \& Astronomy, Faculty of Physical Sciences and Engineering, University of Southampton, Southampton, SO17 1BJ, UK}
\affiliation{Gymnasium Schwarzenbek, 21493 Schwarzenbek, Germany}

\author[0000-0002-4226-8959]{Fiona~A.~Harrison}
\affiliation{Cahill Center for Astrophysics, California Institute of Technology, 1216 East California Boulevard, Pasadena, CA 91125, USA}

\author[0000-0003-2931-0742]{Ji\v{r}\'{i}~Svoboda}
\affiliation{Astronomical Institute, Academy of Sciences, Bo\v{c}n\'{i} II 1401, CZ-14131 Prague, Czech Republic}

\author[0000-0002-7719-5809]{Claire~Greenwell}
\affiliation{Centre for Extragalactic Astronomy, Department of Physics, Durham University, South Road, Durham, DH1 3LE, UK}
\affiliation{Department of Physics \& Astronomy, Faculty of Physical Sciences and Engineering, University of Southampton, Southampton, SO17 1BJ, UK}

\author[0000-0002-7998-9581]{Michael~J.~Koss}
\affiliation{Eureka Scientific, 2452 Delmer Street Suite 100, Oakland, CA 94602-3017, USA}

\author[0000-0002-5896-6313]{David~M.~Alexander}
\affiliation{Centre for Extragalactic Astronomy, Department of Physics, Durham University, South Road, Durham, DH1 3LE, UK}

\author[0000-0003-0387-1429]{Adlyka~Annuar}
\affiliation{Department of Applied Physics, Faculty of Science and Technology, Universiti Kebangsaan Malaysia, 43600 UKM Bangi, Selangor, Malaysia}
\affiliation{Centre for Extragalactic Astronomy, Department of Physics, Durham University, South Road, Durham, DH1 3LE, UK}

\author[0000-0002-8686-8737]{Franz~Bauer}
\affiliation{Instituto de Astrof{\'{\i}}sica and Centro de Astroingenier{\'{\i}}a, Facultad de F{\'{i}}sica, Pontificia Universidad Cat{\'{o}}lica de Chile, Casilla 306, Santiago 22, Chile}
\affiliation{Millennium Institute of Astrophysics (MAS), Nuncio Monse{\~{n}}or S{\'{o}}tero Sanz 100, Providencia, Santiago, Chile}
\affiliation{Space Science Institute, 4750 Walnut Street, Suite 205, Boulder, CO 80301, USA}

\author[0000-0002-0167-2453]{William~N.~Brandt}
\affiliation{Department of Astronomy and Astrophysics, The Pennsylvania State University, 525 Davey Lab, University Park, PA 16802, USA}
\affiliation{Institute for Gravitation and the Cosmos, The Pennsylvania State University, University Park, PA 16802, USA}
\affiliation{Department of Physics, 104 Davey Lab, The Pennsylvania State University, University Park, PA 16802, USA}

\author[0000-0002-8147-2602]{Murray~Brightman}
\affiliation{Cahill Center for Astrophysics, California Institute of Technology, 1216 East California Boulevard, Pasadena, CA 91125, USA}

\author[0000-0002-2115-1137]{Francesca~Civano}
\affiliation{NASA Goddard Space Flight Center, Greenbelt, MD, USA}

\author[0000-0002-4945-5079]{Chien-Ting~J.~Chen}
\affiliation{Science and Technology Institute, Universities Space Research Association, Huntsville, AL 35805, USA}
\affiliation{Astrophysics Office, NASA Marshall Space Flight Center, ST12, Huntsville, AL 35812, USA}

\author[0000-0003-1748-2010]{Duncan~Farrah}
\affiliation{Department of Physics and Astronomy, University of Hawai`i at M\={a}noa, 2505 Correa Rd., Honolulu, HI 96822, USA}
\affiliation{Institute for Astronomy, University of Hawai`i, 2680 Woodlawn Dr., Honolulu, HI 96822, USA}

\author[0000-0001-5800-5531]{Karl~Forster}
\affiliation{Cahill Center for Astrophysics, California Institute of Technology, 1216 East California Boulevard, Pasadena, CA 91125, USA}

\author[0000-0002-1984-2932]{Brian~Grefenstette}
\affiliation{Cahill Center for Astrophysics, California Institute of Technology, 1216 East California Boulevard, Pasadena, CA 91125, USA}

\author[0000-0002-6353-1111]{Sebastian~F.~H{\"o}nig}
\affiliation{Department of Physics \& Astronomy, Faculty of Physical Sciences and Engineering, University of Southampton, Southampton, SO17 1BJ, UK}

\author[0000-0003-3470-4834]{Adam~B.~Hill}
\affiliation{Department of Physics \& Astronomy, Faculty of Physical Sciences and Engineering, University of Southampton, Southampton, SO17 1BJ, UK}
\affiliation{ComplyAdvantage, 2nd Floor, Fetter Yard, Fetter Lane, London EC4A 1AD, United Kingdom}

\author[0000-0002-0273-218X]{Elias~Kammoun}
\affiliation{Dipartimento di Matematica e Fisica, Universit\`{a} Roma Tre, via della Vasca Navale 84, I-00146 Rome, Italy }
\affiliation{INAF -- Osservatorio Astrofisico di Arcetri, Largo Enrico Fermi 5, I-50125 Firenze, Italy}

\author[0000-0002-5328-9827]{George~Lansbury}
\affiliation{European Southern Observatory, Karl-Schwarzschild str. 2, D-85748 Garching bei M{\"u}nchen, Germany}
\affiliation{Institute of Astronomy, University of Cambridge, Madingley Road, Cambridge, CB3 0HA, UK}

\author[0000-0002-3249-8224]{Lauranne~Lanz}
\affiliation{Department of Physics, The College of New Jersey, 2000 Pennington Road, Ewing, NJ 08628, USA}

\author[0000-0002-5907-3330]{Stephanie~LaMassa}
\affiliation{Space Telescope Science Institute, 3700 San Martin Drive, Baltimore, MD 21218, USA}

\author[0000-0003-1252-4891]{Kristin~Madsen}
\affiliation{CRESST and X-ray Astrophysics Laboratory, NASA Goddard Space Flight Center, Greenbelt, MD 20771, USA}

\author[0000-0001-5544-0749]{Stefano~Marchesi}
\affiliation{Dipartimento di Fisica e Astronomia (DIFA) Augusto Righi, Università di Bologna, via Gobetti 93/2, I-40129 Bologna, Italy}
\affiliation{Department of Physics and Astronomy, Clemson University, Kinard Lab of Physics, Clemson, SC 29634, USA}
\affiliation{INAF - Osservatorio di Astrofisica e Scienza dello Spazio di Bologna, Via Piero Gobetti, 93/3, 40129, Bologna, Italy}

\author[0000-0002-8183-2970]{Matthew~Middleton}
\affiliation{Department of Physics \& Astronomy, Faculty of Physical Sciences and Engineering, University of Southampton, Southampton, SO17 1BJ, UK}

\author[0000-0001-5649-938X]{Beatriz~Mingo}
\affiliation{School of Physical Sciences, The Open University, Walton Hall, Milton Keynes MK7 6AA, UK}

\author[0000-0002-8466-7317]{Michael~L.~Parker}
\affil{Optibrium Ltd., Blenheim House, Cambridge Innovation Park, Denny End Rd, Waterbeach, Cambridge CB25 9GL}

\author[0000-0001-7568-6412]{Ezequiel~Treister}
\affiliation{Instituto de Astrof{\'{\i}}sica and Centro de Astroingenier{\'{\i}}a, Facultad de F{\'{i}}sica, Pontificia Universidad Cat{\'{o}}lica de Chile, Casilla 306, Santiago 22, Chile}

\author[0000-0001-7821-6715]{Yoshihiro~Ueda}
\affiliation{Department of Astronomy, Kyoto University, Kitashirakawa-Oiwake-cho, Sakyo-ku, Kyoto 606-8502, Japan}

\author[0000-0002-0745-9792]{C.~Megan~Urry}
\affiliation{Yale Center for Astronomy \& Astrophysics, 219 Prospect Street, New Haven, CT 06511, USA}
\affiliation{Department of Physics, Yale University, P.O. Box 208120, New Haven, CT 06520, USA}

\author[0000-0002-4205-6884]{Luca~Zappacosta}
\affiliation{INAF - Osservatorio Astronomico di Roma, via di Frascati 33, 00078 Monte Porzio Catone, Italy}



\begin{abstract}
Hard X-ray-selected samples of Active Galactic Nuclei (AGN) provide one of the cleanest views of supermassive black hole accretion, but are biased against objects obscured by Compton-thick gas column densities of $N_{\rm H}$\,$>$\,10$^{24}$\,cm$^{-2}$. To tackle this issue, we present the \textit{NuSTAR} Local AGN $N_{\rm H}$ Distribution Survey (NuLANDS)---a legacy sample of 122 nearby ($z$\,$<$\,0.044) AGN primarily selected to have warm infrared colors from \textit{IRAS} between 25\,--\,60\,$\mu$m. We show that optically classified type 1 and 2 AGN in NuLANDS are indistinguishable in terms of optical [O\,\textsc{iii}] line flux and mid-to-far infrared AGN continuum bolometric indicators, as expected from an isotropically selected AGN sample, while type 2 AGN are deficient in terms of their observed hard X-ray flux. By testing many X-ray spectroscopic models, we show the measured line-of-sight column density varies on average by $\sim$\,1.4 orders of magnitude depending on the obscurer geometry. To circumvent such issues we propagate the uncertainties per source into the parent column density distribution, finding a directly measured Compton-thick fraction of 35\,$\pm$\,9\%. By construction, our sample will miss sources affected by severe narrow-line reddening, and thus segregates sources dominated by small-scale nuclear obscuration from large-scale host-galaxy obscuration. This bias implies an even higher intrinsic obscured AGN fraction may be possible, although tests for additional biases arising from our infrared selection find no strong effects on the measured column-density distribution. NuLANDS thus holds potential as an optimized sample for future follow-up with current and next-generation instruments aiming to study the local AGN population in an isotropic manner.
\end{abstract}

\keywords{galaxies: active, X-rays --- surveys --- galaxies: Seyfert}


\section{Introduction}\label{sec:intro}
\subsection{AGN Growth and Obscuration}
The integrated X-ray emission from accreting supermassive black holes (log\,$M_{\rm BH}\,/\,M_{\odot}$\,$\sim$\,6\,--\,9.5), or Active Galactic Nuclei (AGN), across cosmic time dominates the Cosmic X-ray Background spectrum (CXB; \citealt{Giacconi62}) across the wide energy range E\,$\sim$\,1\,--\,300\,keV (e.g., \citealt{Setti89,Comastri95,Mushotzky00,Gandhi03,Gilli07,Treister09,Ueda14,Brandt15,RamosAlmeida17b,Ananna19,Ananna20,Civano24}). Hence, the evolution of and accretion onto supermassive black holes is preserved in the broadband CXB. In particular, AGN in the Seyfert luminosity range ($L_{2-10\,{\rm keV}}$\,$\sim$\,10$^{42}$\,--\,10$^{44}$\,erg\,s$^{-1}$) completely dominate the CXB emissivity to beyond $z$\,=\,1, and contribute between $\approx$\,30\,--\,50\% of the emissivity between $z$\,=\,1\,--\,5 \citep{Ueda14,Buchner15,Aird15a}. These sources are thus crucial to understand, but are best probed in detail in the local universe where fluxes are brighter overall and off-nuclear contaminants can be resolved in the lowest luminosity sources.

Disentangling the evolution and growth of AGN responsible for the observed CXB spectrum (known as population synthesis) requires knowledge of the obscuring neutral hydrogen column density (\nh) distribution, which is predicted to co-evolve with accreting supermassive black holes (e.g., \citealt{Ueda14,Buchner15,Brandt15,Hickox18,Brandt22}). This is particularly important because the {\em majority} of AGN are known to be obscured, typically defined as having \nh\,$\gtrsim 10^{22}$\,\nhunit \citep{Risaliti99b,Burlon11,Ricci15,Koss17}. Absorption can occur over a broad range of host-galaxy spatial scales, but the $\sim$\,parsec-scale obscuring torus\footnote{Throughout this paper, our use of the term \lq torus\rq\ does not refer to the specific geometry or shape of the obscuring structure. We use this term generically as a label to describe the anisotropic equatorial obscuring structure that gives rise to the optical type~1/2 dichotomy of AGN.} of AGN unification schemes plays a key role in AGN detectability and classification (e.g., \citealt{Antonucci93,Urry95,Alexander12,Brandt15,Netzer15,RamosAlmeida17b,AlonsoTetilla24}). Hard X-ray ($E>$\,10\,keV) photons rarely interact with obscuring gas of \nh\,$\lesssim10^{23.5}$\,\nhunit (see e.g., Figure~1 of \citealt{Boorman18}). Hard X-ray observations are thus a very effective means of sampling AGN populations, unbiased by mild obscuration (\nh\,$\lesssim10^{24}$\,\nhunit; e.g., the \textit{BeppoSAX} High-Energy Large Area Survey: \citealt{Fiore98}, the \integral AGN sample: \citealt{Malizia09}, the \bat Surveys: \citealt{Ricci17_bassV}, the \nustar Serendipitous Surveys: \citealt{Lansbury17,Greenwell24_nss80}).

\begin{figure*}
\centering
\includegraphics[width=0.99\textwidth]{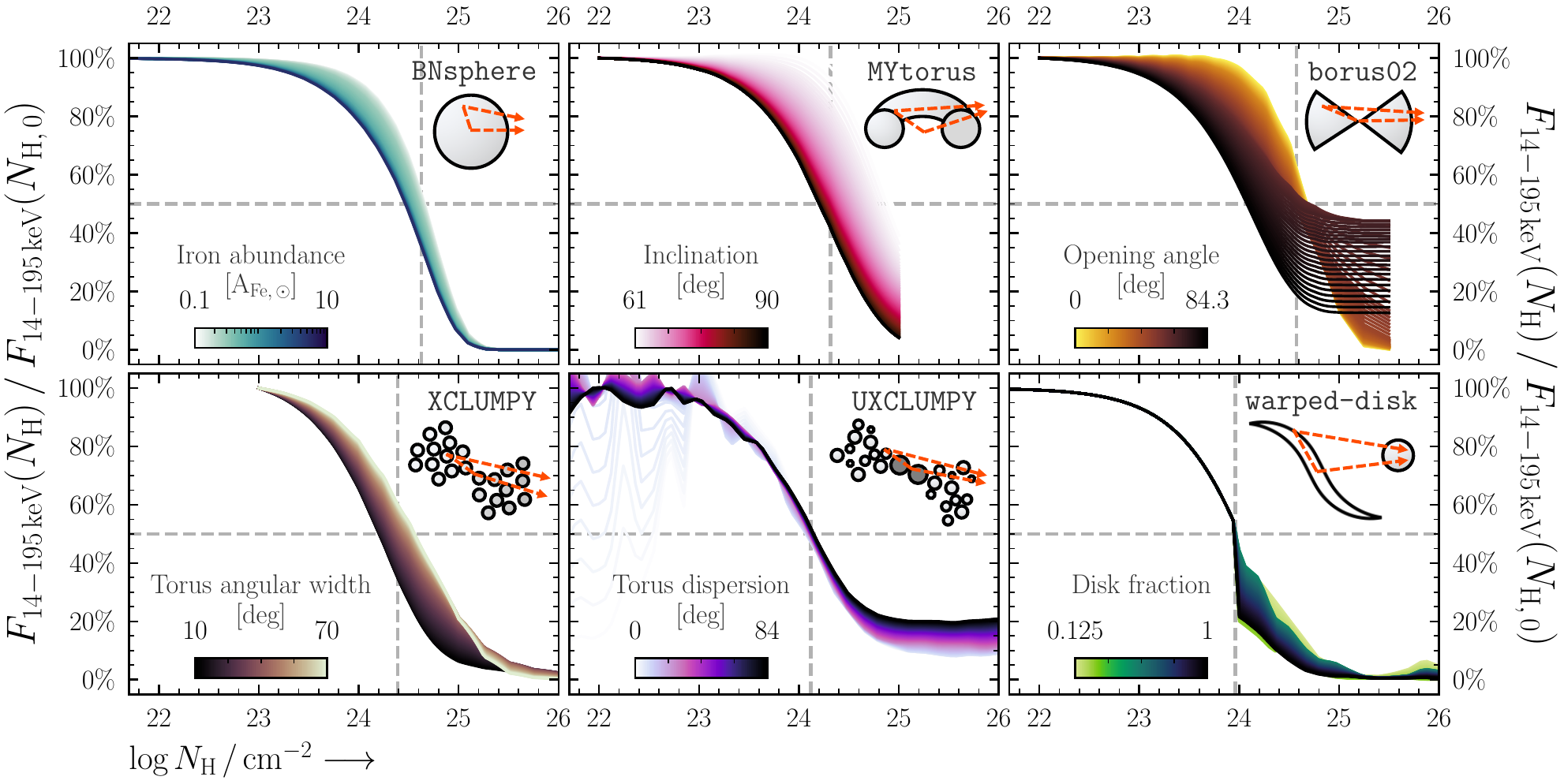}
\caption{\label{fig1_batflux_ratios} For a given line-of-sight column density, each panel shows the fraction of detected hard 14\,--\,195\,keV flux relative to the escaping flux at the minimum column density allowed by each model ($N_{\rm H,\,0}$). The top right of each panel shows a cartoon for the geometry assumed in each model. For visual clarity, the approximate column density corresponding to 50\% depletion is shown with dashed lines for each model, showing a substantial drop in detected hard X-ray flux at high column densities ($N_{\rm H}$\,$\gtrsim$\,$10^{24.5}$\,cm$^{-2}$). We additionally show the spread in predicted depletion factors relative to an additional parameter per model. From the upper left to lower left panels in a clockwise direction, the models are: \texttt{BNsphere} \citep{Brightman11b}, 
coupled \texttt{MYtorus} \citep{Murphy09}; coupled \texttt{borus02} \citep{Balokovic18}; \texttt{warped-disk} \citep{Buchner21b}; \texttt{UXCLUMPY} \citep{Buchner19}; \texttt{XCLUMPY} \citep{Tanimoto18}. Note for \texttt{warped-disk}, the disk fraction corresponds to the warp extent where 0 and 1 represent flat and strong warp geometries, respectively. To reproduce $N_{\rm H}$\,$<$\,$10^{24}$\,cm$^{-2}$ fluxes accurately with \texttt{warped-disk}, we freeze \texttt{NHLOS} to 0.01, and include an optically-thin absorber separately -- see the \texttt{warped-disk} documentation for more information.}
\end{figure*}

For column densities in excess of the inverse Thomson cross-section (\nh\,$\sim 1.5\times10^{24}$\,\nhunit), gas becomes optically thick to Compton scattering and is referred to as Compton-thick\footnote{The definitive \nh\ column density threshold for the Compton-thick regime depends on additional factors such as elemental abundances -- see Section~2 of the \mytorus manual, available at: \url{http://mytorus.com/mytorus-manual-v0p0.pdf}.}. The directly transmitted component of X-ray flux is diminished via photo-electric absorption $\lesssim$\,10\,keV with Compton scattering dominating above $\gtrsim$\,10\,keV (\citealt{Lightman88,Reynolds99}). At \nh$\gg10^{25}$\,\nhunit the transmitted/unabsorbed fraction of hard X-ray photons through the obscuration is negligible relative to the Compton-scattered photons (see Figure~\ref{fig1_batflux_ratios}). Instead, the observed flux is almost entirely due to scattering of photons off the optically-thick surface of circumnuclear material directed into the line-of-sight. These \lq reprocessing-dominated\rq\footnote{Here, \lq reprocessing\rq\ refers to scattered emission through any angle. For this reason, we refer to X-ray reprocessing and scattering interchangeably in this work.} Compton-thick AGN are the most difficult population to detect in X-ray surveys (e.g., \citealt{Matt00,Comastri15,Padovani17}), and even in the very local universe, X-ray flux-limited surveys are biased against Compton-thick AGN detection (e.g., Figure~3 of \citealt{Ricci15}, \citealt{Koss16b}, Annuar et al., in prep.). Compounding these detection/selection biases is the issue of source characterization/classification (see Section~\ref{sec:representative}). Additional confusion can arise from comparatively strong reprocessed components from the intrinsic accretion disk vs. typical torus geometries (e.g., \citealt{Gandhi07,Treister09,Vasudevan16,AvirettMackenzie19}). Conversely, heavily obscured AGN signatures may be swamped by other spectral components resulting in erroneous low-column estimates in the case of low signal-to-noise or band-limited X-ray data (as demonstrated in e.g., \citealt{Civano15,Gandhi17,Boorman18}).

The result is that the fraction of Compton-thick AGN remains uncertain, and hotly debated. X-ray surveys of the local universe typically find an observed Compton-thick fraction of $\lesssim15$\% (e.g., \citealt{Masini18,Georgantopoulos19,TorresAlba21_clemsonVI}), and is increased to intrinsic fractions of $\sim$10\,--\,30\% after applying X-ray-specific bias corrections (e.g., \citealt{Burlon11,Ricci15,Koss16b}). But higher Compton-thick fractions are often discussed in the literature; for example, \citet{Ananna19} require 50\,$\pm$\,9\% of all AGN within $z$\,$\sim$\,0.1 to be Compton-thick based on observed survey flux distributions and the CXB. The local \nh distribution is the $z$\,=\,0 boundary condition imposed in CXB synthesis models (e.g., \citealt{Ueda14}), so accurate determination of the number of obscured and Compton-thick AGN locally is crucial. Independent selection strategies are required at different wavelengths that are not biased against detection of highly obscured AGN in the same way as X-ray flux-limited surveys.

\subsection{Representative AGN Selection}\label{sec:isotropic_selection}
Complementary methodologies for isotropic AGN selection\footnote{We use isotropic here to mean the selection of optically-classified type~1 and~2 AGN in an unbiased manner.} include (i) optical narrow-line emission, (ii) radio low-frequency surveys, (iii) mid-infrared narrow-line emission, and (iv) mid-to-far infrared continuum emission (e.g., reviewed in \citealt{Hickox18}). These techniques are effective because the unified scheme of AGN ascribes observed differences between AGN classes primarily to the orientation of the torus relative to the line-of-sight, whereas all the above methods probe AGN emission on much larger scales.
\begin{enumerate}

\item[(i)] {\em Optical narrow-line emission:} Originating in the narrow-line region, optical emission lines have been successfully used to select \lq optimally\rq\ matched samples of X-ray obscured and unobscured sources (e.g., \citealt{Risaliti99b,Kammoun20}), by treating forbidden transitions such as the [O{\sc iii}]\,$\lambda$\,5007\,\AA\ emission line as \lq bolometric\rq\ estimators of AGN power. The $\sim$\,kpc-scale extension of the narrow-line region means its emission should be relatively insensitive to $\sim$\,pc-scale torus orientation angles. While certainly an effective means of identifying AGN that are heavily obscured in X-rays, this technique requires intensive spectroscopic surveys of large sky areas. In addition, geometric variations, time variability and dust content can cause significant inherent scatter in the \oiii power, implying that it is not a strict and accurate estimator of the current bolometric AGN power (e.g., \citealt{Saunders89,Boroson92,Netzer93,Zakamska03,Berney15,Ueda15,Finlez22}). Additional complications would arise from selecting AGN encompassed by very high -- or even 4$\pi$ -- covering factors of obscuration and/or significant host galaxy obscuration, as the narrow-line region would not be illuminated at all (see discussion in \citealt{Boroson92,Goulding09,Koss10,Greenwell24} for more information). 

\item[(ii)] {\em Radio/mm interferometric surveys:} Radio continuum luminosity can serve as an isotropic indicator of the time-averaged intrinsic AGN power (e.g., \citealt{Wilkes13,Kuraszkiewicz21}). This is true for the unbeamed extended jet component, which is best probed at low frequencies (e.g., \citealt{Orr82,Giuricin90,Singh13,Gurkan19}), and increasing evidence suggests that radio jets are likely a common component of radiatively efficient accretion (e.g., \citealt{Jarvis19}). However, the majority of radio surveys to date have only been able to resolve large-scale jets, which so far have been found to be uncommon in radiatively efficient AGN (e.g., \citealt{Panessa16}). This means the source yield per unit volume with such surveys is currently typically small. However, ongoing surveys (with e.g., the LOw-Frequency ARray; LOFAR, Australian Square Kilometre Array Pathfinder; ASKAP, Meer Karoo Array Telescope; MeerKAT) as well as future surveys (with e.g., the Square Kilometer Array; SKA) will find ever-increasing numbers of radiatively efficient AGN with lower-power/small-scale jets, enabling the selection of statistically large samples of such AGN directly in the radio \citep{Jarvis07,Baldi18,Hardcastle19,Sabater19,Hardcastle20,Williams22,Igo24,Mazzolari24}. At low radio power, the observed radio emission may originate from non-jet sources \citep{Panessa19} and such future surveys will thus help to shed light on the large intrinsic scatter around the correlation between jet power and AGN luminosity (\citealt{Hickox14,Mingo16}), ultimately allowing improved determination of intrinsic AGN power from radio observations. Promising developments have also been made recently with high-frequency ($>$\,10\,GHz) radio emission, with indications that with sufficient sensitivity the compact corona-powered radio emission can be almost universally detected for both radio-loud and radio-quiet AGN even at extremely high line-of-sight column densities (e.g., $N_{\rm H}$\,$\sim$\,10$^{27}$\,cm$^{-2}$; \citealt{Smith20,Kawamuro22,Ricci23_alma}, Magno et al., in prep.).

\item[(iii)] {\em Mid-infrared line emission:} Complementary to optical lines, narrow-line emission can also be excited in the infrared by AGN activity. Commonly used lines include the [Ne{\sc v}]\,$\lambda$\,14/24\,$\mu$m and [O{\sc iv}]\,$\lambda$\,26\,$\mu$m emission lines, both of which have been adopted for isotropic AGN selection and tests of unification (e.g., \citealt{Goulding09,Gilli10,Dicken14,Yang15}). Their high ionization potentials imply that they are better than the optical lines at disentangling AGN activity from star-formation. Additionally, their low optical depths allow probing through high extinction levels from both the host galaxy and nuclear regions. However, intensive infrared spectroscopic surveys are even more sparse than in the optical, and these lines are subject to substantial scatter (e.g., \citealt{Melendez08,Rigby09,DiamondStanic09,LaMassa10,Asmus19,Cleri23,Barchiesi24}). Also even though the line fluxes are expected to be relatively immune to high column densities, the photo-ionising photons required to excite the lines can be obscured in high covering factor scenarios. Ongoing work with the \textit{James Webb Space Telescope} (\textit{JWST}; \citealt{Gardner23}) Mid-Infrared Instrument (MIRI; \citealt{Rieke15,Wells15,Wright15}) holds strong potential for probing the physical origin of various mid-infrared spectral features in nearby Seyfert AGN (e.g., \citealt{GarciaBernete24,Davies24,HermosaMunoz24,Haidar24}).

\item[(iv)] {\em Mid-to-far infrared continuum:} Infrared AGN continuum emission partly arises from dust-reprocessing in the torus, and is thus associated with much larger size scales than the $\sim$\,10$^{-5}$\,--\,10$^{-4}$\,pc-scale X-ray emitting corona. In addition, a significant infrared component sometimes arises on even larger scales from \lq polar\rq\ dust in the inner narrow-line region \citep{Hoenig13,Asmus16,Fuller19,Asmus19}. A combination of low absorption optical depth in the infrared, together with the extended physical scales results in the emission appearing largely isotropic of AGN optical type (e.g., \citealt{Buchanan06,Levenson09,Hoenig11}). This has been shown to be the case for optically classified type~1 and~2 AGN, including X-ray classified heavily obscured and Compton-thick AGN (e.g., \citealt{Gandhi09,Asmus15}). Longer (mid-to-far infrared) wavelengths appear to be more isotropic than the near-infrared (e.g., \citealt{Hoenig11,Mateos15}). Deviations from isotropy are mild in the mid-infrared, estimated to be a factor of $\lesssim$\,1.4 at 12\,$\mu$m, relative to the intrinsic 2\,--\,10\,keV X-ray emission \citep{Asmus15}. The scatter of the correlation between the infrared and X-ray powers is also relatively small, at $\approx$\,0.35\,dex (e.g., \citealt{Asmus15}). This is clearly a complex region with emission occurring over multiple nuclear scales, and there is much debate regarding its nature. For our purposes discussed below, the important factor is the emission isotropy from different AGN obscuration classifications, and we expand our definition of the torus to include the entire nuclear structure including the classical toroidal obscurer and any polar component. Whilst an almost isotropic selector of AGN type (e.g., \citealt{Alexander01}), aperture-dependent infrared continuum selection is not 100\% reliable, and contaminating host-galaxy emission in particular needs to be considered (e.g., \citealt{Lacy07,Stern12,Mateos12,Assef17,Hickox18,Asmus20}).

\end{enumerate}

While each of the above techniques provides a means of isotropic AGN selection to first order, each has associated advantages and disadvantages. So how should one quantify the effectiveness of any one strategy relative to another? This is best done by comparing sample properties according to multiple bolometric power indicators, which requires the construction of samples with several of the above indicators.

We aim to devise a survey strategy in a way that is more representative of the underlying range of AGN obscuration than X-ray flux-limited selection (see Figure~\ref{fig1_batflux_ratios}), especially in the extreme \nh regime. This work adopts infrared continuum selection. We are guided to this choice because (i) of the availability of legacy all-sky infrared imaging surveys, which (ii) allow collation of substantial AGN sample sizes with (iii) follow-up optical source classification already available. In this way, we can compare sample properties in the parameter space of infrared continuum, optical  emission line, and hard X-ray bolometric indicators. Finally, we require high-quality broadband X-ray spectral characterization for \nh measurement.

Our survey is the \textit{NuSTAR} Local AGN $N_{\rm H}$ Distribution Survey (NuLANDS), and is a continuation of the Warm \iras Sample \citep{deGrijp85,deGrijp87}. The sample derivation follows in Section~\ref{sec:sample}, with Section~\ref{sec:representative} presenting updated optical classifications to highlight its AGN type isotropy. Sections~\ref{sec:data},~\ref{sec:5_method} and \ref{sec:6_results} report the X-ray data analysis, method, and results of the \nustar Legacy Survey, respectively. We then present the NuLANDS\ \nh distributions in Section~\ref{sec:discussion}, before summarizing our findings in Section~\ref{sec:summary}. The cosmology adopted in this paper is $H_{0}$\,=\,67.3\,km\,s$^{-1}$\,Mpc$^{-1}$, ${\rm \Omega}_{\Lambda}$\,=\,0.685 and ${\rm \Omega}_{\rm M}$\,=\,0.315 \citep{Planck14}.

\section{The NuLANDS Sample}\label{sec:sample}

\begin{figure*}
\centering
\includegraphics[width=0.99\textwidth]{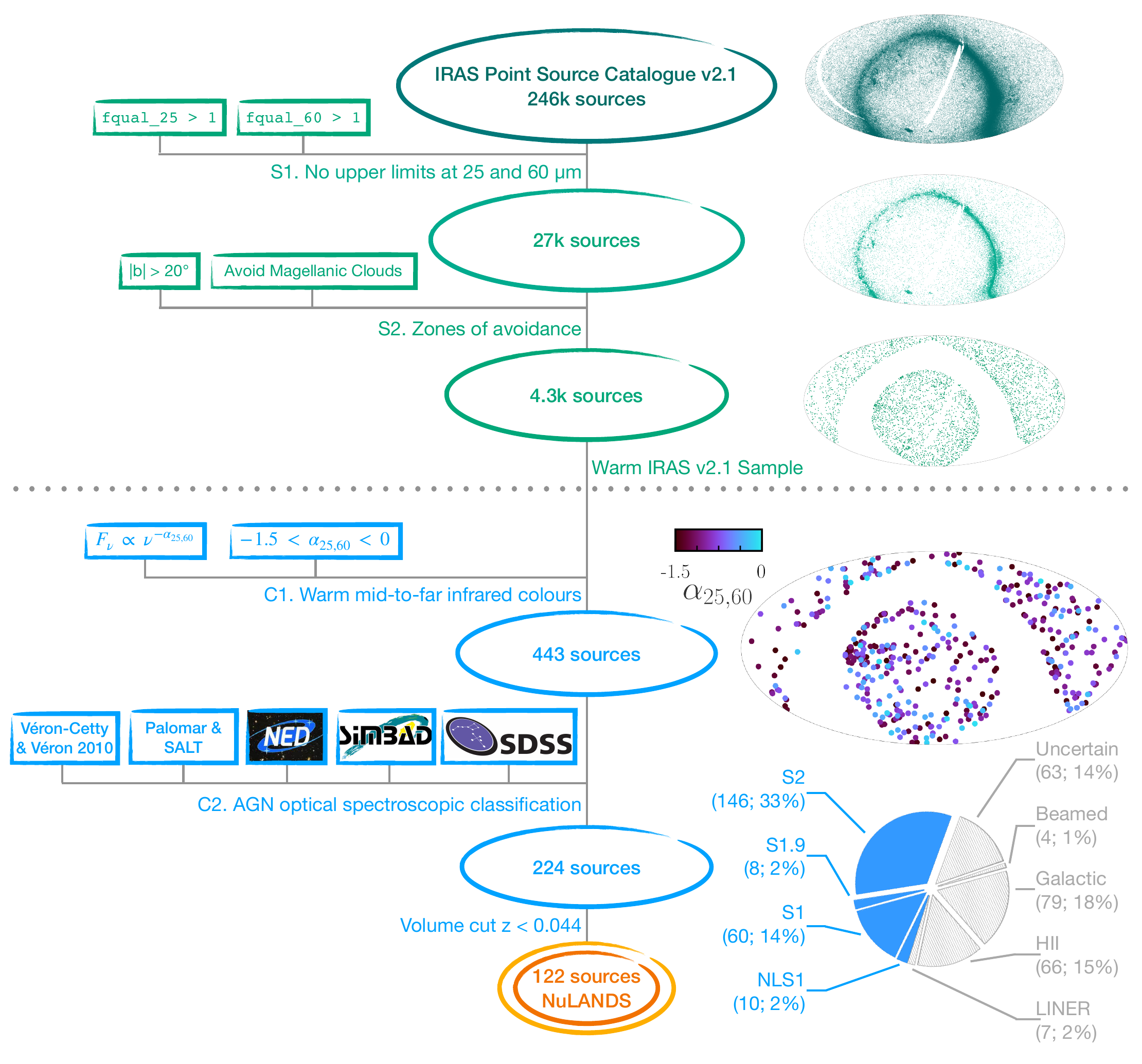}
\caption{\label{fig2_classification_flow} Flow diagram highlighting the selection and classification processes, starting from the top, used to generate the NuLANDS sample. The selection strategy, based on robust detections by \textit{IRAS} in 25 and 60\,$\mu$m, is shown above the horizontal dotted line, together with the corresponding skymaps of sources at each stage shown along the right. There are then 443 sources classified as having warm \textit{IRAS} colors by the cut of \citet{deGrijp85}, followed by 224 AGN confirmed via optical spectroscopic classifications. After finally adopting the volume cut of $z$\,$<$\,0.044, we arrive at the base NuLANDS sample for this paper which consists of 122 sources (87 type~1.9\,--\,2 and 35 type~1\,--\,1.8\,$+$\,Narrow Line Seyfert 1s).}
\end{figure*}

\subsection{Sample Collation}\label{subsec:selection}
The isotropy of mid-to-far infrared torus continuum emission lies at the heart of the NuLANDS selection strategy. Specifically, we use the studies of \citet{deGrijp85}, \citet[\citetalias{deGrijp87} hereafter]{deGrijp87} and \citet[\citetalias{Keel94} hereafter]{Keel94}, who selected objects from the \textit{InfraRed Astronomy Satellite} (\iras\ -- \citealt{Neugebauer84}) all-sky Point Source Catalog version~1 \citep{Beichman85}. \iras performed an all-sky (96\% of the total sky) survey at 12, 25, 60 and 100\,\micron, with positional uncertainties of 2\,--\,16\,arcsec depending on the scan mode. We update the selection method of \citetalias{deGrijp87} to the latest version 2.1 of the Point Source Catalog (PSCv2.1; 245,889 sources total). PSCv2.1 has sensitivity limits of $\sim$\,0.4, 0.5, 0.6 and 1\,Jy at 12, 25, 60 and 100\,\micron, respectively. Importantly, \citetalias{Keel94} showed that there is no significant difference between optical type~1 and~2 AGN in their sample, when comparing the narrow line region and infrared fluxes as bolometric indicators. This congruence is what motivated us to use their sample collation strategy as a starting point, and is as follows:

\begin{enumerate}

\item[S1.] {\em Detections at 25~and~60\,$\mu$m in the \iras PSCv2.1.} Detections at two wavelengths were required for source classification, discussed below, resulting in 27,090 sources.
\item[S2.] {\em Source coordinates restricted to Galactic latitude |$b$|\,>\,20$^{\circ}$ and outside the Magellanic Clouds.} To minimise confusion in dense stellar fields. The coordinate regions of the Magellanic Clouds were:
\begin{itemize}
\item[-] {\em Large Magellanic Cloud:}
\begin{center}
$69.8380^{\circ}<$\,R.\,A.\,[J2000]\,$<88.7937^{\circ}$\\
$-71.9044^{\circ}<$\,Dec.\,[J2000]\,$<-64.9940^{\circ}$
\end{center}
\item[-] {\em Small Magellanic Cloud:}
\begin{center}
$9.4859^{\circ}<$\,R.\,A.\,[J2000]\,$<21.0869^{\circ}$\\
$-73.7252^{\circ}<$\,Dec.\,[J2000]\,$<-71.7400^{\circ}$
\end{center}
\end{itemize}

These coordinate restrictions resulted in 4,253 sources. Assuming the \textit{IRAS} PSCv2.1 covered $\sim$\,96\% of the sky, we estimate that these selections cover $\sim$\,63\% of the sky which amounts to $\sim$\,26,000\,deg$^{2}$.
\end{enumerate}

While capable of isotropically selecting AGN, the above criteria will also pick up sources from a variety of other contaminant classes, including Galactic objects and star-forming galaxies, so the sample must be pruned to isolate AGN via classification. Thus we apply additional selection criteria to generate the warm \iras v2.1 sample as depicted in the upper part of Figure~\ref{fig2_classification_flow} and described in the next section.

\subsection{Sample Classification}\label{subsec:classification}
We follow the source classification of the original warm \iras sample performed by \citetalias{deGrijp87} and subsequently refined by \citet[\citetalias{deGrijp92} hereafter]{deGrijp92} as follows.

\begin{enumerate}
\item[C1.] {\em Warm color cut classification with \iras\ 25 to 60\,$\mu$m\ spectral index\footnote{$F_{\nu}\propto\nu^{-\alpha}$, where $F_{\nu}$ is the flux density, $\nu$ is the frequency and $\alpha$ is the spectral index.} $\alpha_{25,60}$ lying in the range $-1.5$\,$<$\,$\alpha_{25,60}$\,$<$\,$0$.\footnote{Corresponding to 0.27\,$<$\,$F_{60\,\mu{\rm m}}$\,/\,$F_{25\,\mu{\rm m}}$\,$<$\,1.}} This color cut favours the selection of AGN because star-formation-dominated galaxies are typically characterised by cooler dust temperatures of the interstellar medium ($\alpha_{25,60}\gtrsim-2.5$), peaking at $\gtrsim$\,50\,$\mu$m (e.g., \citetalias{deGrijp85,deGrijp87}; \citealt{Elvis94,Dale01,Mullaney11,Harrison14,Ichikawa19,Auge23}). In addition, selecting sources with $\alpha_{25,60}<0$ restricts the selection of particularly blue objects, such as stars with blackbody spectra that peak in the ultraviolet\,--\,optical wavebands. By propagating the uncertainties with Monte Carlo simulations from the flux densities in the \iras PSCv2.1, we select sources that have $-1.5$\,$<$\,$\alpha_{25,60}$\,$<$\,$0$ with $\geq$\,50\% probability. Applying this color cut resulted in a sample of 443 sources, which we refer to as the warm \iras v2.1 sample. We note that increasing the $\alpha_{25,60}$ probability threshold to $\geq$\,68\% would correspond to a reduction in the sample size of $\geq$\,12\%, though at the cost of removing AGN such as NGC\,424, NGC\,7469 and NGC\,2410. Only four sources are not in the warm \iras sample presented in \citetalias{deGrijp87}, which we match to \simbad as the following -- * Upsilon Pavonis (a Galactic source), IRAS\,05215--0352 (a far-IR source), and two galaxies: NGC\,5993 and NGC\,5520.

\item[C2.] {\em Optical spectroscopic emission-line diagnostic classification.} The warm \iras color cut alone is not fully reliable for AGN isolation, i.e., it can still include contaminants such as stars and star-forming galaxies that lie in the warm tail of the distribution of dust temperatures. Thus to further confirm the nature of the warm \iras v2.1 sample, we performed a literature search for optical spectroscopic classifications of all 443 sources. We collate optical spectroscopic classifications and redshifts for all sources where available from the NASA/IPAC Extragalactic Database (\ned), \simbad, the Sloan Digital Sky Survey (SDSS), \citet{VeronCetty10}, the 6dFGS catalogue, in addition to the optical classifications in \citetalias{deGrijp92}. For 14 sources, we have acquired updated Palomar DoubleSpec spectroscopy from new observations with the 200\,inch Hale telescope. The atlas of optical spectra for all sources will be provided in a future publication. Based on these classifications, we reject with high confidence 79 objects as local Galactic sources. A further 63 are found to have uncertain/ambiguous classifications or are listed as an unknown source of emission in \ned or \simbad. Of the 63 uncertain classifications, 34 were found to be Galactic sources by \citet{deGrijp92} with an additional source (IRAS\,05215--0352) found to be an \textit{IRAS} far-infrared source from \simbad. The remaining 28 likely extragalactic sources were found to all have a probable \textit{WISE} match, though 10 lacked a redshift and the remainder had conflicting optical spectroscopic classifications from \ned, \simbad or SDSS and no classification from \citet{VeronCetty10}.\footnote{Of the remaining 18 sources discussed, 12 had a redshift outside the main redshift cut employed for the sample analyzed in this work.} This leaves 301 confident extragalactic sources with an optical spectroscopic classification. We additionally exclude 66 HII galaxies (though note that these may still contain hidden AGN -- see e.g., \citealt{Goulding09} and Section~\ref{sec:representative}), as well as seven Low Ionisation Nuclear Emission Region (LINER) sources\footnote{Of the seven sources classified as LINERs, two lie outside the redshift cut of NuLANDS: 2MASX\,J00283779--0959532 and 2MASX\,J04303327--0937446. The remaining five are 2MASX\,J04282604--0433496, CGCG\,074--129, UGC\,12163, NGC\,1052 and NGC\,7213.} and four beamed sources\footnote{The four beamed sources are [HB89]\,0420--014, 3C\,345, B2\,1732+38A, UGC\,11130, and all lie beyond the redshift cut of NuLANDS.}. We exclude sources with an archival LINER classification from the X-ray analysis presented in this paper since such sources are thought to contribute very little to the overall CXB flux (see Figure~12 of \citealt{Ueda14}). This leaves 224 optically-confirmed Seyfert galaxies in the warm \iras v2.1 sample.

\end{enumerate}

{\em Volume cut.} Lastly, we applied a redshift cut of $z$\,$<$\,0.044 ($D$\,$\lesssim$\,200\,Mpc). This volume restriction was applied as a compromise between achieving sufficient sensitive hard X-ray follow-up with \textit{NuSTAR} whilst having a large enough sample to provide a significant constraint on the Compton-thick fraction in the \nh distribution. After applying this cut, we are left with 122 optically-confirmed warm \iras Seyfert galaxies for the NuLANDS sample analysed in this paper, which includes 87 type~1.9\,--\,2 and 35 type~1\,--\,1.8\,$+$\,Narrow Line Seyfert 1s.

A flow diagram illustrating the selection and classification stages leading to the NuLANDS sample is given in Figure \ref{fig2_classification_flow}.
\section{Sample Biases and Representativeness}\label{sec:representative}

Here, we estimate the effectiveness and any bias of NuLANDS in sampling the parameter space of isotropic luminosity indicators through their corresponding flux ratios. We emphasize that for a luminosity indicator to be isotropic here means that the observed flux with that indicator is largely unaffected by the AGN obscuration class (i.e., type~1~vs.~type~2). As such, our strategy is to derive analytical population distributions to specific flux ratio distributions for the type~1s~vs.~type~2s to quantify the level of isotropy in NuLANDS.

\begin{figure*}
\centering
\includegraphics[width=0.99\textwidth]{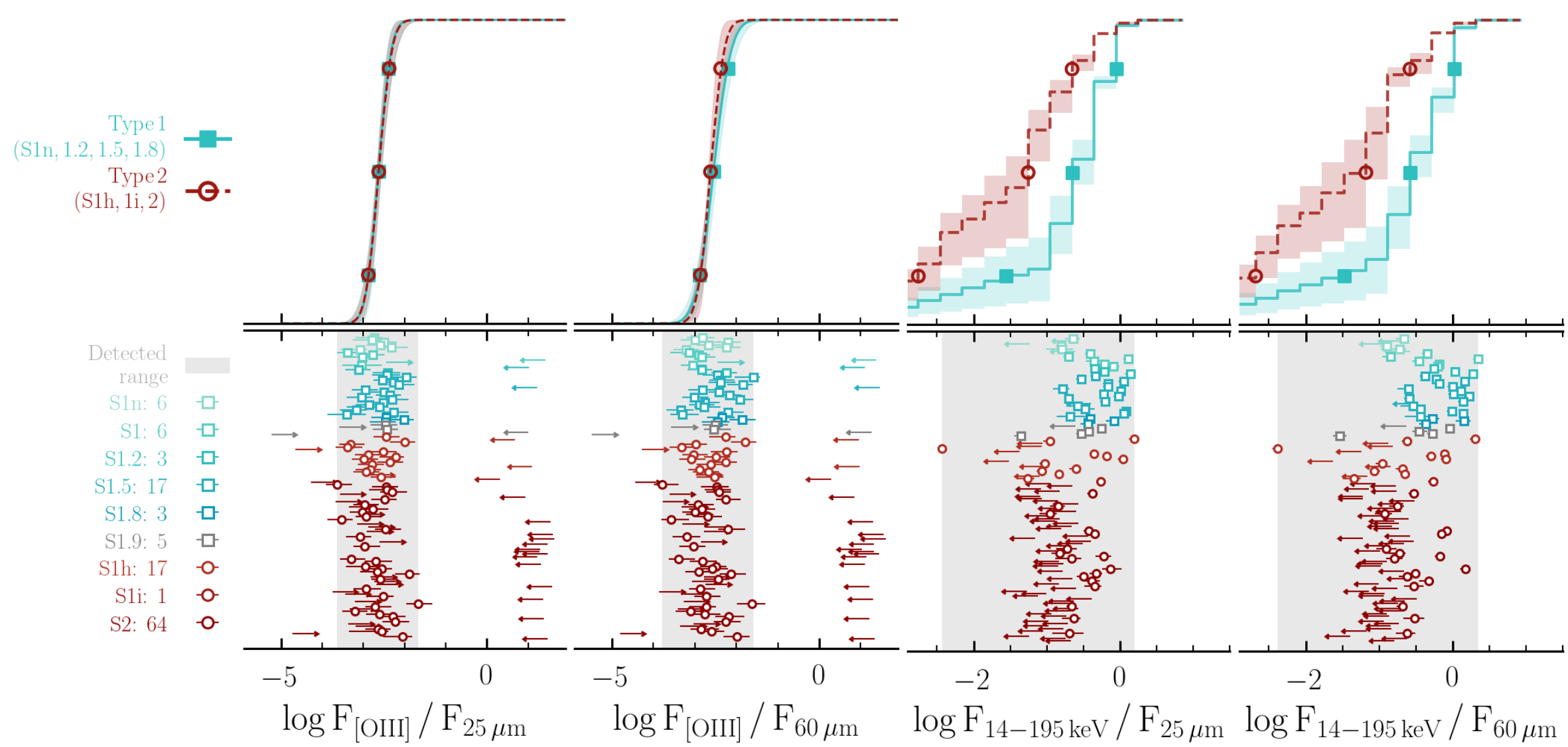}
\caption{\label{fig:representativeness} Flux ratio distributions segregated by AGN optical classification. For each plot (left to right), the lower and upper panels show the individual source flux ratios and cumulative distribution function for the parent population from which the measured flux ratios and error bars are consistent with being drawn from, respectively. The 16th, 50th, and 84th percentiles are marked on each cumulative distribution with large circles/squares for optically obscured/unobscured (i.e., type~1/type~2) objects, respectively. The mid-to-far infrared flux ratios with \oiii (left two panels) have almost identical parent distributions between the type~1 and type~2 AGN. In contrast, the flux ratios between hard 14\,--\,195\,keV X-ray flux from the 105-month BAT survey and the mid-to-far infrared in the right two panels show considerably different best-fit distributions.}
\end{figure*}

Combining the uncertainties on individual flux measurements to place constraints on the parent flux ratio distribution is a problem ideally suited for Hierarchical Bayesian modelling. Traditionally, this would involve solving the per-object constraints simultaneously to the parameters of the parent population, which would be a very high-dimensional problem. To address the issue of dimensionality, we follow the approach of \citet{Baronchelli20} with importance sampling to solve the problem numerically. Our process is as follows: (i) we calculate flux ratios and corresponding uncertainties for \oiii\,/\,60\,$\mu$m, \oiii\,/\,25\,$\mu$m, 14\,--\,195\,keV\,/\,60\,$\mu$m and 14\,--\,195\,keV\,/\,25\,$\mu$m with 25 and 60\,$\mu$m fluxes from the \textit{IRAS} PSC v2.1, \oiii fluxes from \citet{deGrijp92} and 14\,--\,195\,keV observed X-ray fluxes from the 105-month \textit{Swift}/BAT survey \citep{Oh18}. Individual flux measurement probability distributions were estimated as either a two-piece Gaussian distribution (to allow asymmetric error bars) or uniform distributions for limits. We assigned 0.3\,dex uncertainties to all \oiii flux measurements and assigned upper limits of 10$^{-9}$\,erg\,s$^{-1}$\,cm$^{-2}$ for all sources lacking an \oiii measurement. Hard X-ray non-detections were assigned an upper limit corresponding to the 90\% sky sensitivity upper limit of the 105-month \bat survey ($F_{14-195\,\textrm{keV}}$\,$<$\,$8.40\times10^{-12}$\,\fluxunit). The resulting flux ratio values of each source are plotted in the lower panel of Figure~\ref{fig:representativeness}. (ii) The flux ratios were segregated by their optical classifications and 1000 points were sampled from the distribution associated with each value. (iii) We then used UltraNest \citep{Buchner21_ultranest} to sample the likelihood associated with a number of different parent population models for the logarithmic flux ratio distributions of type~1 (defined as any Narrow Line Seyfert 1; S1n, S1.2, S1.5, S1.8) vs. type~2 (any S1i, S1h, S2)\footnote{S1i and S1h refer to sources that are type~2 AGN with broad Paschen lines observed in the infrared or broad polarized Balmer lines identified. See \citet{VeronCetty10} for more information on these classifications.} AGN separately. The resulting cumulative population distributions for the logarithmic flux ratios are shown in the upper panel of Figure~\ref{fig:representativeness}.

\subsection{[O\,\textsc{iii}] to Infrared Flux Ratios}\label{subsec:fluxratios}
The two left-hand plots in Figure~\ref{fig:representativeness} show the \oiii to infrared (25~and~60\,$\mu$m) flux ratio distributions segregated by AGN optical classification. After testing a variety of parent population models, we used a Gaussian parent model to describe the \oiii\,/\,60\,$\mu$m and \oiii\,/\,25\,$\mu$m flux ratio distributions, since the results with a Student's $t$ distribution, asymmetric Gaussian and Gaussian with a constant outlier contribution were all consistent with the results from a symmetric single Gaussian distribution. The parameters describing each parent distribution were the mean $\mu$ logarithmic flux ratio and standard deviation $\sigma$, which were assigned uniform priors $\mathcal{U}(-5, 5)$ and log\,$\mathcal{U}(-4, 2)$, respectively. The parent Gaussian distribution fit results are shown in Table~\ref{tab:flux_rat}. As was found by \citetalias{Keel94}, we quantitatively show that type~1 AGN are indistinguishable from type~2 AGN when comparing the narrow line region (traced by \oiii) to infrared continuum fluxes. We note that Figure~\ref{fig:representativeness} compares the \textsc{[oiii]}-to-infrared flux distributions between the type~1s and type~2s with reddened \textsc{[oiii]} fluxes. However, the Balmer decrements reported by \citet{deGrijp92} for the 77/122 NuLANDS targets with optical H$\alpha$, H$\beta$ and \oiii line flux detections were also well-matched between the type~1s and type~2s with medians and encompassing 68\% interquartile ranges of 0.69$^{+0.12}_{-0.13}$ and 0.72\,$\pm$\,0.13, respectively. As a consistency check, in Table~\ref{tab:flux_rat} we report the population parameter values for the 77/122 NuLANDS sources after correcting the \oiii fluxes for reddening with the de-reddening relation reported by \citet{Bassani99}. As expected, the de-reddened flux ratios are higher than the reddened values. However, both the \oiii\,/\,60\,$\mu$m and \oiii\,/\,25\,$\mu$m flux ratios are still entirely consistent between the type~1 and type~2 AGN, reinforcing our finding that the NuLANDS sample selection is isotropic.

\begin{deluxetable}{ccc}
\tablewidth{0pt}
\tablecaption{Parent Gaussian distribution parameters fit to the distribution of flux ratios amongst the type~1 and type~2 AGN. The lower portion reports the population parameter values found for the 77/122 sources with optical H$\alpha$, H$\beta$ and \oiii line flux detections after correcting the \oiii fluxes for reddening using the \citet{Bassani99} relation.\label{tab:flux_rat}}
\tablehead{
\colhead{Parameter} &
\colhead{log\,([O\,\textsc{iii}]\,/\,25\,$\mu$m)} &
\colhead{log\,([O\,\textsc{iii}]\,/\,60\,$\mu$m)}
}
\tableheadfrac{0.1}
\startdata
\hline
\multicolumn{3}{c}{Reddened \oiii sample (122 sources total)}\\
\hline
Type~1 $\mu$ & $-2.63$\,$\pm$\,0.07 & $-2.62$\,$\pm$\,0.06\\
Type~2 $\mu$ & $-2.63$\,$\pm$\,0.06 & $-2.55$\,$\pm$\,0.08\\
Type~1 log\,$\sigma$ & $-0.63$\,$\pm$\,0.14 & $-0.67$\,$\pm$\,0.28\\
Type~2 log\,$\sigma$ & $-0.61$\,$\pm$\,0.12 & $-0.48$\,$\pm$\,0.12\\[0.1cm]
\hline
\multicolumn{3}{c}{De-reddened \oiii sample (77 sources total)}\\
\hline
Type~1 $\mu$ & $-2.05$\,$\pm$\,0.09 & $-1.97$\,$\pm$\,0.10\\
Type~2 $\mu$ & $-1.94$\,$\pm$\,0.08 & $-1.92$\,$\pm$\,0.09\\
Type~1 log\,$\sigma$ & $-0.40^{+0.09}_{-0.10}$ & $-0.32$\,$\pm$\,0.08\\
Type~2 log\,$\sigma$ & $-0.33$\,$\pm$\,0.07 & $-0.30$\,$\pm$\,0.07\\[0.1cm]
\hline
\enddata
\tablecomments{For each flux ratio, $\mu$ and $\sigma$ represent the mean and standard deviation of the parent Gaussian distribution, respectively. For information regarding the parent model fitting, see Section~\ref{subsec:fluxratios}.}
\end{deluxetable}

\subsection{Observed Hard X-ray to Infrared Flux Ratios}
The individual flux ratio values for 14\,--\,195\,keV\,/\,60\,$\mu$m and 14\,--\,195\,keV\,/\,25\,$\mu$m shown in the lower panels of the two right-hand columns of Figure~\ref{fig:representativeness} illustrate a large number of hard X-ray flux upper limits that do not agree with the approximate distributions of the detected points. Thus, we could not use a Gaussian parent model and instead used a histogram model with a Dirichlet prior\footnote{See the description of the model here: \url{https://github.com/JohannesBuchner/PosteriorStacker}.} to fit the parent distribution. The histogram model is much more flexible than a single analytic model since every histogram bin is a free parameter and can vary independently of one another with the self-consistent requirement that all bin heights sum to unity.

The upper panels of the two right-hand plots show the corresponding parent cumulative distribution functions derived individually for type~1 and type~2 AGN. The observed type~1 vs. type~2 distributions are considerably different, with type~2 AGN skewed towards considerably lower observed \bat fluxes.

\subsection{Other Important Factors}
These tests imply that the NuLANDS sample is well matched between AGN classes in terms of the optical \oiii$\lambda$5007 narrow line with the infrared 60\,$\mu$m bolometric luminosity indicators, but {\em not} in terms of observed hard X-ray flux. Under the orientation-based unification scheme of AGN tori, such a hard X-ray deficit is expected if the hard X-ray fluxes are diminished due to line-of-sight obscuration in heavily obscured and Compton-thick nuclear tori which does not affect the narrow line region optical classification. Any positive bias in terms of measured hard X-ray flux from unobscured AGN due to Compton scattering in the obscurer (e.g., \citealt{Sazonov15}) would only exasperate the flux ratio offset. NuLANDS can detect candidate heavily obscured AGN missed by hard X-ray flux-limited selection. However, there are several important issues to consider if one is to place these results in proper context, as discussed below.

\begin{enumerate}
    \item[1.] {\em Incompleteness:} NuLANDS is not designed to provide a {\em complete} flux or volume-limited sampling of AGN. Firstly, the optical spectroscopic classifications of the Warm \iras 2.1 Sample are themselves incomplete, with $\sim$\,13\% of sources optically unclassified or ambiguous. Instead, we aim to collate a sample that is as representative of AGN circumnuclear obscuration as possible, in the sense of minimizing the impact of selection and classification effects that preferentially favor or disfavor sources at any given \nh affecting the nuclear X-ray emission.
    \item[2.] {\em Infrared-faint AGN:} There is a class of \lq hot-dust-poor\rq\ AGN with observed mid-infrared emission lower than expected from their near-infrared fluxes, suggestive of low torus covering factors (e.g., \citealt{Hao10}). Such sources could be missed from our warm infrared color selection (see C1 in Section~\ref{subsec:classification}). However, \citeauthor{Hao10} find that this class comprises just $\sim$\,6\% of the AGN population at $z$\,$<$\,2. In the local Universe, the prevalence of these AGN remains unclear, with only one AGN---NGC\,4945---known to show a significant infrared deficit relative to the canonical infrared--X-ray luminosity relation \citep{Asmus15}. Another potential source, though with a milder deficit, is NGC\,4785 \citep{Gandhi15a}. Interestingly, both sources are Compton-thick AGN, though these small numbers are insufficient to indicate a strong bias.
    \item[3.] {\em Host Galaxy Biases:} Two potential host galaxy biases are most relevant here:
    \begin{enumerate}
    \item[(i)] {\em Infrared Classification:} The warm \iras color cut will favor infrared-bright AGN that can dominate above host-galaxy emission, especially at 25\,$\mu$m. Conversely, AGN that are infrared-weak relative to stellar emission will end up being classified as having \lq cool\rq\ infrared colors and will drop out from the sample. Under the torus-based unification scheme, this bias should act uniformly for both Type~1 and~2 classes and be independent of \nh.
    
    However, there is evidence (albeit controversial) suggesting that obscured AGN preferentially occur in star-forming galaxies (e.g., \citealt{Villarroel17,Andonie24}, though see \citealt{Zou19} for contradictory findings on star formation rate). If so, this bias would imply a higher intrinsic prevalence of X-ray obscured AGN in \lq infrared--cool\rq\ systems, which NuLANDS would preferentially miss.
    
    \item[(ii)] {\em Optical Classification:} In a similar fashion, optical AGN classification requires the presence of AGN permitted and/or forbidden lines that can stand out above the host galaxy continuum. Host emission flux can swamp such narrow line region tracers, rendering them harder to detect (e.g., \citealt{Moran02,Goulding09,Jones16}). Both type~1 and type~2 AGN could be impacted by this bias, depending upon spectroscopic signal to noise and the fraction of host galaxy vs. AGN flux captured through the spectroscopic aperture. However, recent work suggests that dilution cannot solely explain AGN misclassification satisfactorily \citep{Agostino19}. For example, large-scale host galaxy dust extinction would preferentially adversely influence the detection of narrow lines, with sources ultimately being classified as inactive, weakly active (e.g., LINER) or as HII galaxies (e.g., \citealt{Rigby06}). Including such sources in the primary NuLANDS sample would skew the \oiii to infrared flux ratios of type~2s to lower values than type~1s, and the classes would no longer be well matched. The \oiii-to-infrared flux ratios in Figure~\ref{fig:representativeness} demonstrate that our sample does not suffer from this bias significantly.
    \end{enumerate}
\end{enumerate}

The classification biases discussed above would all act towards preferentially missing type~2 (and possibly also Compton-thick) AGN. If so, our results below should be taken as providing a lower limit to the intrinsic obscured fraction.
\section{X-ray Data}\label{sec:data}
Overall, we identify 122 optically-classified Seyferts in the warm \iras v2.1 sample within $z$\,$\lesssim$\,0.044 ($D_{\rm L}\,\lesssim\,200$\,Mpc), which form the core sample for our study. X-ray data for 102 of these are analyzed in this paper, using soft X-ray data from \textit{Swift}, \textit{XMM-Newton}, \textit{Suzaku} or \textit{Chandra}; 84 of these sources have \textit{NuSTAR} observations as of January 2021.
The X-ray spectral analysis was carried out by combining \nustar data (Section~\ref{subsubsec:nustar}), where available, with \xrt (Section~\ref{subsubsec:xrt}), \xmm (Section~\ref{subsubsec:xmm}), \chandra (Section~\ref{subsubsec:chandra}) and \xis (Section~\ref{subsubsec:suzakuxis}). 

\subsection{X-ray Data Acquisition}\label{subsec:xraydataselection}
The primary aim of this paper is to use \nustar's unique sensitivity above 10\,keV to enable the direct measurement of the \nh distribution in the local universe unbiased by high columns. This was our primary consideration when extracting data for X-ray spectral fitting; namely to have all secondary soft X-ray observations relative to uniquely sampled \nustar observations. To do this, we carefully selected the longest \textit{NuSTAR} observation per source that was available with quasi-simultaneous (less than $\sim$1\,day where possible) soft X-ray data. The quasi-simultaneity was incorporated to minimise the effects of flux and spectral variability. All sources with \textit{NuSTAR} data had quasi-simultaneous soft X-ray observations available at the time of analysis. For sources in our sample without \nustar data, we searched for the longest X-ray exposure from \xrt, \xmm, \xis, or \chandra.

\subsubsection{\nustar}\label{subsubsec:nustar}

The \textit{Nuclear Spectroscopic Telescope ARray} (\nustar; \citealt{Harrison13}) is the first and currently only hard X-ray imaging telescope in orbit capable of focusing hard X-ray photons with energies in the range $\sim$3--79\,keV. The \nustar data for both Focal Plane Modules (FPMA and FPMB) were processed using the \nustar Data Analysis Software (\textsc{nustardas v1.9.2}) package within \textsc{HEAsoft v6.28}. We checked the South Atlantic Anomaly reports for both FPMA and FPMB per observation to choose filters for optimising the background level. The task \textsc{nupipeline} was then used with the corresponding \textsc{caldb v20200712} files and our selected filters to generate cleaned events files. Spectra and response files for both FPMs were generated using the {\sc nuproducts} task for circular source regions with 20 pixels\,$\approx$\,49\farcs2 radii. Background spectra were extracted from off-source circular regions as large as possible on the same detector as the source while avoiding serendipitous sources or regions of greater background flux. Initially, we adopted the same source coordinates as reported in the AllWISE Source Catalog\footnote{All sources were matched to \textit{WISE} through visual inspection of the \textit{WISE} images available at \url{http://irsa.ipac.caltech.edu/Missions/wise.html} and also ESA Sky at \url{https://sky.esa.int/esasky/}.} before manually recentering the source extraction region by eye to account for any astrometric offsets. All offsets were within the typical values found for \textit{NuSTAR} relative to \textit{Chandra} in the \textit{NuSTAR} Serendipitous Survey (see Figure~4 of \citealt{Lansbury17}).

\subsubsection{\swift}\label{subsubsec:xrt}
A total of 53 sources in NuLANDS had snapshots from the \textit{Swift} \citep{Gehrels04} X-ray Telescope (XRT; \citealt{Burrows04}) to provide sensitive soft X-ray constraints down to $\sim$0.3\,keV (note four of these observations were analyzed without \nustar). For each observation, we run \texttt{xrtpipeline} to create cleaned XRT event files, which were then used to create images with \texttt{XSELECT}. Source spectra were then extracted from circular regions of 50$^{\prime\prime}$, and background spectra from annular regions of inner/outer radii of 142/260$^{\prime\prime}$. Both regions were manually re-sized to ensure no obvious contaminating sources were present. Effective area files were then created with \texttt{xrtmkarf} and the corresponding recommended response matrix was copied from the relevant \texttt{CALDB} directory.

To increase the signal-to-noise ratio in four sources (ID\,37: 2MASX\,J01500266--0725482, ID\,72: NGC\,1229, ID\,141: 2MASX\,J04405494--0822221 and ID\,263: KUG\,1021+675), we co-added all available spectra per source using the online \xrt Products tool.\footnote{Available from: \url{http://www.swift.ac.uk/user_objects/index.php}.}

\subsubsection{\xmm}\label{subsubsec:xmm}
The \xmm \citep{Jansen01} EPIC/PN \citep{Struder01} data were analyzed using the Scientific Analysis System (\textsc{sas}; \citealt{Gabriel04}) \textsc{v.16.0.0}. Observation Data Files were processed using the \textsc{sas} commands \textsc{epproc} to generate calibrated and concatenated events files. Intervals of background flaring activity were filtered via a 3$\sigma$ iterative procedure and visual inspection of the light curves in energy regions recommended in the \textsc{sas} threads.\footnote{For more information, see \url{https://www.cosmos.esa.int/web/xmm-newton/sas-thread-epic-filterbackground}.} Corresponding images for the PN detector were generated using the command \textsc{evselect}, and source spectra were extracted from circular regions of radius 35$^{\prime\prime}$. Background regions of similar size to the source regions were defined following the \xmm Calibration Technical Note XMM-SOC-CAL-TN-0018 \citep{Smith16}, ensuring the distance from the readout node was similar to that of the source region, which in turn ensures comparable low-energy instrumental noise in both regions. EPIC/PN source and background spectra were then extracted with \textsc{evselect} in the PI range 0--20479\,eV with patterns less than 4. Finally, response and ancillary response matrices were created with the \textsc{rmfgen} and \textsc{arfgen} tools. We use \xmm/PN spectra for a total of 36 sources, 12 of which lack \nustar data.

To improve the computation time associated with simultaneously fitting arbitrarily complex X-ray models to many spectra, we do not include the less sensitive EPIC/MOS \citep{Turner01} data in our spectral fitting.

\subsubsection{\chandra}\label{subsubsec:chandra}
The \chandra \citep{Weisskopf00} Advanced CCD Imaging Spectrometer (ACIS; \citealt{Garmire03}) data were reduced using \ciao~v4.11 \citep{Fruscione06} following standard procedures. Observation data were downloaded and reprocessed using the \textsc{chandra\_repro} command to apply the latest calibrations for \ciao and the \caldb. The level~2 events files were then used to create circular source and annular background regions centered on the source. The source regions were chosen to be 10$^{\prime\prime}$ in radius, and the background annuli were created to be as large as possible whilst still lying on the same chip as the source. Source, background and response spectral files were then generated with the \textsc{specextract} command in \ciao. We used \chandra/ACIS data for a total of five sources (four in conjunction with \nustar data).

\subsubsection{\xis}\label{subsubsec:suzakuxis}
Data from the \textit{Suzaku} \citep{Mitsuda07} X-ray Imaging Spectrometer (XIS; \citealt{Koyama07}) were used for a total of 8 targets (one of which did not use \nustar). First images in the 0.3\,--\,10\,keV energy range were made with \textsc{ximage}\footnote{https://heasarc.gsfc.nasa.gov/xanadu/ximage/ximage.html} by summing over the cleaned event files for each \xis camera. Next, source counts were extracted from a circular region of radius 3\farcm4, with background counts extracted from an annular region of inner radius 4\farcm2 and outer radius 8\farcm7. Exclusion regions were additionally created for any obvious sources in the corresponding images. \textsc{xselect} was then used to extract a spectrum for each XIS detector cleaned event file using the source and background regions defined above. To enable simultaneous background fitting of the \xis data, individual front-illuminated spectra were not co-added and we chose instead to fit only XIS3 for all sources. All XIS3 spectra in the energy range 1.7\,--\,1.9\,keV and 2.1\,--\,2.3\,keV were ignored due to instrumental calibration uncertainties (see Section~5.5.9 of the \suzaku ABC Guide).

\subsection{Observed signal-to-noise ratio}\label{sec:significance}
\begin{figure*}
\centering
\includegraphics[width=0.99\textwidth]{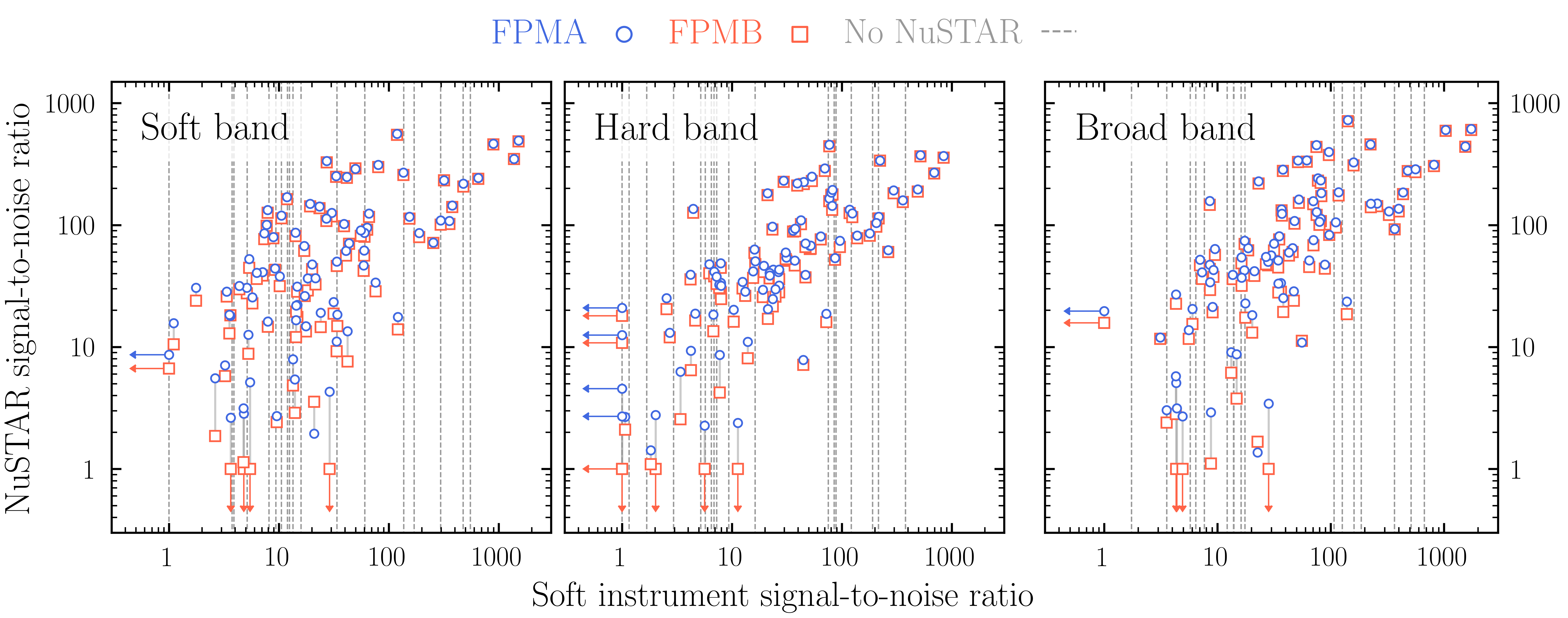}
\caption{\label{fig:significance} Observed signal-to-noise ratios for each of the 102 sources with soft X-ray data used in this work (84 of which had \textit{NuSTAR} data -- see Section~\ref{sec:data} for the full breakdown). Each panel shows the soft X-ray instrument vs. \textit{NuSTAR} signal-to-noise ratio in their respective soft (left panel), hard (middle panel) and broad (right panel) passbands. For the passband definitions per instrument, see Table~\ref{tab:instrument_bands}.}
\end{figure*}

To assess the fraction of sources with low signal-to-noise ratio X-ray data, we calculate the signal-to-noise ratio for each source in the soft, hard and broad bands per instrument (see Table~\ref{tab:instrument_bands} for instrument band definitions used in this work). We follow the formalism of \citet{Li83} which accounts for the Poisson nature of the source and background count values, implemented as the \texttt{poisson\_poisson} function in the \texttt{gv\_significance} library of \citet{Vianello18}.\footnote{\url{https://github.com/giacomov/gv_significance}} Figure~\ref{fig:significance} presents the signal-to-noise ratios for FPMA, FPMB and the soft data per source. No targets are found to have signal-to-noise ratios $<$\,1, with both the soft X-ray instrument and \nustar. We also find that on average, the \nustar/FPMA signal-to-noise ratio is higher than that for \nustar/FPMB. On inspection, we see that this may be caused by the shadow created from the optics bench systematically increasing the background on FPMB by a factor of $\sim$\,2 relative to FPMA for a given observation. For an in-depth analysis of the \nustar background, see \citet{Wik14}.

\startlongtable
\begin{deluxetable}{rcc}
% \tabletypesize{\scriptsize}
\tablewidth{0pt}
\tablecaption{Band definitions used for detection significance.\label{tab:instrument_bands}}
\tablehead{
\colhead{Instrument} &
\colhead{Band} &
\colhead{Energy range\,/\,keV}
}
\tableheadfrac{0.1}
\startdata
\hline
\multirow{3}{*}{\nustar} & Soft & 3--8\,keV\\%[0.2cm]
  & Hard & 8--78\,keV\\%[0.2cm]
  & Broad & 3--78\,keV\\%[0.2cm]
\hline
\multirow{3}{*}{\xrt} & Soft & 0.5--2\,keV\\%[0.2cm]
  & Hard & 2--10\,keV\\%[0.2cm]
  & Broad & 0.5--10\,keV\\%[0.2cm]
\hline
\multirow{3}{*}{\xmm/EPN} & Soft & 0.5--2\,keV\\%[0.2cm]
  & Hard & 2--10\,keV\\%[0.2cm]
  & Broad & 0.5--10\,keV\\%[0.2cm]
\hline
\multirow{3}{*}{\suzaku/XIS3} & Soft & 0.5--1.7 1.9--2\,keV\\%[0.2cm]
  & Hard & 2--2.1 2.3--10\,keV\\%[0.2cm]
  & Broad & 0.5--1.7 1.9--2.1 2.3--10\,keV\\%[0.2cm]
\hline
\multirow{3}{*}{\chandra/ACIS} & Soft & 0.5--1.2\,keV\\%[0.2cm]
  & Hard & 1.2--8\,keV\\%[0.2cm]
  & Broad & 0.5--8\,keV\\%[0.2cm] 
\hline
\enddata
\tablecomments{
For \textit{Suzaku}/XIS, we additionally ignored 1.7\,--\,1.9\,keV and 2.1\,--\,2.3\,keV due to calibration uncertainties associated with silicon and gold edges \footnote{\url{http://heasarc.gsfc.nasa.gov/docs/suzaku/analysis/abc/node8.html}}.
% $^{a}$\,$\mathcal{G}(\mu,\,\sigma)$ and $\mathcal{U}(\textrm{min}, \textrm{max})$ denote Gaussian and uniform priors, respectively. For Gaussian priors, we include the full parameter range in square brackets beneath each.
% %
% $^{b}$\,The cross-calibration constants were varied in log space from the \citealt{Madsen15} values, relative to FPMA when \nustar data was present. In the event that only soft data was available, only one dataset was fit and so no variable cross-calibration was used.
% %
}
\end{deluxetable}

\begin{figure*}
\centering
\includegraphics[width=0.99\textwidth]{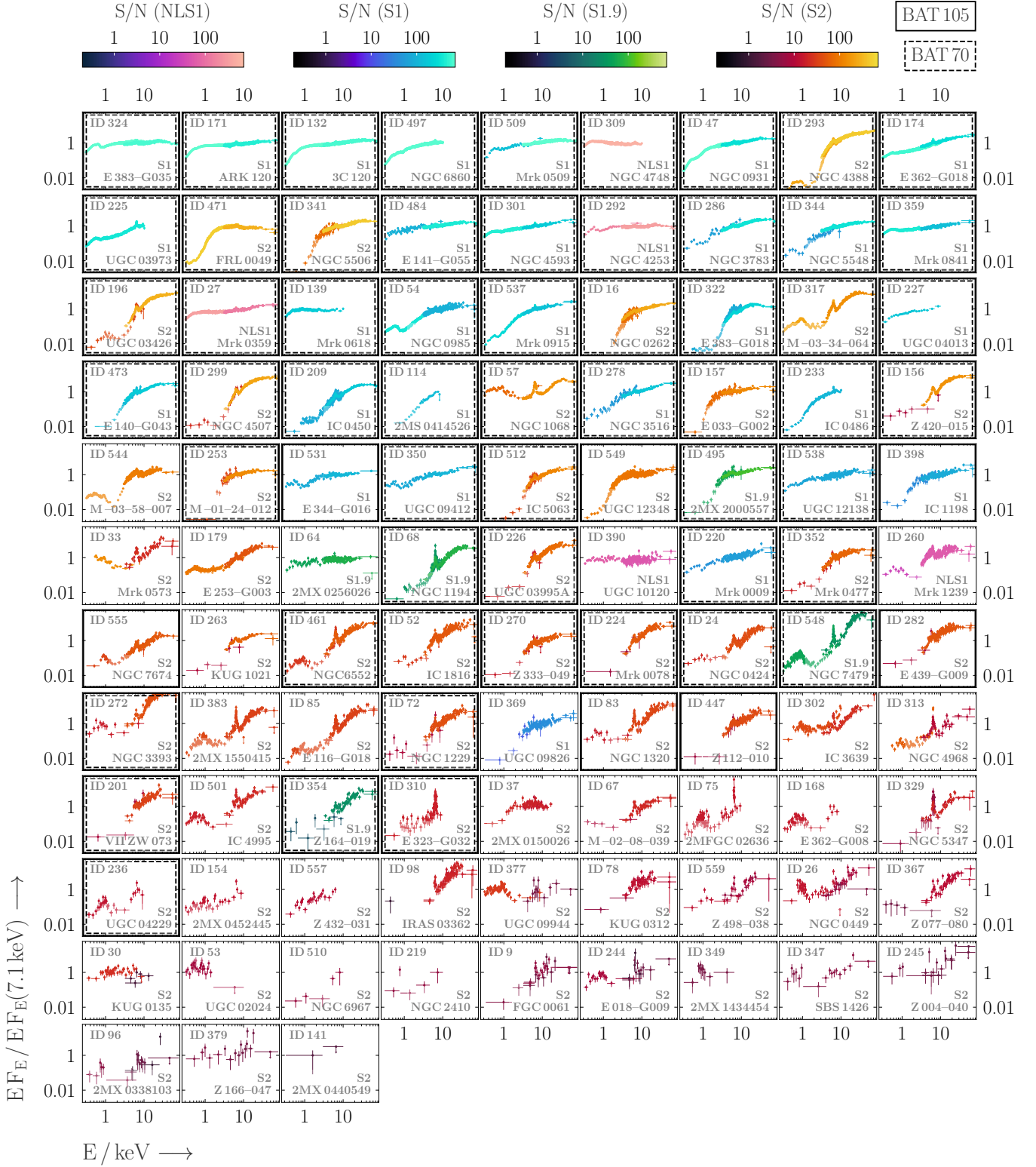}
\caption{\label{fig:allspec} X-ray spectra for every source in NuLANDS that were fit in this work. Each spectrum has been unfolded with a photon index\,=\,2 powerlaw and normalised by its own flux at 7.1\,keV for visual purposes. The spectra are also colored in terms of observed signal-to-noise ratio according to four different color schemes that depend on the optical spectroscopic classification of a given source (see color bars at the top of the figure and individual labels for spectral classifications of each source). Each source is bounded by an additional solid and/or dashed line dependent on detections in the 70-month or 105-month BAT catalogues, respectively. Since the sources are ordered by average broadband signal-to-noise ratio, it is clear that as the observed X-ray signal-to-noise ratio decreases, a significantly larger number of type~2 AGN are present with strong spectral signatures of reprocessing. Per panel, source names have been abbreviated such that `ESO'\,$\equiv$\,`E', `MCG'\,$\equiv$\,`M', `2MASSJ'\,$\equiv$\,`2MS', `2MASXJ'\,$\equiv$\,`2MX' and `CGCG'\,$\equiv$\,`Z'.}
\end{figure*}

Figure~\ref{fig:allspec} presents the spectra for all NuLANDS sources that were used for subsequent spectral fitting. The spectra have been unfolded with a photon index\,=\,2 powerlaw and normalised by the flux at 7.1\,keV for visual purposes. The spectra are additionally ordered by the average signal-to-noise ratio per source, showing that significantly more type~2 objects exist at lower overall observed X-ray signal-to-noise ratios. We also show sources detected in the 70-month and 105-month BAT catalogues with thick solid and dashed borders, respectively. Figure~\ref{fig:allspec} thus additionally highlights the complimentary nature of NuLANDS by identifying a higher proportion of obscured objects than hard X-ray flux-limited selection.

\section{X-ray Spectral Analysis}\label{sec:5_method}

\subsection{Spectral Fitting Procedure}
It is currently popular to explore the parameter space of X-ray spectral models with local optimisation algorithms such as Levenberg-Marquardt (\citealt{Levenberg44,Marquardt63}), which iteratively explores the parameter space from a pre-defined starting point. Such methods are not guaranteed to converge when using the Poisson statistic, for nonlinear models, when dealing with parameter bounds, or for multimodal likelihoods with many local optima (see discussion in \citealt{Buchner23,Barret24}). Parameter error estimation then also relies on probability distributions being approximately Gaussian or parameter degeneracies being minimal, which are often not acceptable assumptions.

A natural alternative is Bayesian X-ray spectral analysis \citep{vanDyk01}, which for a model $M$ which consists of an $N$-dimensional parameter space $\boldsymbol{\theta} = \{\theta_{1}, \theta_{2}, \ldots, \theta_{N}\}$ and dataset $D$ (comprising the source and background spectrum) are described by Bayes' Theorem:

\begin{equation}
    P(\boldsymbol{\theta} | D) = \frac{\pi(\boldsymbol{\theta} | M) \cdot \mathcal{L}(D | \boldsymbol{\theta})}{Z(D | M)}
\end{equation}

$P(\boldsymbol{\theta} | D)$ represents the posterior distribution (or the probability of model parameters $\boldsymbol{\theta}$ given the data), $\pi(\boldsymbol{\theta} | M)$ represents the prior knowledge on the parameters, $\mathcal{L}(D | \boldsymbol{\theta})$ is the likelihood (or the probability of the data given the model parameters) and $Z(D | M)$ is the unconditional probability of the data, also known as the Bayesian evidence. Whilst the likelihood is the basis for standard X-ray spectral fitting, the prior and posterior are unique to Bayesian analysis and allow prior information to be updated according to information contained in the observed data.

Through Monte Carlo sampling the probability posterior distribution can be estimated, and as a result, parameter optimisation and uncertainty estimation are achieved self-consistently. Many different Markov Chain Monte Carlo (MCMC) methods have been developed to do this by testing on a point-by-point (or sample of points) basis in the parameter space against new random points sampled from the prior. Undoubtedly a powerful technique, MCMC algorithms, however, are typically unsuitable for sampling multi-modal posteriors. Furthermore, quantifying the level of convergence of a given chain and ultimately deciding when to terminate the algorithm can be very difficult. More specifically to our use case, fitting low signal-to-noise data for heavily obscured AGN with complex physically-motivated models in an MCMC framework can result in unrealistically small uncertainties on crucial parameters such as the intrinsic source brightness, even for extremely long chain lengths.

An alternative Monte Carlo sampling algorithm to circumvent these issues is nested sampling \citep{Skilling04,Buchner14,Buchner23_review}. Whilst the majority of MCMC techniques work by sampling the posterior, nested sampling directly estimates the Bayesian evidence through numerical integration of the likelihood multiplied by the prior:

\begin{equation}
    Z(D | M) = \int \mathcal{L}(D | \boldsymbol{\theta}) \cdot \pi(\boldsymbol{\theta} | M) d^{N}\boldsymbol{\theta} ~.
    \label{eq:evidence}
\end{equation}

\noindent
The posterior is then treated as an ancillary component which can be sampled post facto from the evidence calculation. The Bayesian evidence is the average likelihood over the prior, meaning that $Z$ is larger if more of a considered parameter space is likely and would not necessarily prefer highly-peaked likelihoods if a large enough portion of the parameter space had low likelihood values. \textsc{MultiNest} \citep{FerozHobson08,Feroz09} provides an efficient numerical approximation to the multi-dimensional integral in Equation~\ref{eq:evidence}. At the start, the algorithm samples a set of \lq live\rq\ points from the full initial prior. On every iteration, the likelihood of the remaining points is calculated and sorted before the lowest likelihood is replaced by a point within the remaining set of likelihood values until some threshold is met for convergence. In particular, \textsc{MultiNest} was designed with an ellipsoidal clustering algorithm to efficiently traverse multi-dimensional parameter spaces with multi-modal distributions and inter-parameter degeneracies.

For this paper, we used the Bayesian X-ray Analysis software package v3.4.2 (BXA; \citealt{Buchner14}), to connect the Python implementation of \textsc{MultiNest} (\textsc{PyMultiNest}; \citealt{Buchner14}) to the X-ray spectral fitting package Sherpa v4.12 \citep{Freeman01}. All sources were initially fit with $W$-statistics (also known as modified $C$-statistics \citep{Wachter79}), and each source spectrum was iteratively binned in Sherpa by integer counts per bin until either all background bins contained $\geq$\,1 count or the spectrum had been binned with effectively $\geq$\,20 counts per bin. Such minimal binning is required for $W$-statistics due to the piece-wise model that is used to describe the background spectrum with number of parameters equal to the number of bins in the background. Despite a number of the NuLANDS sources being well-known bright AGN in X-rays (see Figure~\ref{fig:significance}), we fit all sources using BXA with $W$-statistics instead of significantly binning our data and using $\chi^2$ statistics. By definition, binning the data removes information that may be critical for a robust determination of line-of-sight \nh and using $\chi^2$ statistics can be biased even for high-count spectra (e.g., \citealt{Humphrey09}). Whilst $W$-statistics do still require binning, the binning is minimal compared to $\chi^{2}$ in the majority of cases and provides an alternative to pure $C$-statistics \citep{Cash79} in which the additional requirement for a background model can be computationally expensive if performing large numbers of fits.

Since even the minimal binning required by $W$-statistics can remove vital information for very faint targets, we fit the fainter sources in this paper with pure $C$-statistics by using the non-parametric background models from \citet{Simmonds18}. To qualify for background modeling, we select any sources with an X-ray dataset that was found to have signal-to-noise $<$\,4. For these sources, we use $C$-statistics \citep{Cash79} in our spectral fitting by modeling the contribution from each background spectrum in an automated fashion. We use the \texttt{auto\_background} function in BXA, which uses pre-defined Principal Component Analysis (PCA) templates of background spectra taken from different stacked blank-sky observations for each instrument. The Akaike Information Criterion (AIC; \citealt{Akaike74}) is then used to quantify the inclusion of more and more PCA components so long as the extra model complexity is required by the observed background spectrum. Whilst fitting a source spectrum with BXA, the background model is added to the source model with all PCA components fixed, apart from the total normalization, which is left free to vary along with the source model parameters in the fit. For more details on the \texttt{auto\_background} fitting process in BXA, see Appendix~A of \citet{Simmonds18}. We note that fitting the higher signal-to-noise datasets with $W$-statistics is not expected to cause a significant discrepancy with the datasets fit by $C$-statistics due to the increased source flux relative to the background in the former.

\subsection{Model Comparison}\label{subsec:modcomparison}
Our main strategy for model comparison is to use the ratio of two independent models' Bayesian evidences, often referred to as the Bayes factor $B_{12}$\,=\,$Z_{1}$\,/\,$Z_{2}$ for any two models 1 and 2. Values of $B_{12}$\,$>$\,1 indicate that model 1 is supported by the evidence, though Bayes factors do not follow a strict scale. A popular scale to interpret Bayes factors is the Jeffreys scale \citep{Jeffreys98}, in which $B_{12}$\,$=$\,100 is treated as an unconditional rejection of model 2. We thus adhere to the threshold of $B_{12}$\,$=$\,100 for performing model comparison, but note that such thresholds are not a guarantee. For example, \citet{Buchner14} performed simulations to quantify the Bayes Factor threshold for a sample of AGN detected in several different \textit{Chandra} deep fields, finding a Bayes Factor of $\sim$\,10 to be sufficient for a false selection rate below 1\%. On the other hand, \citet{Baronchelli18} find that a signal-to-noise ratio\,$>$\,7 was required for \textit{Chandra} fitting results to contribute meaningful information to the Bayes factors. Since the X-ray data in this paper come from a variety of different instruments, and observed signal-to-noise ratios, accurate simulation-based calibration of a definitive Bayes Factor threshold is computationally unfeasible. Thus, to be conservative, we select all models that satisfy $B_{12}$\,$<$\,100 relative to the highest Bayesian evidence to select models per each individual source.

\subsection{Model Checking}\label{subsec:verification}

Model comparison alone cannot quantify the quality of a fit, and should thus be combined with some form of model checking (aka goodness-of-fit) to verify that the selected model(s) can explain the data to a satisfactory level. For this paper, we use both qualitative and quantitative model-checking techniques using the PyXspec \citep{Gordon21}, the Python implementation of Xspec v12.11.1 \citep{Arnaud96}.

For qualitative checks we use a variety of visual inspection strategies as an initial sanity check for the automated fitting. Our checks included plotting fit residuals and Quantile-Quantile (Q-Q) plots to visualize the goodness-of-fit. For Q-Q plots, the cumulative observed data counts is typically plotted against the cumulative predicted model counts. A perfect model would be a one-to-one correlation, and a common signature of missing components in the data and/or model are \lq S\rq\ shape curves that are analogous to the cumulative distribution of a Gaussian function (see e.g., \citealt{Buchner14,Buchner23} and references therein). As an additional check, we manually performed spectral fits interactively with Xspec and cross-checked with the results found with our automated fitting pipeline.

For quantitative goodness-of-fit measures, we use simulations to perform posterior predictive checks. Whilst BXA robustly estimates the posterior probability of a given model with a given dataset, this does not account for stochastic changes in the observed data expected for a given observation, nor does this inform us as to model components that are needed/missing. One solution is to compare the observed data to the distribution of data spectra simulated from the best-fit posterior. The method is akin to that of the \texttt{goodness} command in Xspec\footnote{\href{https://heasarc.gsfc.nasa.gov/xanadu/xspec/manual/node83.html}{https://heasarc.gsfc.nasa.gov/xanadu/xspec/manual/node83.html}} and provides a comparison metric for an entire dataset being considered. We perform posterior predictive checks for the highest Bayes Factor model per source to check if the models used on average are sufficiently complex to explain the general shape of the observed data. We additionally note that the method is very useful for discovering outliers in which fits are incorrect or missing many features in the observed data.

\subsection{X-ray Spectral Components}\label{subsec:speccomps}
Here we describe the individual components used in our X-ray spectral models, and the free parameters in each.

\subsubsection{Intrinsic X-ray Continuum}
For the majority of models\footnote{For the MYtorus model, we use the \texttt{xszpowerlw} model in Sherpa for the Zeroth order continuum since high-energy exponential cut-offs are not included in the Monte Carlo simulations used to generate the table models -- see \url{http://mytorus.com/mytorus-instructions.html} for more information.}, the intrinsic coronal emission is approximated by a power law with high-energy exponential cut-off (\texttt{xscutoffpl} in Sherpa), which we refer to as the intrinsic powerlaw; IPL. The free parameters of this model are the photon index ($\Gamma$), the high-energy cut-off ($E_{\rm cut}$) and the normalisation ($\mathcal{A}_{\rm ipl}$). Although the high-energy cut-off can be constrained with physical modeling of the underlying Compton scattered continuum in bright high signal-to-noise ratio data (e.g., \citealt{Garcia15}), the vast majority of targets in our sample are at low enough signal-to-noise ratio to cause significant degeneracies between reprocessing parameters (e.g., global column density) and the high-energy cut-off (see discussion in \citealt{Buchner21b} for more information of how such degeneracies can affect inference in obscured AGN). For this reason, we froze $E_{\rm cut}$ to 300\,keV for all fits, in conjunction with the median found by \citet{Balokovic20}.

\subsubsection{Absorption}
Absorption along the line of sight occurs as a result of not just photo-electric absorption, but also Compton scattering which cannot be neglected for $N_{\rm H}$\,$\gtrsim$\,10$^{23}$\,cm$^{-2}$ (e.g., \citealt{Yaqoob97}). Thus, for phenomenological absorption models, we use the Sherpa model components \texttt{xsztbabs*xscabs}, assuming the abundances of \citet{Wilms00}. Note that this approximation does not account for energy downshifting from multiple scatterings, which will become increasingly important at higher column densities. However, we also use a wide range of publicly-available Monte Carlo reprocessing models that do account for this, meaning the effects of absorption should be covered for unobscured, obscured and Compton-thick sources.

\subsubsection{Compton-scattered Continuum}\label{sec:comptonscattering}
X-ray photons recoil from the material, lose energy, and change direction due to Compton scattering. Two prominent sources of such Compton scattering that are modeled in X-rays are (i) the accretion disk, sufficiently distant from the central engine to not require ionizing or relativistic effects, and (ii) the distant obscurer, which has been found to require substantial scale heights (and hence covering factors) from different dedicated studies (e.g., \citealt{Ricci15,Balokovic18,Buchner19}). We neglect relativistic effects on the Compton-scattered continuum, since the majority of our observations lack the sensitivity required to detect such spectral features that are often degenerate with Compton scattering in the circum-nuclear obscurer (e.g., \citealt{Tzanavaris21} and references therein).

For accretion disk Compton scattering, we use the \texttt{pexrav} model \citep{Magdziarz95}, which assumes a cold semi-infinite slab with infinite optical depth which Compton scatters incident photons from an exponential cut-off power law. To reduce the computation time involved with fitting, we generate a table model for the pure reprocessed portion of \texttt{pexrav} by creating a grid of \texttt{PhoIndex}, \texttt{rel\_refl} and \texttt{cosIncl} from \texttt{pexrav}, whilst assuming solar abundances and again freezing the high-energy cutoff to 300\,keV. We decouple the reprocessed spectrum from the incident one by only including negative \texttt{rel\_refl} values in the range [-100,\,-0.1]. We refer to our table model approximation of the \texttt{pexrav} model as \texttt{texrav} hereafter.

We used various models for Compton scattering from cold neutral material in the circumnuclear obscurer. Though slab-based models are not appropriate for modeling torus reprocessed emission (especially in the Compton-thick regime), some insight is attainable by comparing best-fit parameters (such as \texttt{rel\_refl}) to previous slab-based fits. For this reason, we first fit each source with a variety of \texttt{texrav}-based {\em obscured} geometry models, in which the Compton-scattered spectrum is disentangled from the column density (i.e. the reprocessed spectrum is not absorbed; see \citealt{Ricci17_bassV} for more details). We then follow the \texttt{texrav} modeling with a large library of physically-motivated torus models assuming several different geometries and parameter spaces. Such torus models are typically created with Monte Carlo radiative transfer simulations of X-ray propagation through a certain geometry of neutral cold gas, whilst accounting for photoelectric absorption, fluorescence, and Compton scattering self-consistently. Thus the column density self-consistently impacts not just the absorption but also the Compton-scattering, in contrast to \texttt{pexrav}.

\subsubsection{Fluorescence}
Fluorescence emission lines are commonly observed in the X-ray spectra of AGN, with the features arising from Fe\,K$\alpha$ at 6.4\,keV often being the strongest due to the combination of cosmic abundances and fluorescent yield (e.g., \citealt{Krause79,Anders89,Mushotzky93,Shu10}). The broad component of the Fe\,K$\alpha$ feature likely arises from relativistic reprocessing in the innermost parts of the accretion disk in some sources (e.g., \citealt{Fabian89,Fabian00,Brenneman06,Dovciak14}), but others may arise from the distortion effects associated with more complex ionized absorption (e.g., \citealt{Turner09,Miyakawa12}). The second component observed in the Fe\,K$\alpha$ feature is narrow, and may arise from the broad line region (e.g., \citealt{Bianchi08,Ponti13}), a small region between the broad line region and dust sublimation radius (e.g., \citealt{Gandhi15b,Minezaki15,Uematsu21}), the circumnuclear obscurer (e.g., \citealt{Ricci14,Boorman18}), or from much more extended material at $>$\,10\,pc (e.g., \citealt{Arevalo14,Bauer15,Fabbiano17}, but see \citealt{Andonie22}). In our phenomenological models that do not self-consistently model fluorescence of atoms within some assumed geometry, we solely aim to reproduce the more common narrow component of the 6.4\,keV Fe\,K$\alpha$ feature with a single narrow redshifted Gaussian (\texttt{xszgauss} in Sherpa). The Gaussian line has fixed line centroid and width of 6.4\,keV and 1\,eV, respectively, whilst having variable log-normalization.

\subsubsection{Soft Excess}
A common and important feature in unobscured and obscured AGN is an excess above the observed X-ray continuum $\lesssim$\,2\,keV -- the so-called \lq soft excess\rq. For unobscured AGN, the soft excess is observed to peak at $\sim$\,1\,--\,2\,keV. The current models to explain the soft excess tend to include relativistic blurring of soft emission lines produced from X-ray reprocessing in the accretion disk (e.g., \citealt{Crummy06,Zoghbi08,Fabian09,Walton13}), Comptonization of accretion disk photons by a cool corona situated above the disk that is optically-thicker and cooler than the primary X-ray source (e.g., \citealt{Czerny87,Middleton09,Jin09,Done12}) or relativistically-smeared ionized absorption in a wind from the inner accretion disk (e.g., \citealt{Gierlinski04,Middleton07,Parker22}). Previous works have tried to decipher the correct scenario by considering soft and hard X-ray data (e.g., \citealt{Vasudevan14,Boissay16,Garcia19}, Adegoke et al., in prep.), but the origin of the soft excess in unobscured AGN remains uncertain. For our purposes, we model the soft excess in unobscured objects simply with a black body (\texttt{xsbbody} in Sherpa). Though not physically-motivated, our simplistic modelling is chosen as a computationally-efficient way to phenomenologically account for the soft excess whilst estimating the (likely low) neutral line-of-sight column densities present in such objects.

For obscured AGN, another soft excess is observed. This is often suggested to arise from some combination of collisionally-ionised gas possibly correlated with circumnuclear star formation (e.g., \citealt{Guainazzi09,Iwasawa11}), photoionized emission powered by the central AGN (e.g., \citealt{Bianchi06,Guainazzi07}) and Thomson scattering of the intrinsic X-ray continuum by diffuse ionized gas of much lower column than the circumnuclear obscurer (often called the \lq warm mirror\rq; e.g., \citealt{Ueda07,Matt19}). First, for all sources, we include a Thomson-scattered component, which manifests as some fraction of the intrinsic transmitted spectrum. Some physically-motivated torus models (e.g., \texttt{UXCLUMPY} and \texttt{warped-disk}) include a self-consistent Thomson-scattered component in the list of available tables. In most cases, however, we simply include an additional power-law component pre-multiplied by a constant. The power law is tied to the intrinsic powerlaw in the model, and the pre-multiplying constant, \texttt{fscat}, is allowed to vary from 0.001\,--\,10\% in agreement with the bounds recommended on the XARS webpages\footnote{\href{https://github.com/JohannesBuchner/xars?tab=readme-ov-file}{https://github.com/JohannesBuchner/xars?tab=readme-ov-file}}. Concerning ionised gas emission, it can be extremely difficult to differentiate between the two with CCD-level spectral resolution. To test the effects of using collisionally-ionized vs. photo-ionized models to phenomenologically account for the soft excess in obscured AGN whilst trying to constrain the neutral column density, we include two different model components. First, for collisionally-ionized gas, we use the \texttt{apec} model (Astrophysical Plasma Emission Code, v.12.10.1; \citealt{Smith01}), with fixed solar abundances and variable normalization and temperature. Second, for photo-ionized gas we use an Xspec table model version of the SPEX \citep{Kaastra96} photo-ionized model \texttt{PION} \citep{Miller15}. \texttt{PION} calculates the photoionised emission from a slab\footnote{See \url{https://var.sron.nl/SPEX-doc/manualv3.05/manualse72.html} for more information.}, though we solely use the model to reproduce photo-ionized emission features (for details of the model creation, see \citealt{Parker19}). The free parameters in the \texttt{PION} table model are the column density, the ionization parameter, and normalization.

\subsection{X-ray Spectral Models}\label{subsec:models}

\begin{figure*}
\centering
\includegraphics[width=0.99\textwidth]{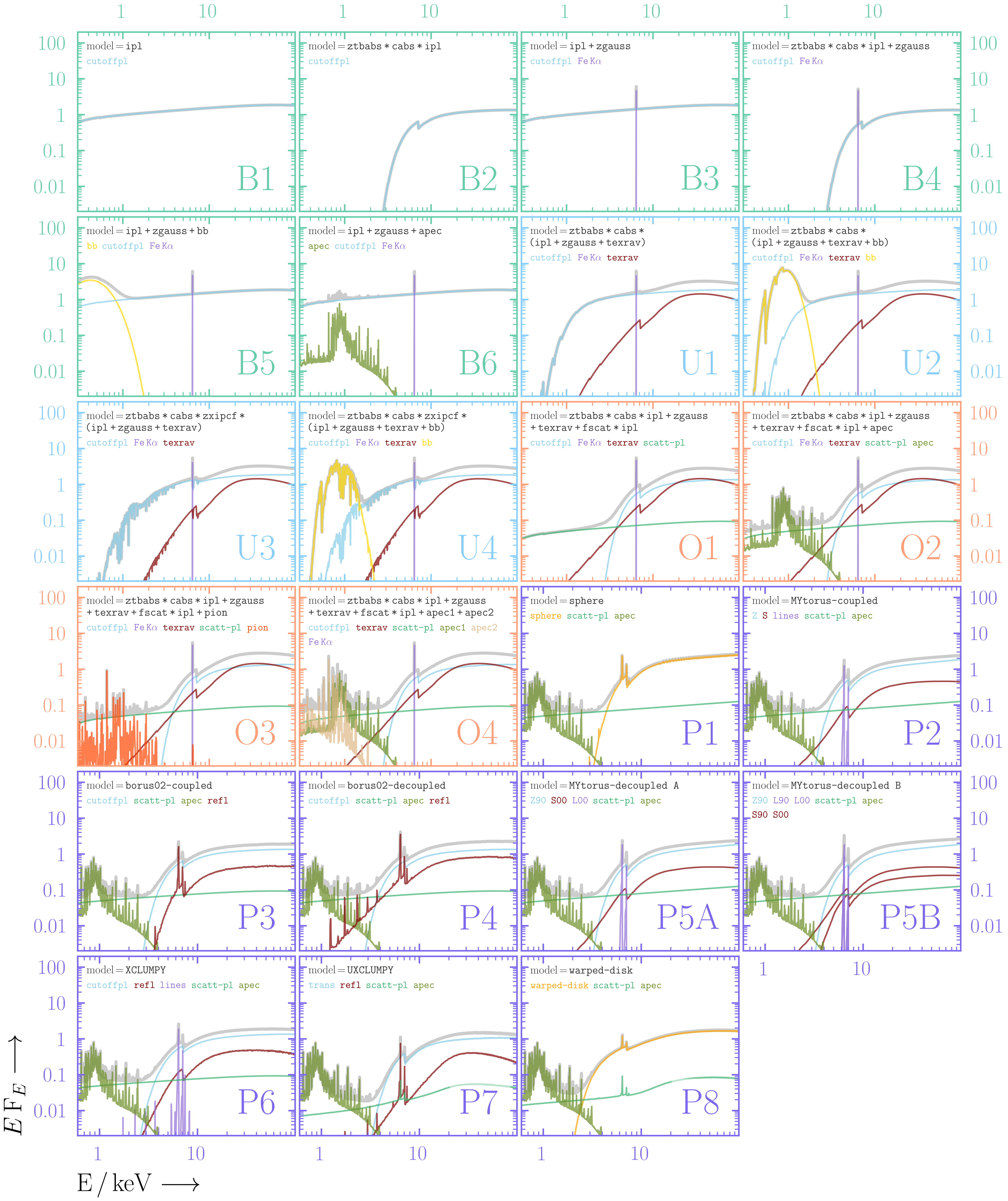}
\caption{\label{fig:xraymodels} The spectral models fit to each source using representative parameter values (see Tables~\ref{tab:bmods}, \ref{tab:umods}, \ref{tab:omods} and \ref{tab:pmods} for the full allowable parameter spaces per model). (B)asic, (U)nobscured, (O)bscured and (P)hysically-obscured models are shown in green, blue, orange and purple frames, respectively. In all panels, the photon index was set to 1.8, intrinsic normalisation was set to unity, line-of-sight column density was set to 5\,$\times$\,10$^{23}$\,cm$^{-2}$ (apart from the U models in which the line-of-sight column density is assumed to be 10$^{22}$\,cm$^{-2}$) and the Thompson-scattered fraction was set to 5\%. For all decoupled P models (i.e. P4, P5A and P5B), the global column density was set to 10$^{24}$\,cm$^{-2}$. For P2 and P6, we set the geometrical obscurer parameters such that the global column density was 10$^{24}$\,cm$^{-2}$ given the line-of-sight column density was 5\,$\times$\,10$^{23}$\,cm$^{-2}$.} For details of each model, including the full parameter spaces allowed during fitting, see Section~\ref{subsec:models}.
\end{figure*}

We fit three classes of model to all sources: Basic (\lq B\rq\ models hereafter), phenomenological (Unobscured and Obscured or \lq U\rq\ and \lq O\rq\ models, respectively hereafter) and Physically-motivated obscured (\lq P\rq\ models hereafter). The B models do not include a component for reprocessing, and are instead designed to provide insight into the observed spectral shape of a given source rather than any intrinsic properties. The U and O models feature the \texttt{texrav} model described in Section~\ref{sec:comptonscattering}, which provide a parametric and systematic modeling structure to compare unobscured and obscured AGN spectral shapes. The primary difference between the implementation of \texttt{texrav} in the U and O models is that the Compton-scattered continuum (assumed to arise from the accretion disk) is self-consistently obscured in U models but not in the O models (see \citealt{Ricci17_bassV} for more information regarding this implementation). The P models each define a unique physically-motivated obscurer geometry and properly account for multiple scatterings whilst self-consistently illuminating the geometry with Monte Carlo radiative transfer simulations. The model syntax used, free parameters, parameter priors, and parameter units are given in Tables~\ref{tab:bmods},~\ref{tab:umods},~\ref{tab:omods}~and~\ref{tab:pmods} for the B, U, O and P models, respectively. All models included a cross-calibration constant that was free to vary for every dataset. Each cross-calibration was given a log-Gaussian prior centered on the logarithm of the \citet{Madsen17} values if \nustar data were present, and zero (i.e. unity in linear units) if not. The log-Gaussian prior was given a broad standard deviation of 0.15 in logarithmic space. All datasets were optimally chosen to be quasi-simultaneous or to show minimal spectral variability, allowing us to assume the cross-calibration between \textit{NuSTAR} and the other X-ray instruments to agree with \citet{Madsen17} for the majority of cases. Example photon spectra for each implemented model are shown in Figure~\ref{fig:xraymodels}.

\subsection{Parameter Sample Distributions}\label{subsec:sampledists}
Several sources in NuLANDS are within the low-count regime (e.g., signal-to-noise ratio\,$<$\,3, see Figure~\ref{fig:significance}). Parameter constraints are often broad for such faint sources, and inter-parameter degeneracies can be substantial. With this in mind, we combine individual parameter posterior distributions into parameter sample distributions with Hierarchical Bayesian modeling, much like the histogram model available in \texttt{PosteriorStacker} (also see description in Section~\ref{sec:representative}). For our analysis, we preferentially use the histogram model to derive flexible sample distributions without assuming a priori specific sample distribution model shapes.

For a given parameter, we generate a sample distribution as follows: (1) for parameter posteriors that were generated from a non-uniform prior (e.g., photon index, $\Gamma$ in this work - see Tables~\ref{tab:bmods},~\ref{tab:umods},~\ref{tab:omods}~and~\ref{tab:pmods}), we first re-sample the posterior via the inverse of the prior. (2) Next, 1000 posterior samples are drawn randomly. We sample from the cumulative distribution function of the existing parameter posterior rows for sources with fewer posterior samples. (3) \texttt{PosteriorStacker} then computes a likelihood as a function of the sample distribution parameters (assuming that all objects are described by the same sample distribution). For the histogram model, a flat Dirichlet prior is used for the individual bin heights, self-consistently ensuring that all the bin heights sum to unity. Note that for bin widths not equal to unity, one must divide by the bin widths to derive the histogram.
\startlongtable
\begin{deluxetable*}{clccc}
\tabletypesize{\scriptsize}
\tablewidth{0pt}
\tablecaption{\label{tab:bmods} B ({\em Basic phenomenological}) model parameter information}
\tablehead{
\colhead{Model} &
\colhead{Components} &
\colhead{Free Parameters} &
\colhead{Priors$^{a}$} &
\colhead{Units}
}
\tableheadfrac{0.1}
\startdata
\hline
\multicolumn{5}{c}{\em Global model form} \\[0.2cm]
\hline
 & \multirow{1}{*}{\scriptsize $\mathcal{C}_{\textsc{cal}}*\texttt{tbabs}*(\textsc{U\{1,\,2,\,3,\,4\}})$} & log\,$\mathcal{C}_{\textsc{cal}}$ & \makecell{$\mathcal{G}(\mu=\textrm{log\,M17},\,\sigma=0.15)^{b}$ \\ $[-2,\,2]$} & -- \\[0.2cm]
\hline
\multicolumn{5}{c}{\em Basic (B) phenomenological models} \\[0.2cm]
\hline
\multirow{2}{*}{\textsc{B1}} & \multirow{2}{*}{\parbox{6cm}{\scriptsize $\texttt{cutoffpl}$}} & $\Gamma\,[\textsc{ipl}]$ & \makecell{$\mathcal{G}(\mu=1.8,\,\sigma=0.2)$ \\ $[-1,\,3]$} & -- \\[0.2cm]
& & log\,$\mathcal{A}\,[\textsc{ipl}]$ & $\mathcal{U}(-8, 2)$ & ph\,keV$^{-1}$\,cm$^{-2}$\,s$^{-1}$ at 1\,keV \\[0.2cm]
\hline
\rule{0pt}{1.5\normalbaselineskip}
\multirow{3}{*}{\textsc{B2}} & \multirow{3}{*}{\parbox{6cm}{\scriptsize $\texttt{ztbabs}*\texttt{cabs}*\texttt{cutoffpl}$}} & $\Gamma\,[\textsc{ipl}]$ & \makecell{$\mathcal{G}(\mu=1.8,\,\sigma=0.2)$ \\ $[-1,\,3]$} & -- \\[0.2cm]
& & log\,$\mathcal{A}\,[\textsc{ipl}]$ & $\mathcal{U}(-8, 2)$ & ph\,keV$^{-1}$\,cm$^{-2}$\,s$^{-1}$ at 1\,keV \\
& & log\,$N_{\rm H, Z}$ & $\mathcal{U}(20, 25)$ & cm$^{-2}$ \\[0.2cm]
\hline
\rule{0pt}{1.5\normalbaselineskip}
\multirow{3}{*}{\textsc{B3}} & \multirow{3}{*}{\parbox{6cm}{\scriptsize $\texttt{cutoffpl}+\texttt{zgauss}$}} & $\Gamma\,[\textsc{ipl}]$ & \makecell{$\mathcal{G}(\mu=1.8,\,\sigma=0.2)$ \\ $[-1,\,3]$} & -- \\[0.2cm]
& & log\,$\mathcal{A}\,[\textsc{ipl}]$ & $\mathcal{U}(-8, 2)$ & ph\,keV$^{-1}$\,cm$^{-2}$\,s$^{-1}$ at 1\,keV \\
& & log\,$\mathcal{A}\,[\texttt{zgauss}]$ & $\mathcal{U}(-8, 2)$ & ph\,keV$^{-1}$\,cm$^{-2}$\,s$^{-1}$ at 1\,keV \\[0.2cm]
\hline
\rule{0pt}{1.5\normalbaselineskip}
\multirow{4}{*}{\textsc{B4}} & \multirow{4}{*}{\parbox{6cm}{\scriptsize $\texttt{ztbabs}*\texttt{cabs}*\texttt{cutoffpl}+\texttt{zgauss}$}} & $\Gamma\,[\textsc{ipl}]$ & \makecell{$\mathcal{G}(\mu=1.8,\,\sigma=0.2)$ \\ $[-1,\,3]$} & -- \\[0.2cm]
& & log\,$\mathcal{A}\,[\textsc{ipl}]$ & $\mathcal{U}(-8, 2)$ & ph\,keV$^{-1}$\,cm$^{-2}$\,s$^{-1}$ at 1\,keV \\
& & log\,$N_{\rm H, Z}$ & $\mathcal{U}(20, 25)$ & cm$^{-2}$ \\
& & log\,$\mathcal{A}\,[\texttt{zgauss}]$ & $\mathcal{U}(-8, 2)$ & ph\,keV$^{-1}$\,cm$^{-2}$\,s$^{-1}$ at 1\,keV \\[0.2cm]
\hline
\rule{0pt}{1.5\normalbaselineskip}
\multirow{5}{*}{\textsc{B5}} & \multirow{5}{*}{\parbox{6cm}{\scriptsize $\texttt{cutoffpl}+\texttt{zgauss}+\texttt{bbody}$}} & $\Gamma\,[\textsc{ipl}]$ & \makecell{$\mathcal{G}(\mu=1.8,\,\sigma=0.2)$ \\ $[-1,\,3]$} & -- \\[0.2cm]
& & log\,$\mathcal{A}\,[\textsc{ipl}]$ & $\mathcal{U}(-8, 2)$ & ph\,keV$^{-1}$\,cm$^{-2}$\,s$^{-1}$ at 1\,keV \\
& & log\,$\mathcal{A}\,[\texttt{zgauss}]$ & $\mathcal{U}(-8, 2)$ & ph\,keV$^{-1}$\,cm$^{-2}$\,s$^{-1}$ at 1\,keV \\
& & log\,$kT$ & $\mathcal{U}(-2, 0)$ & keV \\
& & log\,$\mathcal{A}\,[\texttt{bbody}]$ & $\mathcal{U}(-8, 2)$ & ph\,keV$^{-1}$\,cm$^{-2}$\,s$^{-1}$ at 1\,keV \\[0.2cm]
\hline
\rule{0pt}{1.5\normalbaselineskip}
\multirow{5}{*}{\textsc{B6}} & \multirow{5}{*}{\parbox{6cm}{\scriptsize $\texttt{cutoffpl}+\texttt{zgauss}+\texttt{apec}$}} & $\Gamma\,[\textsc{ipl}]$ & \makecell{$\mathcal{G}(\mu=1.8,\,\sigma=0.2)$ \\ $[-1,\,3]$} & -- \\[0.2cm]
& & log\,$\mathcal{A}\,[\textsc{ipl}]$ & $\mathcal{U}(-8, 2)$ & ph\,keV$^{-1}$\,cm$^{-2}$\,s$^{-1}$ at 1\,keV \\
& & log\,$\mathcal{A}\,[\texttt{zgauss}]$ & $\mathcal{U}(-8, 2)$ & ph\,keV$^{-1}$\,cm$^{-2}$\,s$^{-1}$ at 1\,keV \\
& & log\,$kT$ & $\mathcal{U}(-2, \textrm{log}\,2)$ & keV \\
& & log\,$\mathcal{A}\,[\texttt{apec}]$ & $\mathcal{U}(-8, 2)$ & ph\,keV$^{-1}$\,cm$^{-2}$\,s$^{-1}$ at 1\,keV \\[0.2cm]
\hline
\hline
\enddata
\tablecomments{
$^{a}$\,$\mathcal{G}(\mu,\,\sigma)$ and $\mathcal{U}(\textrm{min}, \textrm{max})$ denote Gaussian and uniform priors, respectively. For Gaussian priors, we include the full parameter range in square brackets beneath each. $^{b}$\,The cross-calibration constants were varied in log space from the \citet{Madsen17} values, relative to FPMA when \nustar data was present. In the event that only soft data was available, only one dataset was fit and so no variable cross-calibration was used. The parameter symbol definitions are: power law photon index ($\Gamma$), a given model component normalisation ($\mathcal{A}$) and the line-of-sight column density ($N_{\rm H,\,Z}$).
}
\end{deluxetable*}
\startlongtable
\begin{deluxetable*}{clccc}
\tabletypesize{\scriptsize}
\tablewidth{0pt}
\tablecaption{\label{tab:umods} U ({\em Unobscured phenomenological}) model parameter information}
\tablehead{
\colhead{Model} &
\colhead{Components} &
\colhead{Free Parameters} &
\colhead{Priors$^{a}$} &
\colhead{Units}
}
\tableheadfrac{0.1}
\startdata
\hline
\multicolumn{5}{c}{\em Global model form} \\[0.2cm]
\hline
 & \multirow{1}{*}{\scriptsize $\mathcal{C}_{\textsc{cal}}*\texttt{tbabs}*(\textsc{U\{1,\,2,\,3,\,4\}})$} & log\,$\mathcal{C}_{\textsc{cal}}$ & \makecell{$\mathcal{G}(\mu=\textrm{log\,M17},\,\sigma=0.15)^{b}$ \\ $[-2,\,2]$} & -- \\[0.2cm]
\hline
\multicolumn{5}{c}{\em Unobscured (U) phenomenological models} \\[0.2cm]
\hline
\rule{0pt}{1.5\normalbaselineskip}
\multirow{5}{*}{\textsc{U1}} & \multirow{5}{*}{\parbox{6cm}{\scriptsize $\texttt{ztbabs}*\texttt{cabs}*(\texttt{cutoffpl}+\texttt{texrav}+\texttt{zgauss})$}} & $\Gamma\,[\textsc{ipl}]$ & \makecell{$\mathcal{G}(\mu=1.8,\,\sigma=0.2)$ \\ $[1.4,\,2.8]$} & -- \\[0.2cm]
& & log\,$\mathcal{A}\,[\textsc{ipl}]$ & $\mathcal{U}(-8, 2)$ & ph\,keV$^{-1}$\,cm$^{-2}$\,s$^{-1}$ at 1\,keV \\
& & log\,$N_{\rm H, Z}$ & $\mathcal{U}(20, 25)$ & cm$^{-2}$ \\
& & log\,$|\mathcal{R}\,(<0)|$ & $\mathcal{U}(-2, 1)$ & -- \\
& & log\,$\mathcal{A}\,[\texttt{zgauss}]$ & $\mathcal{U}(-8, 2)$ & ph\,keV$^{-1}$\,cm$^{-2}$\,s$^{-1}$ at 1\,keV \\[0.2cm]
\hline
\rule{0pt}{1.5\normalbaselineskip}
\multirow{7}{*}{\textsc{U2}} & \multirow{7}{*}{\parbox{6cm}{\scriptsize $\texttt{ztbabs}*\texttt{cabs}*(\texttt{cutoffpl}+\texttt{texrav}+\texttt{zgauss}+\texttt{bbody})$}} & $\Gamma\,[\textsc{ipl}]$ & \makecell{$\mathcal{G}(\mu=1.8,\,\sigma=0.2)$ \\ $[1.4,\,2.8]$} & -- \\[0.2cm]
& & log\,$\mathcal{A}\,[\textsc{ipl}]$ & $\mathcal{U}(-8, 2)$ & ph\,keV$^{-1}$\,cm$^{-2}$\,s$^{-1}$ at 1\,keV \\
& & log\,$N_{\rm H, Z}$ & $\mathcal{U}(20, 25)$ & cm$^{-2}$ \\
& & log\,$|\mathcal{R}\,(<0)|$ & $\mathcal{U}(-2, 1)$ & -- \\
& & log\,$\mathcal{A}\,[\texttt{zgauss}]$ & $\mathcal{U}(-8, 2)$ & ph\,keV$^{-1}$\,cm$^{-2}$\,s$^{-1}$ at 1\,keV \\
& & log\,$kT$ & $\mathcal{U}(-2, 0)$ & keV \\
& & log\,$\mathcal{A}\,[\texttt{bbody}]$ & $\mathcal{U}(-8, 2)$ & ph\,keV$^{-1}$\,cm$^{-2}$\,s$^{-1}$ at 1\,keV \\[0.2cm]
\hline
\rule{0pt}{1.5\normalbaselineskip}
\multirow{8}{*}{\textsc{U3}} & \multirow{8}{*}{\parbox{6cm}{\scriptsize $\texttt{ztbabs}*\texttt{cabs}*\texttt{zxipcf}*(\texttt{cutoffpl}+\texttt{texrav}+\texttt{zgauss})$}} & $\Gamma\,[\textsc{ipl}]$ & \makecell{$\mathcal{G}(\mu=1.8,\,\sigma=0.2)$ \\ $[1.4,\,2.8]$} & -- \\[0.2cm]
& & log\,$\mathcal{A}\,[\textsc{ipl}]$ & $\mathcal{U}(-8, 2)$ & ph\,keV$^{-1}$\,cm$^{-2}$\,s$^{-1}$ at 1\,keV \\
& & log\,$N_{\rm H, Z}$ & $\mathcal{U}(20, 25)$ & cm$^{-2}$ \\
& & log\,$|\mathcal{R}\,(<0)|$ & $\mathcal{U}(-2, 1)$ & -- \\
& & log\,$\mathcal{A}\,[\texttt{zgauss}]$ & $\mathcal{U}(-8, 2)$ & ph\,keV$^{-1}$\,cm$^{-2}$\,s$^{-1}$ at 1\,keV \\
& & log\,$N_{\rm H, ion}$ & $\mathcal{U}(\textrm{log}\,5\times10^{20}, \textrm{log}\,5\times10^{24})$ & cm$^{-2}$ \\
& & log\,$\xi$ & $\mathcal{U}(-3, 6)$ & 10$^{-1}$\,erg\,s$^{-1}$\,m \\
& & $CF$ & $\mathcal{U}(0, 1)$ & -- \\[0.2cm]
\hline
\rule{0pt}{1.5\normalbaselineskip}
\multirow{10}{*}{\textsc{U4}} & \multirow{10}{*}{\parbox{6cm}{\scriptsize $\texttt{ztbabs}*\texttt{cabs}*\texttt{zxipcf}*(\texttt{cutoffpl}+\texttt{texrav}+\texttt{zgauss}+\texttt{bbody})$}} & $\Gamma\,[\textsc{ipl}]$ & \makecell{$\mathcal{G}(\mu=1.8,\,\sigma=0.2)$ \\ $[1.4,\,2.8]$} & -- \\[0.2cm]
& & log\,$\mathcal{A}\,[\textsc{ipl}]$ & $\mathcal{U}(-8, 2)$ & ph\,keV$^{-1}$\,cm$^{-2}$\,s$^{-1}$ at 1\,keV \\
& & log\,$N_{\rm H, Z}$ & $\mathcal{U}(20, 25)$ & cm$^{-2}$ \\
& & log\,$|\mathcal{R}\,(<0)|$ & $\mathcal{U}(-2, 1)$ & -- \\
& & log\,$\mathcal{A}\,[\texttt{zgauss}]$ & $\mathcal{U}(-8, 2)$ & ph\,keV$^{-1}$\,cm$^{-2}$\,s$^{-1}$ at 1\,keV \\
& & log\,$N_{\rm H, ion}$ & $\mathcal{U}(\textrm{log}\,5\times10^{20}, \textrm{log}\,5\times10^{24})$ & cm$^{-2}$ \\
& & log\,$\xi$ & $\mathcal{U}(-3, 6)$ & 10$^{-1}$\,erg\,s$^{-1}$\,m \\
& & $CF$ & $\mathcal{U}(0, 1)$ & -- \\
& & log\,$kT$ & $\mathcal{U}(-2, 0)$ & keV \\
& & log\,$\mathcal{A}\,[\texttt{bbody}]$ & $\mathcal{U}(-8, 2)$ & ph\,keV$^{-1}$\,cm$^{-2}$\,s$^{-1}$ at 1\,keV \\[0.2cm]
\hline
\hline
\enddata
\tablecomments{
$^{a}$\,$\mathcal{G}(\mu,\,\sigma)$ and $\mathcal{U}(\textrm{min}, \textrm{max})$ denote Gaussian and uniform priors, respectively. For Gaussian priors, we include the full parameter range in square brackets beneath each.
$^{b}$\,The cross-calibration constants were varied in log space from the \citet{Madsen17} values, relative to FPMA when \nustar data was present. In the event that only soft data was available, only one dataset was fit and so no variable cross-calibration was used.
The parameter symbol definitions are: power law photon index ($\Gamma$), a given model component normalisation ($\mathcal{A}$), the relative scaling of the Compton-scattered continuum ($|\mathcal{R}(<0)|$), the line-of-sight column density ($N_{\rm H,\,Z}$), ionised column density in \texttt{zxipcf} ($N_{\rm H, ion}$), the ionisation parameter in \texttt{zxipcf} ($\xi$) and the ionised absorber covering factor in \texttt{zxipcf} ($CF$).
}
\end{deluxetable*}
\startlongtable
\begin{deluxetable*}{clccc}
\tabletypesize{\scriptsize}
\tablewidth{0pt}
\tablecaption{\label{tab:omods} O ({\em Obscured phenomenological}) model parameter information}
\tablehead{
\colhead{Model} &
\colhead{Components} &
\colhead{Free Parameters} &
\colhead{Priors$^{a}$} &
\colhead{Units}
}
\tableheadfrac{0.1}
\startdata
\hline
\multicolumn{5}{c}{{\em Global model form}} \\[0.2cm]
\hline
\rule{0pt}{1.5\normalbaselineskip}
 & \multirow{1}{*}{\scriptsize $\mathcal{C}_{\textsc{cal}}*\texttt{tbabs}*(\textsc{O\{1,\,2,\,3,\,4\}})$} & log\,$\mathcal{C}_{\textsc{cal}}$ & \makecell{$\mathcal{G}(\mu=\textrm{log\,M17},\,\sigma=0.15)^{b}$ \\ $[-2,\,2]$} & -- \\[0.2cm]
\hline
\multicolumn{5}{c}{\em Obscured (O) phenomenological models} \\[0.2cm]
\hline
\rule{0pt}{1.5\normalbaselineskip}
\multirow{6}{*}{\textsc{O1}} & \multirow{6}{*}{\parbox{6cm}{\scriptsize $\texttt{ztbabs}*\texttt{cabs}*\texttt{cutoffpl}+\texttt{texrav}+\texttt{zgauss}+\texttt{f}_{\rm scat}*\texttt{cutoffpl}$}} & $\Gamma\,[\textsc{ipl}]$ & \makecell{$\mathcal{G}(\mu=1.8,\,\sigma=0.2)$ \\ $[1.4,\,2.8]$} & -- \\
& & log\,$\mathcal{A}\,[\textsc{ipl}]$ & $\mathcal{U}(-8, 2)$ & ph\,keV$^{-1}$\,cm$^{-2}$\,s$^{-1}$ at 1\,keV \\
& & log\,$N_{\rm H, Z}$ & $\mathcal{U}(20, 25)$ & cm$^{-2}$ \\
& & log\,$|\mathcal{R}\,(<0)|$ & $\mathcal{U}(-2, 1)$ & -- \\
& & log\,$\mathcal{A}\,[\texttt{zgauss}]$ & $\mathcal{U}(-8, 2)$ & ph\,keV$^{-1}$\,cm$^{-2}$\,s$^{-1}$ at 1\,keV \\
& & log\,$f_{\textsc{scat}}$ & $\mathcal{U}(-5, -1)$ & -- \\[0.2cm]
\hline
\rule{0pt}{1.5\normalbaselineskip}
\multirow{8}{*}{\textsc{O2}} & \multirow{8}{*}{\parbox{6cm}{\scriptsize $\texttt{ztbabs}*\texttt{cabs}*\texttt{cutoffpl}+\texttt{texrav}+\texttt{zgauss}+\texttt{f}_{\rm scat}*\texttt{cutoffpl}+\texttt{apec}$}} & $\Gamma\,[\textsc{ipl}]$ & \makecell{$\mathcal{G}(\mu=1.8,\,\sigma=0.2)$ \\ $[1.4,\,2.8]$} & -- \\
& & log\,$\mathcal{A}\,[\textsc{ipl}]$ & $\mathcal{U}(-8, 2)$ & ph\,keV$^{-1}$\,cm$^{-2}$\,s$^{-1}$ at 1\,keV \\
& & log\,$N_{\rm H, Z}$ & $\mathcal{U}(20, 25)$ & cm$^{-2}$ \\
& & log\,$|\mathcal{R}\,(<0)|$ & $\mathcal{U}(-2, 1)$ & -- \\
& & log\,$\mathcal{A}\,[\texttt{zgauss}]$ & $\mathcal{U}(-8, 2)$ & ph\,keV$^{-1}$\,cm$^{-2}$\,s$^{-1}$ at 1\,keV \\
& & log\,$f_{\textsc{scat}}$ & $\mathcal{U}(-5, -1)$ & -- \\
& & log\,$kT$ & $\mathcal{U}(-2, \textrm{log}\,2)$ & keV \\
& & log\,$\mathcal{A}\,[\texttt{apec}]$ & $\mathcal{U}(-8, 2)$ & ph\,keV$^{-1}$\,cm$^{-2}$\,s$^{-1}$ at 1\,keV \\[0.2cm]
\hline
\rule{0pt}{1.5\normalbaselineskip}
\multirow{10}{*}{\textsc{O3}} & \multirow{10}{*}{\parbox{6cm}{\scriptsize $\texttt{ztbabs}*\texttt{cabs}*\texttt{cutoffpl}+\texttt{texrav}+\texttt{zgauss}+\texttt{f}_{\rm scat}*\texttt{cutoffpl}+\texttt{pion}$}} & $\Gamma\,[\textsc{ipl}]$ & \makecell{$\mathcal{G}(\mu=1.8,\,\sigma=0.2)$ \\ $[1.4,\,2.8]$} & -- \\
& & log\,$\mathcal{A}\,[\textsc{ipl}]$ & $\mathcal{U}(-8, 2)$ & ph\,keV$^{-1}$\,cm$^{-2}$\,s$^{-1}$ at 1\,keV \\
& & log\,$N_{\rm H, Z}$ & $\mathcal{U}(20, 26)$ & cm$^{-2}$ \\
& & log\,$|\mathcal{R}\,(<0)|$ & $\mathcal{U}(-2, 1)$ & -- \\
& & log\,$\mathcal{A}\,[\texttt{zgauss}]$ & $\mathcal{U}(-8, 2)$ & ph\,keV$^{-1}$\,cm$^{-2}$\,s$^{-1}$ at 1\,keV \\
& & log\,$f_{\textsc{scat}}$ & $\mathcal{U}(-5, -1)$ & -- \\
& & log\,$\xi$ & $\mathcal{U}(-2, 3)$ & 10$^{-1}$\,erg\,s$^{-1}$\,m \\
& & log\,$N_{\rm H, pion}$ & $\mathcal{U}(-6, -2)$ & 10$^{24}$\,cm$^{-2}$ \\
& & log\,$\mathcal{A}\,[\texttt{pion}]$ & $\mathcal{U}(-8, 2)$ & ph\,keV$^{-1}$\,cm$^{-2}$\,s$^{-1}$ at 1\,keV \\[0.2cm]
\hline
\rule{0pt}{1.5\normalbaselineskip}
\multirow{10}{*}{\textsc{O4}} & \multirow{10}{*}{\parbox{6cm}{\scriptsize $\texttt{ztbabs}*\texttt{cabs}*\texttt{cutoffpl}+\texttt{texrav}+\texttt{zgauss}+\texttt{f}_{\rm scat}*\texttt{cutoffpl}+\texttt{apec1}+\texttt{apec2}$}} & $\Gamma\,[\textsc{ipl}]$ & \makecell{$\mathcal{G}(\mu=1.8,\,\sigma=0.2)$ \\ $[1.4,\,2.8]$} & -- \\
& & log\,$\mathcal{A}\,[\textsc{ipl}]$ & $\mathcal{U}(-8, 2)$ & ph\,keV$^{-1}$\,cm$^{-2}$\,s$^{-1}$ at 1\,keV \\
& & log\,$N_{\rm H, Z}$ & $\mathcal{U}(20, 25)$ & cm$^{-2}$ \\
& & log\,$|\mathcal{R}\,(<0)|$ & $\mathcal{U}(-2, 1)$ & -- \\
& & log\,$\mathcal{A}\,[\texttt{zgauss}]$ & $\mathcal{U}(-8, 2)$ & ph\,keV$^{-1}$\,cm$^{-2}$\,s$^{-1}$ at 1\,keV \\
& & log\,$f_{\textsc{scat}}$ & $\mathcal{U}(-5, -1)$ & -- \\
& & log\,$kT1$ & $\mathcal{U}(-2, \textrm{log}\,2)$ & keV \\
& & log\,$\mathcal{A}\,[\texttt{apec1}]$ & $\mathcal{U}(-8, 2)$ & ph\,keV$^{-1}$\,cm$^{-2}$\,s$^{-1}$ at 1\,keV \\
& & log\,$kT2$ & $\mathcal{U}(-2, \textrm{log}\,2)$ & keV \\
& & log\,$\mathcal{A}\,[\texttt{apec2}]$ & $\mathcal{U}(-8, 2)$ & ph\,keV$^{-1}$\,cm$^{-2}$\,s$^{-1}$ at 1\,keV \\[0.2cm]
\hline
\hline
\enddata
\tablecomments{
$^{a}$\,$\mathcal{G}(\mu,\,\sigma)$ and $\mathcal{U}(\textrm{min}, \textrm{max})$ denote Gaussian and uniform priors, respectively. For Gaussian priors, we include the full parameter range in square brackets beneath each.
$^{b}$\,The cross-calibration constants were varied in log space from the \citet{Madsen17} values, relative to FPMA when \nustar data was present. In the event that only soft data was available, only one dataset was fit and so no variable cross-calibration was used.
The parameter symbol definitions are: power law photon index ($\Gamma$), a given model component normalisation ($\mathcal{A}$), the relative scaling of the Compton-scattered continuum ($|\mathcal{R}(<0)|$), the line-of-sight column density ($N_{\rm H,\,Z}$), Thomson-scattered emission fraction ($f_{\textsc{scat}}$), the ionisation parameter in \texttt{pion} ($\xi$) and ionised column density in \texttt{pion} ($N_{\rm H, pion}$).
}
\end{deluxetable*}
\startlongtable
\begin{deluxetable*}{clccc}
\tabletypesize{\scriptsize}
\tablewidth{0pt}
\tablecaption{\label{tab:pmods} P ({\em Physically-motivated obscured}) model parameter information}
\tablehead{
\colhead{Model} &
\colhead{Components} &
\colhead{Free Parameters} &
\colhead{Priors$^{a}$} &
\colhead{Units}
}
\tableheadfrac{0.1}
\startdata
\hline
\multicolumn{5}{c}{\em Global model form} \\[0.2cm]
\hline
 & \multirow{3}{*}{\scriptsize $\mathcal{C}_{\textsc{cal}}*\texttt{tbabs}*(\texttt{apec}+\textsc{P\{1,\,2,\,3,\,4,\,5,\,6,\,7\}})$} & log\,$\mathcal{C}_{\textsc{cal}}$ & \makecell{$\mathcal{G}(\mu=\textrm{log\,M17},\,\sigma=0.15)^{b}$ \\ $[-2,\,2]$} & -- \\[0.2cm]
& & log\,$kT$ & $\mathcal{U}(-2, \textrm{log}\,2)$ & keV \\
& & log\,$\mathcal{A}\,[\texttt{apec}]$ & $\mathcal{U}(-8, 2)$ & ph\,keV$^{-1}$\,cm$^{-2}$\,s$^{-1}$ at 1\,keV \\[0.2cm]
\hline
\multicolumn{5}{c}{\em Physical obscurer (P) models} \\[0.2cm]
\hline
\rule{0pt}{1.5\normalbaselineskip}
\multirow{3}{*}{\textsc{P1}} & \multirow{3}{*}{\parbox{5cm}{\scriptsize $\texttt{sphere}$}} & $\Gamma\,[\textsc{ipl}]$ & \makecell{$\mathcal{G}(\mu=1.8,\,\sigma=0.2)$ \\ $[1.2,\,2.8]$} & -- \\[0.2cm]
& & log\,$\mathcal{A}\,[\textsc{ipl}]$ & $\mathcal{U}(-8, 2)$ & ph\,keV$^{-1}$\,cm$^{-2}$\,s$^{-1}$ at 1\,keV \\
& & log\,$N_{\rm H, Z}$ & $\mathcal{U}(20, 26)$ & cm$^{-2}$ \\[0.2cm]
\hline
\rule{0pt}{1.5\normalbaselineskip}
\multirow{5}{*}{\textsc{P2}} & \multirow{5}{*}{\parbox{5cm}{\scriptsize $\texttt{mytorus\_zero}*\texttt{zpowerlw}+\texttt{mytorus\_scat}+\texttt{mytorus\_lines}+\texttt{f}_{\rm scat}*\texttt{zpowerlw}$}} & $\Gamma\,[\textsc{ipl}]$ & \makecell{$\mathcal{G}(\mu=1.8,\,\sigma=0.2)$ \\ $[1.4,\,2.6]$} & -- \\[0.2cm]
& & log\,$\mathcal{A}\,[\textsc{ipl}]$ & $\mathcal{U}(-8, 2)$ & ph\,keV$^{-1}$\,cm$^{-2}$\,s$^{-1}$ at 1\,keV \\
& & log\,$N_{\rm H, Z}$ & $\mathcal{U}(22, 25)$ & cm$^{-2}$ \\
& & $\theta_{\rm inc}$ & $\mathcal{U}(0, 90)$ & deg \\
& & log\,$f_{\textsc{scat}}$ & $\mathcal{U}(-5, -1)$ & -- \\[0.2cm]
\hline
\rule{0pt}{1.5\normalbaselineskip}
\multirow{7}{*}{\textsc{P3}} & \multirow{7}{*}{\parbox{5cm}{\scriptsize $\texttt{ztbabs}*\texttt{cabs}*\texttt{cutoffpl}+\texttt{borus02}+\texttt{f}_{\rm scat}*\texttt{cutoffpl}$}} & $\Gamma\,[\textsc{ipl}]$ & \makecell{$\mathcal{G}(\mu=1.8,\,\sigma=0.2)$ \\ $[1.4,\,2.6]$} & -- \\[0.2cm]
& & log\,$\mathcal{A}\,[\textsc{ipl}]$ & $\mathcal{U}(-8, 2)$ & ph\,keV$^{-1}$\,cm$^{-2}$\,s$^{-1}$ at 1\,keV \\
& & log\,$N_{\rm H, Z}$ & $\mathcal{U}(22, 25.5)$ & cm$^{-2}$ \\
& & $\theta_{\rm tor}$ & $\mathcal{U}(0, 84.3)$ & deg \\
& & $\theta_{\rm inc}$ & $\mathcal{U}(18.3, 87)$ & deg \\
& & log\,$A_{\rm Fe}$ & \makecell{$\mathcal{G}(\mu=0,\,\sigma=0.2)$ \\ $[-0.65,\,0.65]$} & $A_{{\rm Fe,}\,\odot}$ \\[0.2cm]
& & log\,$f_{\textsc{scat}}$ & $\mathcal{U}(-5, -1)$ & -- \\[0.2cm]
\hline
\rule{0pt}{1.5\normalbaselineskip}
\multirow{7}{*}{\textsc{P4}} & \multirow{7}{*}{\parbox{5cm}{\scriptsize $\texttt{ztbabs}*\texttt{cabs}*\texttt{cutoffpl}+\texttt{borus02}[\theta_{\rm inc}=87^{\circ}]+\texttt{f}_{\rm scat}*\texttt{cutoffpl}$}} & $\Gamma\,[\textsc{ipl}]$ & \makecell{$\mathcal{G}(\mu=1.8,\,\sigma=0.2)$ \\ $[1.4,\,2.6]$} & -- \\[0.2cm]
& & log\,$\mathcal{A}\,[\textsc{ipl}]$ & $\mathcal{U}(-8, 2)$ & ph\,keV$^{-1}$\,cm$^{-2}$\,s$^{-1}$ at 1\,keV \\
& & log\,$N_{\rm H, Z}$ & $\mathcal{U}(22, 25.5)$ & cm$^{-2}$ \\
& & $\theta_{\rm tor}$ & $\mathcal{U}(0, 84.3)$ & deg \\
& & log\,$A_{\rm Fe}$ & \makecell{$\mathcal{G}(\mu=0,\,\sigma=0.2)$ \\ $[-0.65,\,0.65]$} & $A_{{\rm Fe,}\,\odot}$ \\[0.2cm]
& & log\,$N_{\rm H, S}$ & $\mathcal{U}(22, 25.5)$ & cm$^{-2}$ \\
& & log\,$f_{\textsc{scat}}$ & $\mathcal{U}(-5, -1)$ & -- \\[0.2cm]
\hline
\rule{0pt}{1.5\normalbaselineskip}
\multirow{6}{*}{\textsc{P5A}} & \multirow{6}{*}{\parbox{5cm}{\scriptsize $\texttt{mytorus\_zero}[\theta_{\rm inc}=90^{\circ}]*\texttt{zpowerlw}+A_{\rm S00}*\texttt{mytorus\_scat}[\theta_{\rm inc}=0^{\circ}]+A_{\rm L00}[=\,A_{\rm S00}]*\texttt{mytorus\_lines}[\theta_{\rm inc}=0^{\circ}]+\texttt{f}_{\rm scat}*\texttt{zpowerlw}$}} & \makecell{$\mathcal{G}(\mu=1.8,\,\sigma=0.2)$ \\ $[1.4,\,2.6]$} & -- \\[0.2cm]
& & log\,$\mathcal{A}\,[\textsc{ipl}]$ & $\mathcal{U}(-8, 2)$ & ph\,keV$^{-1}$\,cm$^{-2}$\,s$^{-1}$ at 1\,keV \\
& & log\,$N_{\rm H, Z}$ & $\mathcal{U}(22, 25)$ & cm$^{-2}$ \\
& & log\,$A_{\rm S00}$ & \makecell{$\mathcal{G}(\mu=0,\,\sigma=0.2)$ \\ $[-4,\,4]$} & -- \\[0.2cm]
& & log\,$N_{\rm H, S}$ & $\mathcal{U}(22, 25)$ & cm$^{-2}$ \\
& & log\,$f_{\textsc{scat}}$ & $\mathcal{U}(-5, -1)$ & -- \\[0.2cm]
\hline
\rule{0pt}{1.5\normalbaselineskip}
\multirow{6}{*}{\textsc{P5B}} & \multirow{6}{*}{\parbox{5cm}{\scriptsize $\texttt{mytorus\_zero}[\theta_{\rm inc}=90^{\circ}]*\texttt{zpowerlw}+A_{\rm S00}*\texttt{mytorus\_scat}[\theta_{\rm inc}=0^{\circ}]+A_{\rm L00}[=\,A_{\rm S00}]*\texttt{mytorus\_lines}[\theta_{\rm inc}=0^{\circ}]+A_{\rm S90}*\texttt{mytorus\_scat}[\theta_{\rm inc}=90^{\circ}]+A_{\rm L90}[=\,A_{\rm S90}]*\texttt{mytorus\_lines}[\theta_{\rm inc}=90^{\circ}]+\texttt{f}_{\rm scat}*\texttt{zpowerlw}$}} & \makecell{$\mathcal{G}(\mu=1.8,\,\sigma=0.2)$ \\ $[1.4,\,2.6]$} & -- \\[0.2cm]
& & log\,$\mathcal{A}\,[\textsc{ipl}]$ & $\mathcal{U}(-8, 2)$ & ph\,keV$^{-1}$\,cm$^{-2}$\,s$^{-1}$ at 1\,keV \\
& & log\,$N_{\rm H, Z}$ & $\mathcal{U}(22, 25)$ & cm$^{-2}$ \\
& & log\,$A_{\rm S00}$ & \makecell{$\mathcal{G}(\mu=0,\,\sigma=0.2)$ \\ $[-4,\,4]$} & -- \\[0.2cm]
& & log\,$A_{\rm S90}$ & \makecell{$\mathcal{G}(\mu=0,\,\sigma=0.2)$ \\ $[-4,\,4]$} & -- \\[0.2cm]
& & log\,$N_{\rm H, S}$ & $\mathcal{U}(22, 25)$ & cm$^{-2}$ \\
& & log\,$f_{\textsc{scat}}$ & $\mathcal{U}(-5, -1)$ & -- \\[0.2cm]
\hline
\rule{0pt}{1.5\normalbaselineskip}
\multirow{6}{*}{\textsc{P6}} & \multirow{6}{*}{\parbox{5cm}{\scriptsize $\texttt{ztbabs}*\texttt{cabs}*\texttt{cutoffpl}+\texttt{xclumpy\_refl}+\texttt{xclumpy\_lines}+\texttt{f}_{\rm scat}*\texttt{cutoffpl}$}} & $\Gamma\,[\textsc{ipl}]$ & \makecell{$\mathcal{G}(\mu=1.8,\,\sigma=0.2)$ \\ $[1.5,\,2.5]$} & -- \\[0.2cm]
& & log\,$\mathcal{A}\,[\textsc{ipl}]$ & $\mathcal{U}(-8, 2)$ & ph\,keV$^{-1}$\,cm$^{-2}$\,s$^{-1}$ at 1\,keV \\
& & log\,$N_{\rm H, Z}$ & $\mathcal{U}(23, 26)$ & cm$^{-2}$ \\
& & $\sigma$ & $\mathcal{U}(10, 70)$ & deg \\
& & $\theta_{\rm inc}$ & $\mathcal{U}(18.2, 87.1)$ & deg \\
& & log\,$f_{\textsc{scat}}$ & $\mathcal{U}(-5, -1)$ & -- \\[0.2cm]
\hline
\rule{0pt}{1.5\normalbaselineskip}
\multirow{7}{*}{\textsc{P7}} & \multirow{7}{*}{\parbox{5cm}{\scriptsize $\texttt{uxclumpy-transmit}+\texttt{uxclumpy-reflect}+\texttt{f}_{\rm scat}*\texttt{uxclumpy-omni}$}} & $\Gamma\,[\textsc{ipl}]$ & \makecell{$\mathcal{G}(\mu=1.8,\,\sigma=0.2)$ \\ $[1.2,\,2.8]$} & -- \\[0.2cm]
& & log\,$\mathcal{A}\,[\textsc{ipl}]$ & $\mathcal{U}(-8, 2)$ & ph\,keV$^{-1}$\,cm$^{-2}$\,s$^{-1}$ at 1\,keV \\
& & log\,$N_{\rm H, Z}$ & $\mathcal{U}(20, 26)$ & cm$^{-2}$ \\
& & $\textsc{TORsigma}$ & $\mathcal{U}(0, 84)$ & deg \\
& & $\textsc{CTKcover}$ & $\mathcal{U}(0, 0.6)$ & -- \\
& & $\theta_{\rm inc}$ & $\mathcal{U}(18.2, 87.1)$ & deg \\
& & log\,$f_{\textsc{scat}}$ & $\mathcal{U}(-5, -1)$ & -- \\[0.2cm]
\hline
\rule{0pt}{1.5\normalbaselineskip}
\multirow{7}{*}{\textsc{P8}} & \multirow{7}{*}{\parbox{5cm}{\scriptsize $\texttt{warped-disk}+\texttt{f}_{\rm scat}*\texttt{warped-disk-omni}$}} & $\Gamma\,[\textsc{ipl}]$ & \makecell{$\mathcal{G}(\mu=1.8,\,\sigma=0.2)$ \\ $[1.2,\,2.8]$} & -- \\[0.2cm]
& & log\,$\mathcal{A}\,[\textsc{ipl}]$ & $\mathcal{U}(-8, 2)$ & ph\,keV$^{-1}$\,cm$^{-2}$\,s$^{-1}$ at 1\,keV \\
& & log\,$N_{\rm H, Z}$ & $\mathcal{U}(20, 26)$ & cm$^{-2}$ \\
& & $f_{\rm disk}$ & $\mathcal{U}(0.125, 1)$ & -- \\
& & log\,$N_{\rm H, disk}$ & $\mathcal{U}(24, 25.5)$ & cm$^{-2}$ \\
& & $\theta_{\rm inc}$ & $\mathcal{U}(18.2, 87.1)$ & deg \\
& & log\,$f_{\textsc{scat}}$ & $\mathcal{U}(-5, -1)$ & -- \\[0.2cm]
\hline
\hline
\enddata
\tablecomments{
$^{a}$\,$\mathcal{G}(\mu,\,\sigma)$ and $\mathcal{U}(\textrm{min}, \textrm{max})$ denote Gaussian and uniform priors, respectively. For Gaussian priors, we include the full parameter range in square brackets beneath each.
$^{b}$\,The cross-calibration constants were varied in log space from the \citet{Madsen17} values, relative to FPMA when \nustar data was present. In the event that only soft data was available, only one dataset was fit and so no variable cross-calibration was used.
The parameter symbol definitions are: power law photon index ($\Gamma$), a given model component normalisation ($\mathcal{A}$), the line-of-sight column density ($N_{\rm H,\,Z}$), Thomson-scattered emission fraction ($f_{\textsc{scat}}$), obscurer inclination angle ($\theta_{\rm inc}$), half-opening angle ($\theta_{\rm tor}$), iron abundance ($A_{\rm Fe}$), column density out of the line-of-sight ($N_{\rm H, S}$), relative scaling of the face-on reprocessing component ($A_{\rm S00}$), relative scaling of the edge-on reprocessing component ($A_{\rm S90}$), torus angular width ($\sigma$), torus dispersion ($\textsc{TORsigma}$), covering factor of a Compton-thick inner ring of clouds ($\textsc{CTKcover}$), warp extent ($f_{\rm disk}$) and the column density of the warped disk ($N_{\rm H, disk}$).
The relevant papers for each obscuration model are: \texttt{BNsphere} \citep{Brightman11b},  \texttt{MYtorus} \citep{Murphy09}; \texttt{borus02} \citep{Balokovic18}; \texttt{XCLUMPY} \citep{Tanimoto18}; \texttt{UXCLUMPY} \citep{Buchner19}; \texttt{warped-disk} \citep{Buchner21b}. For a detailed review of decoupled modelling with \texttt{MYtorus}, see \citet{Yaqoob12}.
For \texttt{MYtorus} in coupled mode and \texttt{XCLUMPY}, the line-of-sight column densities we use throughout this paper are calculated as a function of the assumed obscurer geometries and the equatorial column density.
}
\end{deluxetable*}

\section{X-ray Spectral Results}\label{sec:6_results}
So far we have detailed a self-consistent Bayesian framework for the semi-automated fitting of many spectral models to each source with at least one X-ray spectrum available in NuLANDS. Here we detail the results of our fitting method, focusing on the bulk spectral properties of the sample and the process of model comparison to generate an $N_{\rm H}$ distribution with reliable bin heights and associated uncertainties.

\subsection{The Requirement for Multiple Models}\label{subsec:nhdistallmods}
Before performing model comparison on the spectral fits performed, we sought to investigate the effect choosing a model has on the shape of the $N_{\rm H}$ distribution. Of the 23 models we fit to every source, 19 have an available line-of-sight column density parameter. We consider each model in turn and compute the $N_{\rm H}$ distribution assuming the model can represent every observed spectrum well. This is of course an over-simplification, but we justify the test with the assumption that for progressively suppressed signal-to-noise ratio, the best-fit found by BXA would eventually be able to reproduce the data satisfactorily well. We note that the line-of-sight column density posteriors derived from clearly bad fits would be characteristically narrow since it is expected that a relatively smaller subset of spectral shapes could explain the data in a manner that is consistent with the global minimum in fit statistic. The resulting uncertainties on individual $N_{\rm H}$ distribution bins would then be artificially smaller, due to the incorrectly narrow $N_{\rm H}$ posteriors. However, we do not require accurate $N_{\rm H}$ distribution bin heights with this exercise, but rather to look for strong overall differences in $N_{\rm H}$ distribution shapes arising purely from selecting different models.

\begin{figure*}
\centering
\includegraphics[width=0.9\textwidth]{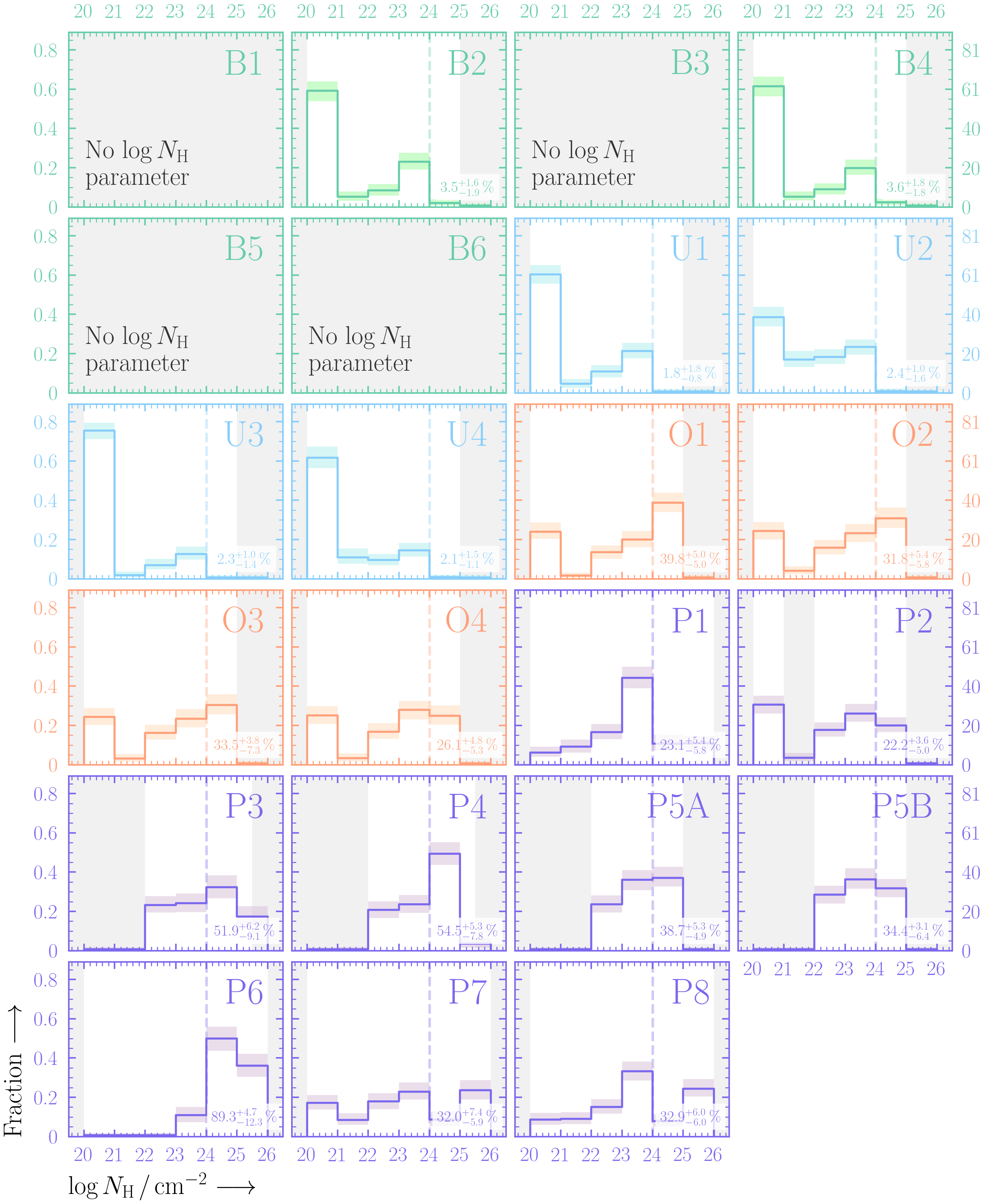}
\caption{\label{fig:modelnhdists} Line-of-sight column density distribution for the entire sample, assuming a different model per panel (for model descriptions, see Tables~\ref{tab:bmods}, \ref{tab:umods}, \ref{tab:omods} and \ref{tab:pmods}). Grey shaded regions show the column densities that are not allowed by each model, and the total Compton-thick fraction is shown in the bottom right corner of each panel. Note no attempt at model selection has been made at this stage, meaning that these column densities are purely to present the $N_{\rm H}$ parameter space attainable with each model setup. The figure showcases the importance of testing many different models when performing bulk X-ray spectral fits to a sample.}
\end{figure*}

We thus produce a column density distribution {\em per model} for all 102 sources with X-ray data. Using a similar hierarchical technique to that described in Section~\ref{sec:representative}, we use the histogram model to constrain the parent column density distribution given the individual source posterior distributions on line-of-sight column density. We show each $N_{\rm H}$ distribution per model in Figure~\ref{fig:modelnhdists} using single dex bins. It is broadly clear from the figure that choosing a model has a stark effect on the resulting $N_{\rm H}$ distribution. Interestingly, a number of the phenomenological model classes reproduce somewhat similar $N_{\rm H}$ distribution shapes. For example the U models, in which the reprocessed component is assumed to arise from the accretion disk which is absorbed by the line-of-sight obscurer, gives a far lower Compton-thick fraction than the O models in which the reflector is decoupled from the line-of-sight absorption. The B models are somewhat similar to the U models in their inability to produce high numbers of Compton-thick AGN, which may arise from their lack of reprocessed component.

The situation becomes more concerning when considering the physically-motivated P models. Despite each model being a physically-motivated prescription for the obscuring environment of AGN, the shape of each $N_{\rm H}$ distribution is uniquely different with drastically different predictions for the Compton-thick fraction. This suggests that considering too few spectral models can have model-dependent effects on the resulting measurements of line-of-sight column density. Similar results have recently been found in the simulation-based study of \citet{Saha22}, in which the authors found that data simulated from a given physical model can be successfully fit with a different obscuration model whilst giving drastically different posteriors for the line-of-sight column density amongst other key spectral parameters. Since any number of the AGN in NuLANDS are capable a-priori of being fit statistically well with any of the models we consider, we sought to include the additional systematic uncertainty associated with the choice of spectral model into the final $N_{\rm H}$ distribution.

\subsection{Model Selection}\label{subsec:modelcomp}
The exercise performed in Section~\ref{subsec:nhdistallmods} is useful to gain insight into the model-dependent uncertainties associated with choosing an obscuration model. However, reproducing a robust $N_{\rm H}$ distribution requires model selection to filter (1) physically incorrect or (2) statistically bad model fits from the sample. Physically incorrect model fits can be difficult to filter from the sample, since it is not implausible that a statistically good fit is acquired for a physically improbable scenario. Our first step is to exclude B models since four of the six do not possess a line-of-sight column density parameter, and the remaining two do not feature any component that can reproduce Compton scattering well.

A further complication arises from reprocessed emission in unobscured AGN. Accretion-disk X-ray reprocessing produces features that can look very similar to obscuration-based X-ray reprocessing with sufficiently low signal-to-noise -- i.e. a soft excess, iron fluorescence and a Compton hump (see e.g., \citealt{Garcia19}). Thus our next step was to filter the models considered for given targets by their optical classification. To do this, we restricted the models attainable by each source to be U models if the source is classified as a type~1\,--\,1.8 or a Narrow Line Seyfert 1, and the O/P models if the source were a type~1.9\,--\,2. We justify our separation based on optical classifications due to the overall very good agreement between optical and X-ray obscuration distinctions in local samples of AGN (e.g., \citealt{Koss17}). The U models feature a line-of-sight column density parameter, such that X-ray obscured, optically unobscured sources are still plausibly allowed with our model selection. Likewise, the O models and a number of P models feature line-of-sight column density parameters capable of $N_{\rm H}$\,$<$\,10$^{22}$\,cm$^{-2}$ such that X-ray unobscured, optically obscured sources are also plausible. However, Figure~\ref{fig:modelnhdists} shows that the U and O/P models do show a tendency for unobscured and obscured sight lines on average, respectively, which indicates some preference being imposed to the line-of-sight column density based on the restriction by optical class. Finally, type~1.9 sources are included with the obscured objects since existing analysis of type~1.9s has found a wide range of possible column densities, including above the Compton-thick limit (e.g., \citealt{Koss17,Shimizu18}).

It is well known that by decoupling the Compton-scattered continuum from the line-of-sight column density in the phenomenological manner of the O models can present difficulties in reproducing the reprocessing-dominated spectra of Compton-thick AGN (see discussion in e.g., \citealt{Balokovic17b}). The net undesirable result is unphysically large reprocessing scaling factors (that control the strength of the underlying reprocessed spectrum) whilst giving an unobscured sight line with artificially hard photon index (e.g., $\Gamma$\,$\lesssim$\,1.4). From initial fit tests, we find that our automated fitting technique can suffer from this issue for sufficiently low signal-to-noise ratio data. As such we refine our model selection to only allow P models for the type~1.9\,--\,2 AGN.

Next we turn our attention to filtering fits that are statistically worse than the highest Bayes Factor fits per source using our Bayes Factor threshold of 100. The line-of-sight column density quantile-based measurements that were selected for the type~1 and type~2 NuLANDS AGN not included in the 70-month BAT sample are shown in Tables~\ref{tab:logNHlos_Umods} and~\ref{tab:logNHlos_Pmods}, respectively. In total, we find that 19/40 of the corresponding type~2 AGN have at least one model giving a lower bound on line-of-sight column density above the Compton-thick limit. However, if we consider any source with at least one model giving a line-of-sight column density upper bound above the Compton-thick limit we find 33/40 sources. The average ratio between the maximum and minimum line-of-sight column density median per source is $\sim$\,1.4 orders of magnitude, but reaches $>$\,2 orders of magnitude in the most extreme cases. Such large differences in measured line-of-sight column density (and corresponding intrinsic luminosity) are relatively common in the literature, especially for Compton-thick AGN (e.g., \citealt{Boorman24}). Even though some of the varying column density medians may be consistent within uncertainties, such large differences confirm the results found in Section~\ref{sec:6_results} and Figure~\ref{fig:modelnhdists} -- namely that the choice of model can impinge significant changes on the column density inference for a given source. By including a large number of models in the analysis presented in this work, the column density constraints we find are expected to encompass a wider range of possibilities than if fewer models were used.

To investigate any possible preference for specific models after applying model selection, we used chord diagrams \citep{Holtan06}. Chord diagrams display inter-relationships between data with samples plotted as arcs on a circle and chords drawn to connect arcs to one another with a thickness and arc length that is proportional to their connections. For our purposes, we plot chord diagrams to show the proportion of models (the arcs) that are statistically well-fit by other models (the chords). The thickness of the chords show how frequently a given pair of models provide a statistically acceptable fit, and we provide chord diagrams for various subsamples of our data (i.e. type~1 and type~2 sources, as a function of signal-to-noise ratio in the 8\,--\,24\,keV band with \textit{NuSTAR}/FPMA.

\begin{figure*}
\centering
\includegraphics[width=0.9\textwidth]{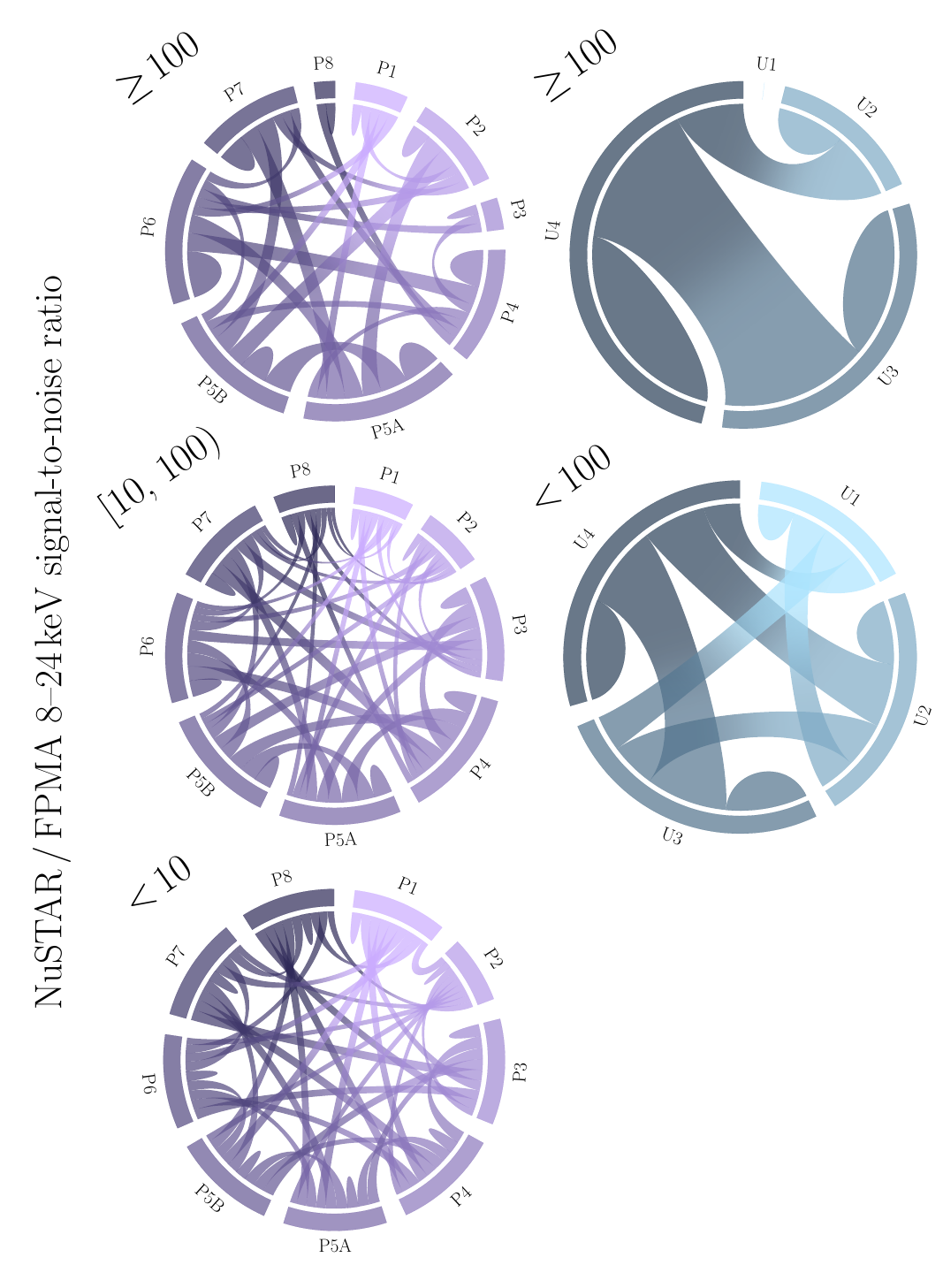}
\caption{\label{fig:chords} NuLANDS model selection chord diagrams binned by signal-to-noise ratio with \textit{NuSTAR}/FPMA in the 8\,--\,24\,keV band. The left and right columns show chord diagrams for the type~2s and type~1s, respectively, and the signal-to-noise ratio increases vertically. Each chord diagram shows the proportion of models that are selected with our model selection criterion as a function of every other model that is simultaneously selected. The length of the arcs shows the relative level to which a given model is favored over another, whereas the thickness of the chords represents the level to which two models can represent the data equally-well. See Section~\ref{subsec:modelcomp} for more details.}
\end{figure*}

Figure~\ref{fig:chords} presents chord diagrams for the type~2s on the left and type~1s on the right, with signal-to-noise ratio increasing from bottom to top. If we consider the lowest signal-to-noise type~2 chord diagram (bottom left panel), all arcs surrounding the circle are approximately the same length, indicating that there is a chord connecting every model to every other model with approximately equal proportion. This implies that all the physical obscuration models fit the data equally-well at low signal-to-noise ratio.

As the signal-to-noise ratio increases, all obscuration models are still selected approximately proportionately, though some trends become evident. For example, the arcs in the highest signal-to-noise ratio type~2 chord plot (upper left panel) are shortest for P1 (spherical obscurer), P3 (coupled borus02) and P8 (warpeddisk), implying these models are less capable of reproducing the observed data. On the other hand, there are only 8 type~2s in the highest signal-to-noise ratio bin, so this may simply reflect the specific obscurer geometry in this small set of sources. Considering all type~2s with signal-to-noise ratio above 10 (middle left and top left panels), some trends do appear. There is a general preference for decoupled models such as P4 (decoupled borus02), P5A (decoupled MYtorus with one reprocessor) and P5B (decoupled MYtorus with two reprocessors), likely due to the additional variable reprocessed component making those models more flexible. The thickness of the arcs for P6 (XCLUMPY) and P7 (UXCLUMPY) highlights the flexibility of those models in reproducing the broadband spectra of obscured AGN. For P7 specifically, the addition of the \texttt{CTKcover} parameter is particularly useful for fitting \textit{NuSTAR} spectra of the Compton hump in local obscured AGN \citep{Buchner19}.

For the type~1s, the situation is more complex. On first glance, the U1 and U2 models are broadly disfavored, especially at high signal-to-noise ratios. However, unlike for the type~2s in which the main difference between models is the obscuration model being used (i.e. the setup is effectively identical between all P models), the type~1 U models have significant component differences. For example, U1 and U2 lack warm absorption\footnote{We use the term `warm absorption' to refer to absorption from ionized material, typically manifesting with observable signatures in soft X-rays (e.g., \citealt{Miller06,Tombesi13}).}. However, we cannot rule out that \texttt{zxipcf} (the ionized absorption model in U3 and U4) is statistically favored due to its greater flexibility in fitting a range of sources. \citet{Ricci17_bassV} found that 22\% of unobscured non-blazar AGN from the 70-month BAT catalogue required including \texttt{zxipcf}, suggesting that warm absorption is required statistically, but not necessarily physically for the unobscured AGN in NuLANDS. Since Figure~\ref{fig:modelnhdists} shows that all U models struggle to reproduce log\,$N_{\rm H}$\,/\,cm$^{-2}$\,$>$\,21, any degeneracy with neutral line-of-sight column density is unlikely to affect the obscured and Compton-thick fractions of the sample. Future high-resolution spectroscopic studies in soft X-rays (e.g., with \textit{XRISM}; \citealt{Tashiro20} and \textit{Athena}/X-IFU; \citealt{Barret23}, see also \citealt{Gandhi22}) will test the need for warm absorption.

Since Figure~\ref{fig:modelnhdists} shows substantial differences in measured line-of-sight column density per model and it appears quite common for almost all models to be selected amongst both the type~1s and type~2s, we require a method that propagates the posterior probabilities of all possible selected models per source into the global line-of-sight $N_{\rm H}$ distribution.

\subsection{Model Verification}\label{subsec:modelverification}
Performing model comparison removes fits that are statistically worse than the most favourable fit per source. However, model comparison does not guarantee a statistically good fit is selected in the first place. Two risks that arise which could result in a systematic bias to the resulting population $N_{\rm H}$ distribution are (1) local minima giving incorrect model fits and (2) models that do not contain enough complexity for the given data quality. An additional complication can also arise from variability between soft and \textit{NuSTAR} exposures but we defer discussion of this to Section~\ref{subsec:var}, in which we show that variability should not be a strong concern in the sample. Local minima are a well-known issue associated with X-ray spectral fitting, and a major advantage of using nested sampling is that the vast majority of model fits are expected to give the global minimum in fit statistic and its associated line-of-sight column density posterior. Since most of our sample are X-ray-bright Seyfert galaxies, a more likely scenario is that the chosen models themselves cannot reproduce the complexity encompassed in the observed data. We note that for this paper, in which the $N_{\rm H}$ distribution is the primary goal, we seek to understand the average quality of our spectral fits to infer any systematic biases that may affect our line-of-sight column density posteriors. As described in Section~\ref{subsec:modelverification}, our strategy is to use posterior predictive checks, and our specific process was as follows:

\begin{enumerate}
    \item Select the highest Bayes Factor models per source.
    \item Per model fit, select 20 random posterior rows and save the real fit statistic after loading the unbinned data, without fitting.
    \item For each posterior row, simulate 20 random observations with each dataset in question and save the simulated fit statistics for the corresponding unbinned data without fitting.
\end{enumerate}

\begin{figure*}
\centering
\includegraphics[width=0.9\textwidth]{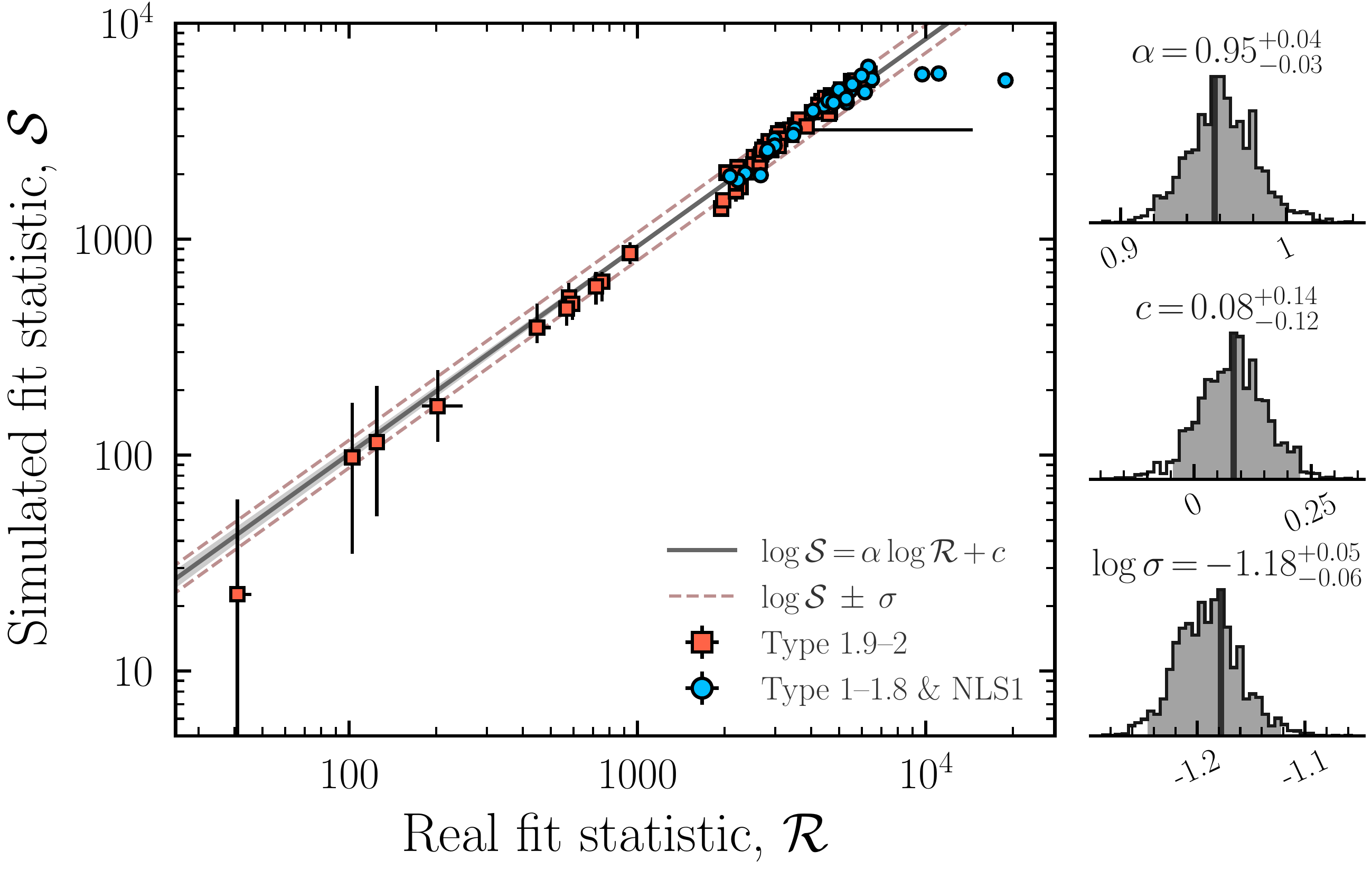}
\caption{\label{fig:ppc} (Left) Posterior predictive checks showing the real fit statistics vs. simulated fit statistics for the highest Bayes Factor models per source with X-ray data, separated into type~1s and~2s with blue and red points, respectively. The straight line shown in dark grey and associated shading is the median posterior model line and associated posterior uncertainty. The dashed lines on either side of the median denote the intrinsic scatter in the vertical direction from the fit. (Right) from top to bottom shows the marginalised posterior for the straight line fit gradient, intercept and intrinsic scatter, respectively. Each marginalised distribution shows the mode with dark grey shading, together with the 95\% highest density interval with light grey shading. The posterior straight line fit is consistent with a one-to-one relation on average for the entire sample, suggesting the majority of model fits are acceptable on average.}
\end{figure*}

We only consider unbinned data in our posterior predictive checks to avoid stochastic uncertainties arising from binning simulated data. Our method provides distributions of both real and simulated fit statistics, arising from the sampled posterior rows and corresponding simulations being performed. The result is shown in the left main panel of Figure~\ref{fig:ppc}, in which the real fit statistic is plotted against the simulated fit statistic (with associated 68\% quantile errorbars) in red and blue for sources fit with P and U models, respectively.

To assess the quality of the spectral fits in an ensemble-averaged manner, we next fit a straight line to the data in logarithmic space. Perfect fits would result in a one-to-one straight-line fit. We use UltraNest to perform fitting with a linear model that includes a slope, intercept and intrinsic scatter in the vertical direction. By comparing the resulting parameter values of the straight-line fit to a perfect one-to-one relation (i.e. slope unity and intercept zero), we can gain insight into the goodness-of-fit of the sample on average. We plot the resulting straight line fit in the left panel of Figure~\ref{fig:ppc} using a dark grey line with light grey shading to denote the median posterior fit and associated posterior uncertainty, respectively. The additional intrinsic scatter is plotted with two dashed lines either side of the relation. The corresponding straight-line fit marginalized parameter posterior distributions are shown on the right side of the plot with the slope, intercept, and intrinsic scatter in the upper, center, and bottom panels, respectively. For each distribution, the mode is shown with dark grey shading, and the 95\% confidence interval is denoted with light grey shading.

We find the slope to be consistent with unity and intercept to be consistent with zero within $\sim$\,95\% probability. Such results indicate that on average the spectral fits in the sample are consistent with being able to reproduce the general spectral shapes contained within the data. Interestingly, the average intercept tends to more negative values, suggesting that our data has additional spectral variations not encompassed by the spectral models being fit. On further investigation, a plausible contributor to this could be the presence of a stronger aperture background component in FPMB spectra relative to FPMA on average (see Figure~\ref{fig:significance}). Since our models are setup with a cross-calibration constant forced to be relative to FPMA, any increased background component in FPMB could give rise to additional scatter in the observed data and hence an overall worse fit statistic for FPMB relative to FPMA.

In addition, the slope found is shifted more in favour of shallower values. Since increased real fit statistics are associated with higher signal-to-noise data, a shallower slope indicates that it is more difficult to fit bright spectra with the relatively simpler model setup chosen for all sources. There are clearly three such cases in Figure~\ref{fig:ppc} given by the three blue points with the highest real fit statistic values on the plot. These fits correspond to Ark\,120, 3C\,120 and MCG\,--06--30--015. All three sources have been studied in extensive detail in the past, revealing complex X-ray spectral and timing properties that require additional model complexity to fit (see e.g., \citealt{Matt14,Marinucci14_mcg06,Rani18,Wilkins19}).

It is outside the scope of this paper to provide physically-motivated fits for these sources. However, it is important to check that the line-of-sight column density posteriors for each does not impact the measured obscured and Compton-thick fractions in the final $N_{\rm H}$ distribution. We thus compare the column density posteriors from our fitting of Ark\,120, 3C\,120 and MCG\,--06--30--015 to the results of \citet{Ricci17_bassV} as a comparison. Of the three sources, the line-of-sight column density posteriors we derive are in agreement with \citeauthor{Ricci17_bassV} for Ark\,120 and MCG\,--06--30--015, finding both to have $N_{\rm H}$\,=\,10$^{20}$\,--\,10$^{21}$\,cm$^{-2}$. In contrast, we find a discrepancy for 3C\,120 with a line-of-sight column density in the range $N_{\rm H}$\,=\,10$^{20}$\,--\,10$^{21}$\,cm$^{-2}$, compared to the range $N_{\rm H}$\,=\,10$^{21}$\,--\,10$^{22}$\,cm$^{-2}$ for \citet{Ricci17_bassV}. However, since the Galactic column density we use for 3C\,120 is 1.94\,$\times$\,10$^{21}$\,cm$^{-2}$, we find that the net line-of-sight column density would agree with \citet{Ricci17_bassV}. For further discussion of the effect Galactic column density has on the lowest bins of the $N_{\rm H}$ distribution, see Section~\ref{subsec:nhdistdiscussion}.

The posterior predictive checks shown in Figure~\ref{fig:ppc} thus indicate that on average our automated fitting method is able to reproduce the bulk shape of the X-ray spectra in the sample.

\subsection{The Column Density Distribution}\label{subsec:nhdist}
Having used a few different metrics to select models in Section~\ref{subsec:modelcomp}, each source is allowed to have N line-of-sight \nh posterior distributions, where N is the number of suitable models per source, given their optical spectroscopic classification. Each accepted model then fits the observed spectra equally well within our assumed model selection thresholds. Incorporating every accepted model posterior per source is important, since individual \nh posterior distributions can differ in terms of not only quantiles but also shape. We specifically refer to the ability of different obscuring models fitting the same source with different posterior distributions as \lq geometry-dependent degeneracies\rq\ (previously discussed in \citealt{Yaqoob12,Brightman15,Lamassa19,Saha22,Kallova24}). However, the ability of BXA to traverse a parameter space globally means that each accepted posterior for a given source should have a negligible if not non-existent effect from local minima. We hence assume that each BXA posterior represents a robust possible solution to explain a given set of observed source X-ray spectra, and attempt to include all possible solutions in the final log\,$N_{\rm H}$ distribution with a Hierarchical Bayesian Model (HBM).

The HBM we use is very similar in form to the histogram model with Dirichlet prior described in Section~\ref{sec:representative}, but with line-of-sight $N_{\rm H}$ posteriors from our X-ray spectral fitting. The parameters of the parent model are the bin heights of the $N_{\rm H}$ distribution in unit dex bins from $N_{\rm H}$\,=\,10$^{20}$\,--\,10$^{24}$\,cm$^{-2}$, and one two-dex wide Compton-thick bin with $N_{\rm H}$\,=\,10$^{24}$\,--\,10$^{26}$\,cm$^{-2}$. We additionally performed a Monte Carlo simulation to incorporate systematic model dependencies (e.g., the choice of model setup, the obscuration geometry, model parameterisations) into the final distribution. We selected one random accepted line-of-sight $N_{\rm H}$ posterior per source and generated the $N_{\rm H}$ distribution 200 times before appending the HBM model chains together. After experimenting with different numbers of repeats, 200 was chosen since all sources were sampled after significantly fewer iterations than this. As detailed earlier, 20 additional type~2 sources in the sample did not have any X-ray spectral constraints. To predict the line-of-sight $N_{\rm H}$ for these sources, we applied our Monte Carlo HBM to just the type~2 sources in the sample with X-ray spectral constraints and used the resulting $N_{\rm H}$ distribution as the predicted posterior for each type~2 lacking X-ray spectra. By including the 20 sources lacking X-ray spectra self-consistently, we assume the remaining type~2s share the same characteristics as the existing type~2s with X-ray spectra. Since the type~2 $N_{\rm H}$ distribution is heavily skewed to $N_{\rm H}$\,$>$\,10$^{23}$\,cm$^{-2}$, the main effect of incorporating sources with no X-ray data is to marginally increase the obscured and Compton-thick fractions.

\begin{figure*}
\centering
\includegraphics[width=0.8\textwidth]{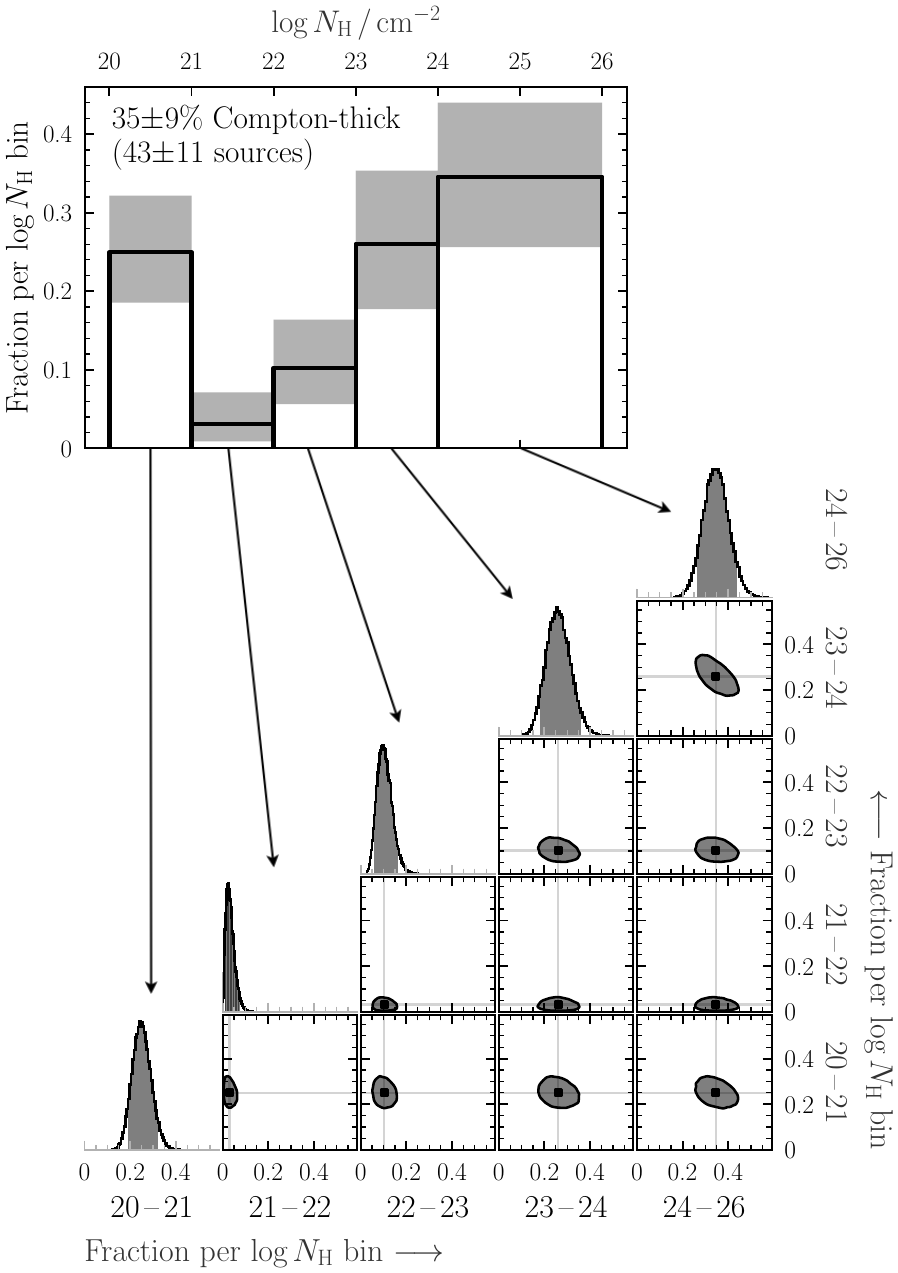}
\caption{\label{fig:nhdist} \textbf{The NuLANDS $N_{\rm H}$ distribution.} The result from the HBM described in Section~\ref{subsec:nhdist} is shown here as a corner plot with each axis representing a different parameter 90\% confidence range from the model (i.e. the bin heights in the $N_{\rm H}$ distribution). Each parameter is not strongly degenerate with one another, apart from a slight negative diagonal dependence between the Compton-thin ($N_{\rm H}$\,=\,10$^{23}$\,--\,10$^{24}$\,cm$^{-2}$) and Compton-thick ($N_{\rm H}$\,=\,10$^{24}$\,--\,10$^{26}$\,cm$^{-2}$) fractions. This highlights the power of our self-consistent method for deriving line-of-sight column densities. For a number of sources consistent with both classifications, a source can become Compton-thick so long as the Compton-thin fraction is reduced and vice-versa. Due to the range in maximum line-of-sight column density values allowed by each model considered, we choose to represent the Compton-thick fraction as a single two dex bin encompassing $N_{\rm H}$\,=\,10$^{24}$\,--\,10$^{26}$\,cm$^{-2}$.
}
\end{figure*}

The corresponding $N_{\rm H}$ distribution we find with our Monte Carlo HBM is shown in Figure~\ref{fig:nhdist}, with the individual fractions per bin given in Table~\ref{tab:nhdist}. By plotting the inter-parameter dependencies of the HBM in Figure~\ref{fig:nhdist}, we show there is no strong degeneracy between any bin fractions apart from the Compton-thin ($N_{\rm H}$\,=\,10$^{23}$\,--\,10$^{24}$\,cm$^{-2}$) and the Compton-thick ($N_{\rm H}$\,=\,10$^{24}$\,--\,10$^{26}$\,cm$^{-2}$) fractions which shows a slight negative correlation. Such a degeneracy indicates that a number of sources have $N_{\rm H}$ posteriors consistent with both obscuration classes, such that the overall Compton-thick fraction can only increase at the detriment of the Compton-thin fraction and vice-versa. This inter-bin fraction degeneracy also shows the benefit of our self-consistent fitting method, and shows the difficulty associated with defining sources as Compton-thick when too few obscurer geometries are considered in the modelling process.

\begin{deluxetable}{cc}
\tablewidth{0pt}
\tablecaption{NuLANDS $N_{\rm H}$ distribution fractions.\label{tab:nhdist}}
\tablehead{
\colhead{$N_{\rm H}$ bin boundaries\,/\,cm$^{-2}$} &
\colhead{Fraction per dex $\times$ bin width}
}
\tableheadfrac{0.1}
\startdata
\hline
10$^{20}$\,--\,10$^{21}$ & $0.25^{+0.07}_{-0.06}$\\
10$^{21}$\,--\,10$^{22}$ & $0.03^{+0.04}_{-0.02}$\\
10$^{22}$\,--\,10$^{23}$ & $0.10^{+0.06}_{-0.05}$\\
10$^{23}$\,--\,10$^{24}$ & $0.26^{+0.09}_{-0.08}$\\
10$^{24}$\,--\,10$^{26}$ & $0.35 \pm 0.09$\\
\hline
\enddata
\tablecomments{All fractions were calculating using the Monte Carlo Hierarchical Bayesian Model method described in Section~\ref{sec:5_method} with results presented in Section~\ref{subsec:nhdist} and Figure~\ref{fig:nhdist}.}
\end{deluxetable}

\section{Discussion} \label{sec:discussion}
% Remember to discuss Severgnini+12: https://ui.adsabs.harvard.edu/abs/2012A%26A...542A..46S/abstract

\subsection{The NuLANDS Column Density Distribution}\label{subsec:nhdistdiscussion}
The $N_{\rm H}$ distribution for the full NuLANDS sample is presented in Figure~\ref{fig:nhdist} as a one-dimensional histogram and a two-dimensional corner plot. The corner plot highlights the structure of the hierarchical model used to construct the $N_{\rm H}$ distribution, in which the fractions in each bin are the free parameters of the model. No strong correlations between individual bin fractions are found for most cases. The most notable parameter dependence between individual bin fractions is a slight negative trend between the Compton-thin ($N_{\rm H}$\,=\,10$^{23}$\,--\,10$^{24}$\,cm$^{-2}$) and Compton-thick ($N_{\rm H}$\,=\,10$^{24}$\,--\,10$^{26}$\,cm$^{-2}$) fractions. Such a trend indicates that a number of the AGN in NuLANDS have line-of-sight $N_{\rm H}$ posteriors from some proportion of the selected model fits that are consistent with both Compton-thin and Compton-thick classifications. Due to the lack of such trends between other bin fractions, the trend between Compton-thin and Compton-thick fractions highlights the overall difficulty, even with \textit{NuSTAR}, to classify the line-of-sight column density with high precision when the source is heavily obscured and a single spectral model is being used for inference. A much weaker anti-correlation is visible between the unobscured ($N_{\rm H}$\,=\,10$^{20}$\,--\,10$^{21}$\,cm$^{-2}$) and Compton-thick fractions. Though unlikely to affect our final column density distribution, a negative trend could suggest a small number of sources with unobscured reprocessing signatures (e.g., from accretion disk and/or outflow-based reprocessing; \citealt{Parker22,Matzeu22}) that are being explained by high column density reprocessing in the circum-nuclear obscurer.

Of the full sample comprising 122 sources, we find a Compton-thick fraction of 35\,$\pm$\,9\% (equivalent to 43\,$\pm$\,11 sources) with $N_{\rm H}$\,=\,10$^{24}$\,--\,10$^{26}$\,cm$^{-2}$ where the fraction has been normalised to unity in the log\,$N_{\rm H}$\,=\,10$^{20}$\,--\,10$^{26}$\,cm$^{-2}$ range. The NuLANDS Compton-thick fraction is thus fully consistent with the value found by \citet{Buchner15} of 38$^{+8}_{-7}$\%, broadly consistent with the value found by \citet{Ananna19} of 50\,$\pm$\,9\% up to redshift 0.1\footnote{Equivalent to a local volume of 464\,Mpc with our assumed cosmological parameters.} within 90\% confidence and the value found by \citet{Ueda14} of $\sim$44\%.

For intermediate obscuration levels, $N_{\rm H}$\,=\,10$^{22}$\,--\,10$^{24}$\,cm$^{-2}$, we find an increase in the fraction of sources from 10$^{+6}_{-5}$\% for $N_{\rm H}$\,=\,10$^{22}$\,--\,10$^{23}$\,cm$^{-2}$ to 26$^{+9}_{-8}$\% for $N_{\rm H}$\,=\,10$^{23}$\,--\,10$^{24}$\,cm$^{-2}$. Interestingly, obscuration is expected to be partly explained by host-galaxy obscurers below $N_{\rm H}$\,$\sim$\,10$^{23}$\,--\,10$^{23.5}$\,cm$^{-2}$ \citep{Buchner17a,Buchner17b,Silverman23}, though with a strong dependence on redshift \citep{Andonie22}. Compton-thick levels are unlikely to be produced by kpc-scale obscurers in all but the most extreme compact and luminous starbursts in the nearby universe (see e.g., \citealt{Gilli22,Andonie24}). Deriving properties of the host-galaxy obscurer and its connection with the central AGN requires comprehensive multi-wavelength contributions (e.g., \citealt{GarciaBurillo21}), but is outside the scope of this paper. We note that if a majority of the NuLANDS obscured AGN were dominated by large-scale host galaxy obscurers, the isotropy tests between the mid-to-far infrared continuum and optical narrow line regions shown in Section~\ref{sec:representative} and Figure~\ref{fig:representativeness} would not be expected to agree so well between the type~1 and type~2 sources. Thus, such isotropy tests indicate the NuLANDS $N_{\rm H}$ distribution in Figure~\ref{fig:nhdist} is dominated by circum-nuclear rather than large-scale obscuration.

The overall shape of the one dimensional $N_{\rm H}$ distribution presents an apparent drop of sources with $N_{\rm H}$\,=\,10$^{21}$\,--\,10$^{22}$\,cm$^{-2}$ from the much higher fraction of unobscured sources. Though a somewhat common feature of previous $N_{\rm H}$ distributions (see Section~\ref{subsec:lognhcomp}), we adhere caution to interpreting a decrease since an unknown fraction of the $N_{\rm H}$\,=\,10$^{20}$\,--\,10$^{21}$\,cm$^{-2}$ sources are $N_{\rm H}$ upper limits. The exact $N_{\rm H}$ upper limits are either limited by the minimum allowed $N_{\rm H}$ per fit (the lowest considered in the models is $N_{\rm H}$\,=\,10$^{20}$\,cm$^{-2}$) or could be degenerate with the Galactic column density that was fixed in each fit. Since the Galactic $N_{\rm H}$ values used in this work (from \citealt{Willingale13}) were distributed between $N_{\rm H}$\,$\sim$\,10$^{20.0}$\,--\,10$^{21.3}$\,cm$^{-2}$ (i.e. entirely encompassing the lowest column density bin of the $N_{\rm H}$ distribution), it is difficult to determine accurately how many sources could have intrinsic line-of-sight column densities $N_{\rm H}$\,$<$\,10$^{20}$\,cm$^{-2}$.

\subsection{NuLANDS Column Density Dependencies}

\subsubsection{Distance}\label{subsec:nH_v_D}
\begin{figure*}
\centering
\includegraphics[width=0.8\textwidth]{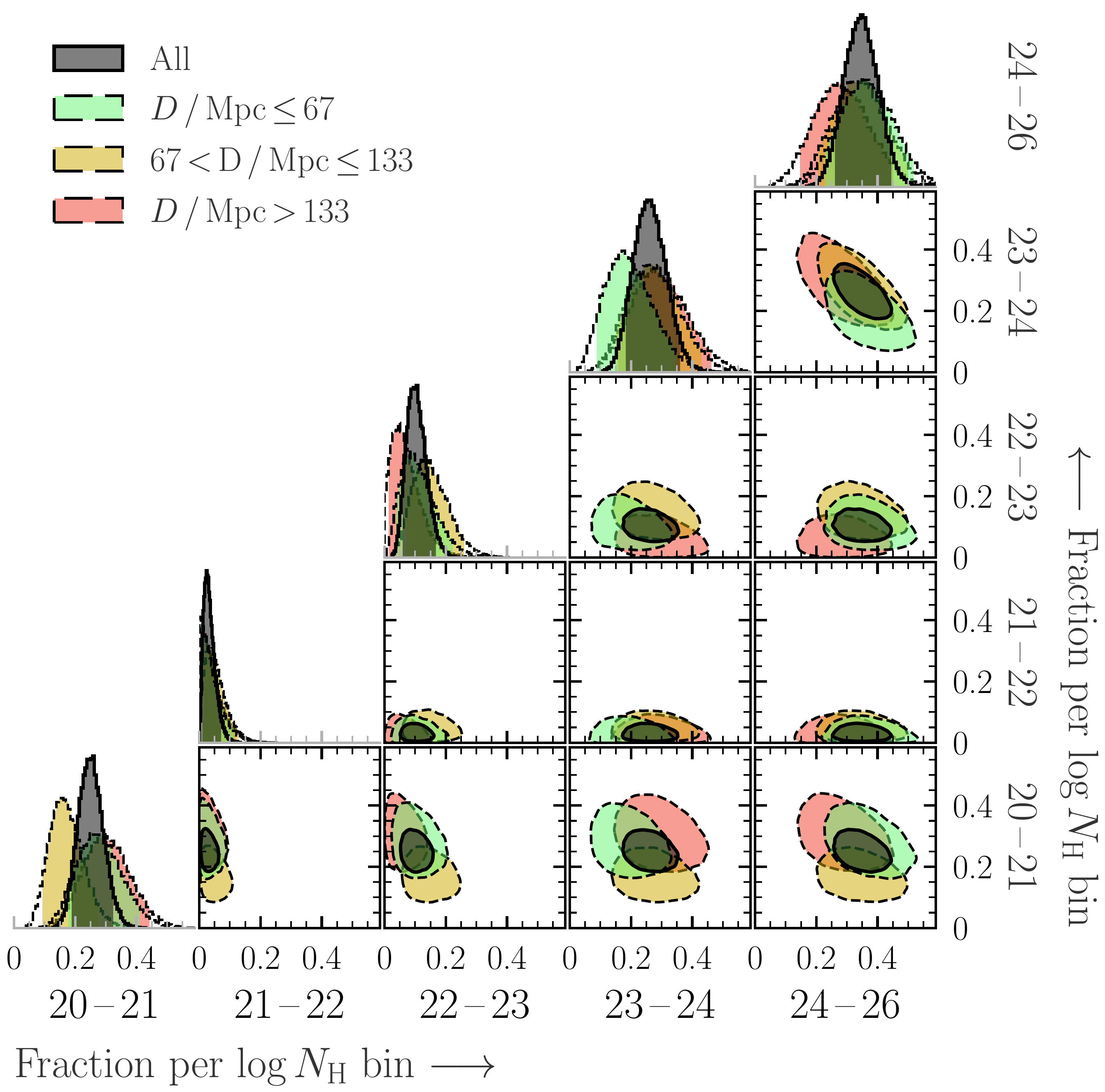}
\caption{\label{fig:nhdist_DMpc}$N_{\rm H}$ distribution hierarchical model corner plot, binned by distance in Mpc. No significant deviations are observed with distance.
}
\end{figure*}

As noted earlier, a clear trend in hard X-ray all-sky flux-limited surveys of AGN is the decrease of the Compton-thick AGN fraction with distance \citep{Ricci15,TorresAlba21_clemsonVI} due to an observational bias against heavily obscured sources. Since NuLANDS was constructed to select Compton-thick AGN with approximately equal efficacy as less-obscured AGN, it is important to test for a similar effect with distance in our sample. In Figure~\ref{fig:nhdist_DMpc} we show the NuLANDS log\,$N_{\rm H}$ distribution for the entire sample (black contours) and for three different bins of distance. Within the 90\% percentile range for each distribution (the shaded regions in the one-dimensional histograms), there is no significant change in the Compton-thick fraction. In addition, all other log\,$N_{\rm H}$ distribution bins are consistent within the 90\% percentile range, suggesting that distance effects do not strongly affect the NuLANDS fitting results.

\subsubsection{Total Infrared Luminosity}\label{subsec:nh_v_lirtot}
Higher galaxy total infrared luminosities can imply large quantities of dust on large galactic scales and/or merging/interacting systems. In the former scenario, large scale host-galaxy dust can lead to enhanced X-ray obscuration of AGN up to $N_{\rm H}$\,$\sim$\,10$^{23.5}$\,cm$^{-2}$ \citep{Buchner17a}. However, for merging/interacting systems, the material is thought to be funneled to the parsec-scale environment surrounding the supermassive black hole, simultaneously triggering star formation (e.g., \citealt{Sanders88}) and enhancing circum-nuclear obscuration (e.g., \citealt{Ricci17_mergers,Ricci21_goals}).

\begin{figure*}
\centering
\includegraphics[width=0.8\textwidth]{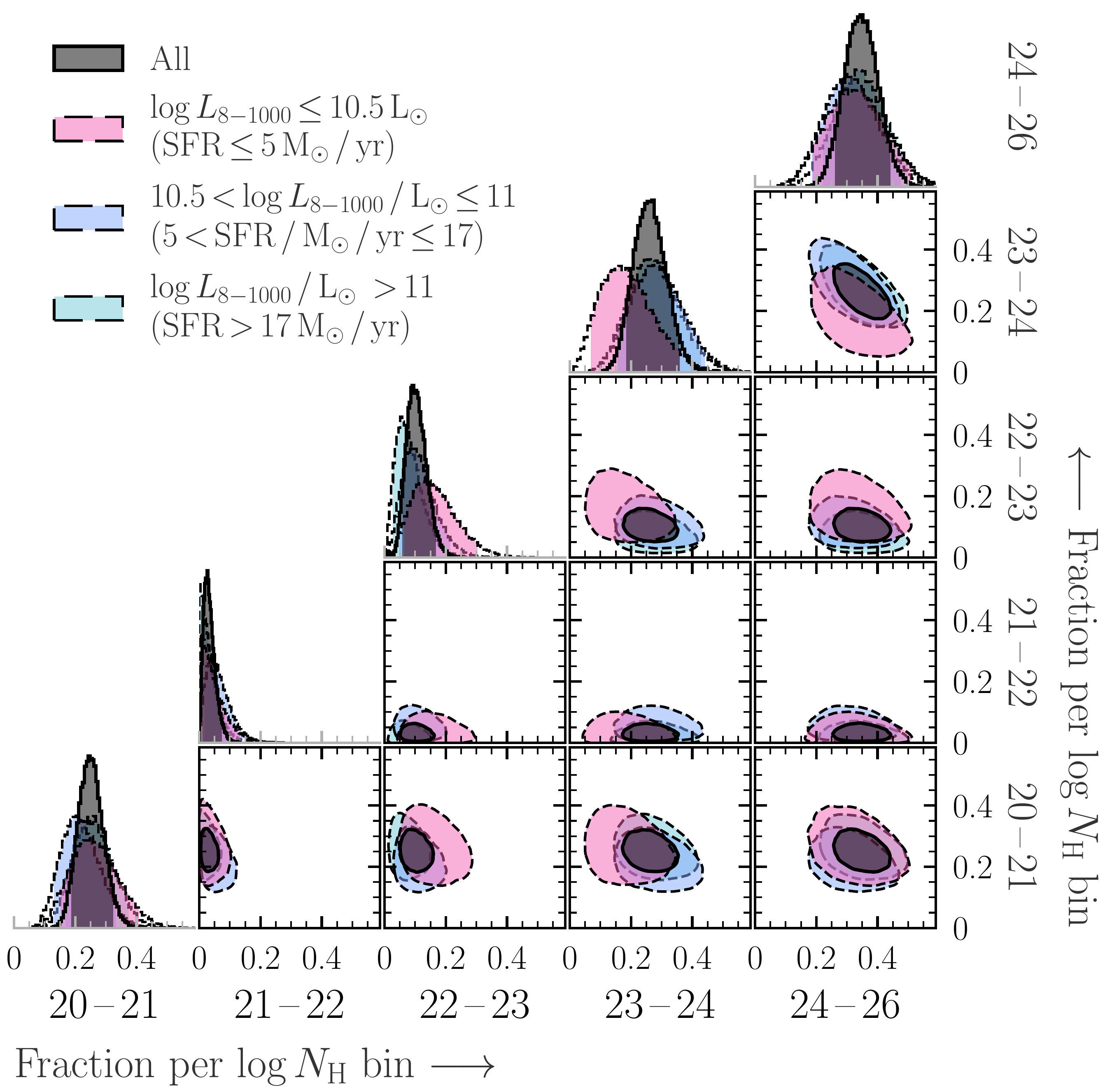}
\caption{\label{fig:nhdist_logL81000}$N_{\rm H}$ distribution hierarchical model corner plot, binned by total infrared luminosity, log\,$L_{8-1000 \mu{\rm m}}$. All $N_{\rm H}$ contours are consistent with the values found for the whole sample.
}
\end{figure*}

As discussed in Section~\ref{sec:representative}, if a substantial fraction of the NuLANDS AGN had enhanced levels of large-scale host-galaxy dust obscuration, the isotropy tests highlighted in Figure~\ref{fig:representativeness} would not be expected to agree between type~1 and~2 AGN so well. As shown in Figure~\ref{fig:nhdist_logL81000}, we find no large disagreements between log\,$N_{\rm H}$ fractions within the 90\% percentile range for the total NuLANDS sample (black) and binned by total infrared luminosity. Note that we also provide a total infrared luminosity-based estimate for star formation rate (SFR) using the relation from \citet{Kennicutt98}.

The lack of large offsets for the $N_{\rm H}$ measurements with different total infrared luminosities provides further evidence that the NuLANDS selection can identify AGN isotropically. We include the corresponding translation from the total infrared luminosity bin edges to star formation rate in the legend of Figure~\ref{fig:nhdist_logL81000} assuming the relation of \citet{Kennicutt98}. Whilst this may indicate no strong relation between star formation rate and line-of-sight column density, we caution the reader that the presence of an AGN in the infrared can dramatically affect star formation rate estimations using that relation.

\subsubsection{Near-to-Mid Infrared Colors}\label{subsec:nh_v_wisecol}
The NuLANDS AGN all satisfy a warm \textit{IRAS} color classification, essentially a relatively steep mid-to-far infrared color in the 25--60\,$\mu$m band. Such an AGN imprint on the observed spectrum above $\sim$25$\mu$m would be expected to correlate with similar infrared color selections at shorter wavelengths, for example in the near-to-mid-infrared. To date, many such near-to-mid infrared color selections exist for e.g., \textit{WISE} \citep{Jarrett11,Stern12,Mateos12,Assef17,Satyapal18}. To investigate such near-to-mid infrared color selections, and their possible effect on identifying log\,$N_{\rm H}$ for a given sample we apply a number of popular \textit{WISE} color selections from the literature to the NuLANDS AGN.

\begin{figure*}
\centering
\includegraphics[width=0.8\textwidth]{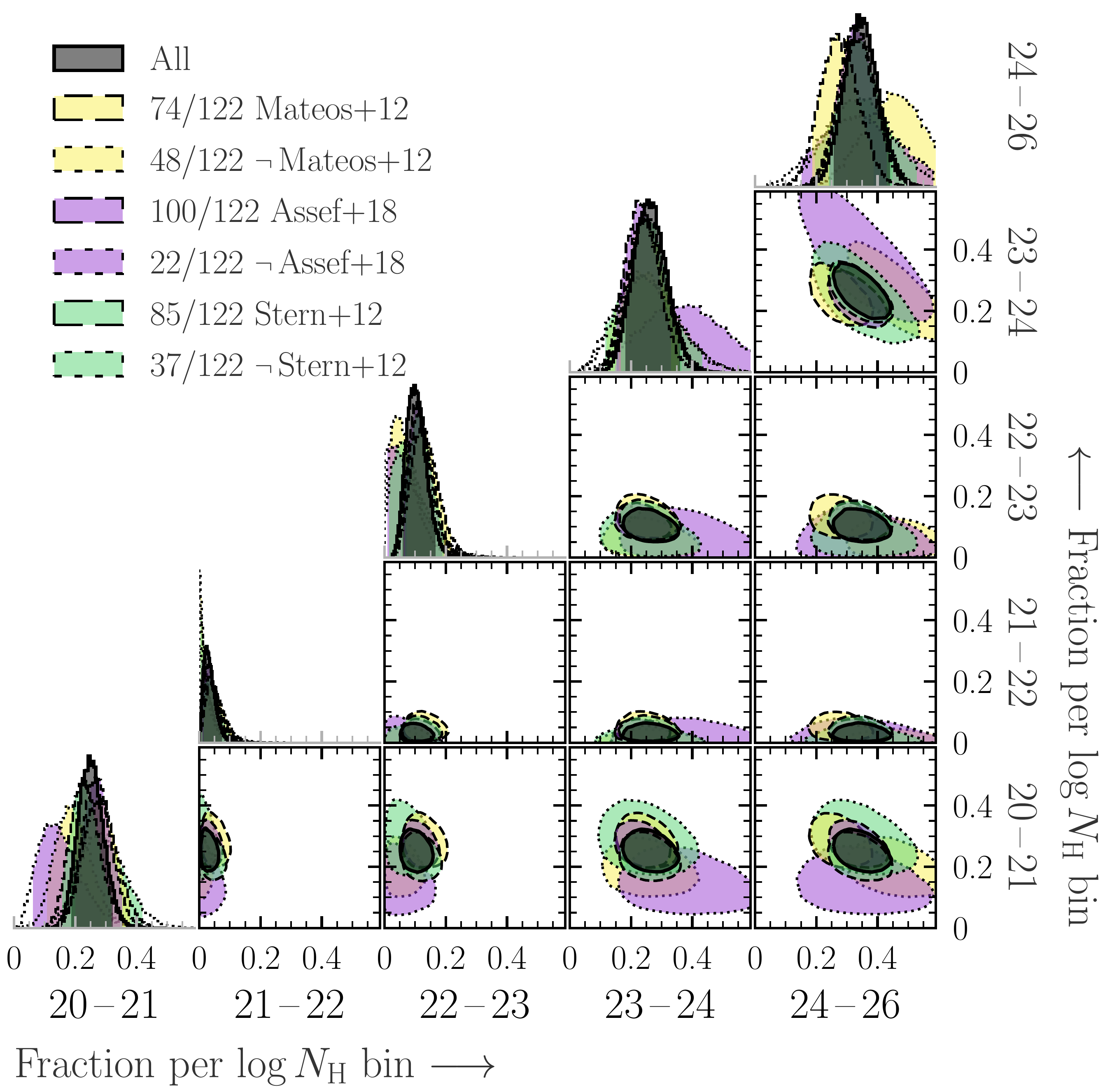}
\caption{\label{fig:nh_wisecols}$N_{\rm H}$ distribution hierarchical model corner plot, binned by near-to-mid infrared \textit{WISE} colors. No strong deviations are observed, whether in (solid lines) or out (dashed lines) of the selection criteria from \citet{Stern12}, \citet{Mateos12}, or \citet{Assef17}, indicating no strong relation between X-ray-derived $N_{\rm H}$ and star-formation contamination (which tends to preferentially affect near-to-mid infrared AGN color selection criteria).
}
\end{figure*}

In Figure~\ref{fig:nh_wisecols} we report the NuLANDS log\,$N_{\rm H}$ distribution for AGN that are and are not selected based on the \textit{WISE} color selections of \citet{Stern12}, \citet{Mateos12} and \citet[we use the R90 selection specifically]{Assef17}. We find that the majority ($>$\,60\%) of NuLANDS AGN are identified as AGN based on all \textit{WISE} color selections chosen, in broad agreement with their confirmed warm \textit{IRAS} colors between 25--60\,$\mu$m. For the AGN not selected, a likely reason is host galaxy dilution caused by star formation and other host galaxy-related processes dominating the bolometric output of the galaxy rather than the AGN (e.g., \citealt{Murphy09_IR,Eckart10,Mateos13,Pfeifle22}).

We find consistent Compton-thick fractions for NuLANDS AGN both in and out of the \textit{WISE} color selections considered. A similar trend was reported in \citet{Gandhi15a} who found no clear preference for bona fide Compton-thick AGN in the local Universe across the \textit{WISE} color-color space that were originally selected via a wide range of multi-wavelength methods. As expected, the largest uncertainties on log\,$N_{\rm H}$ bin fractions are found for AGN not selected by each \textit{WISE} color criterion. This is likely a result of having more sources selected as AGN than not, and also that \textit{WISE} color criteria are known to be more efficient for high X-ray luminosities often associated with AGN-dominated systems (e.g., \citealt{Stern12,Mateos13}). All $N_{\rm H}$ bin fractions are found to be consistent within 90 percentile contours, with some slight offsets observed. The largest log\,$N_{\rm H}$ bin fraction offsets are found for sources not selected as AGN with the 90\% reliability (R90) cut of \citet{Assef17}, though note that of all NuLANDS AGN, the R90 cut is the most effective by selecting 82\% of the NuLANDS sources. A possible link between the NuLANDS near-to-mid infrared spectral shapes and their X-ray column densities would require broadband spectral energy distribution decomposition, which is outside the scope of the current work. However, the general agreement between $N_{\rm H}$ bin fractions for NuLANDS AGN outside and inside a number of different \textit{WISE} color selections provides further evidence for the isotropic nature of the NuLANDS selection.

\subsubsection{AGN Optical Classification}\label{subsec:nh_v_type}

Several previous works have found a correlation between optical spectral classification and X-ray-derived $N_{\rm H}$ above and below $\sim$10$^{22}$\,cm$^{-2}$. For example, \citet{Koss17} find $\sim$94\% agreement between Seyfert~1--1.8 and Seyfert~2 for X-ray $N_{\rm H}$ below and above a boundary of $\sim$\,10$^{21.9}$\,cm$^{-2}$, respectively, in the \textit{Swift}/BAT 70-month sample. \citet{Merloni14} compare the X-ray and optical/UV classifications of the \textit{XMM-Newton} COSMOS survey complete to observed X-ray flux, finding a substantial fraction of sources with unobscured/obscured X-ray classifications but the reverse in the optical/UV. The authors find that optically-classified type~2 AGN that are unobscured in X-rays are likely caused by host galaxy dilution, whereas for optically-classified type~1 AGN that are obscured in X-rays could be caused by dust-free gas within/inside the broad line region. For the latter cross-classification sources, the authors find the majority to be at relatively high intrinsic luminosity, which are likely less relevant in the volume-limited NuLANDS sample considered here. For the former class, this may be a possibility in a subset of the sources.

\begin{figure*}
\centering
\includegraphics[width=0.8\textwidth]{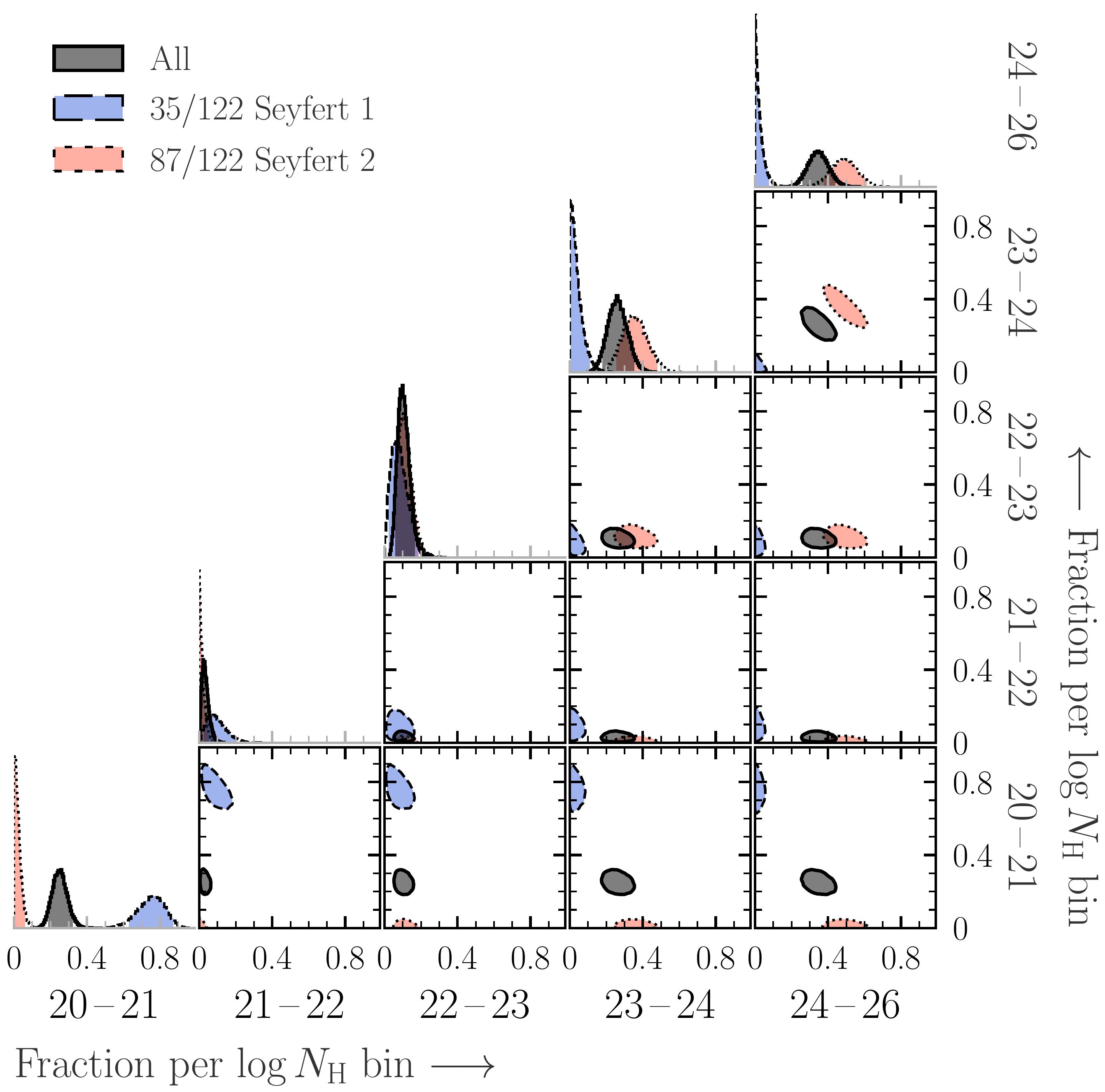}
\caption{\label{fig:nhdist_seyfert_type}$N_{\rm H}$ distribution hierarchical model corner plot, binned by optical spectroscopic classification (Seyfert~1\,=\,type~1--1.8; Seyfert~2\,=\,type~1.9--2). A strong deviation is found between the two classifications, as expected from unification, and consistent with previous works.
}
\end{figure*}

To test the possible presence of type~2 AGN that are X-ray unobscured, we plot the NuLANDS log\,$N_{\rm H}$ distribution for type~1--1.8 and type~1.9--2 in Figure~\ref{fig:nhdist_seyfert_type}. We find negligible fractions of type~2 AGN with unobscured X-ray spectra, though note this may be somewhat by design. Our X-ray model selection does allow neutral obscuration for type~1s, but we restrict the models accessible to either optical spectral class (i.e. physical torus models were only selected for optically-classified type~1.9--2 sources). Interestingly, the type~2 Compton-thick fraction of 49$\pm12$\% is fully consistent with the type~2 Compton-thick fraction of \citet{Kammoun20}, which may suggest that the vast majority of the type~2 AGN in our sample agree with their X-ray obscuration classification. Additionally, there are 20 type~2 AGN with no X-ray coverage that we assume an $N_{\rm H}$ prior for in our $N_{\rm H}$ distribution hierarchical model based on the type~2s with X-ray data. These missing sources may include sources with disagreeing optical and X-ray obscuration classification.

\subsubsection{Intrinsic X-ray Luminosity}\label{subsec:nh_v_l210int}

Whilst the principle aim from our spectral fitting was to constrain the global line-of-sight column density for the local AGN population, the models we use all parameterise the intrinsic coronal X-ray continuum as a powerlaw with intrinsic normalisation and photon index as free parameters.\footnote{The high-energy cut-off was fixed to 300\,keV for all models in which it was an optional free parameter, in agreement with the recent constraints from \citet{Balokovic20}.} We generate intrinsic (i.e. unabsorbed) X-ray luminosity posteriors in the 2--10\,keV band for each source by integrating the equivalent powerlaw flux generated from the intrinsic normalisation and photon index posteriors. As such, all uncertainties are propagated into the intrinsic luminosity posterior.

\begin{figure*}
\centering
\includegraphics[width=0.8\textwidth]{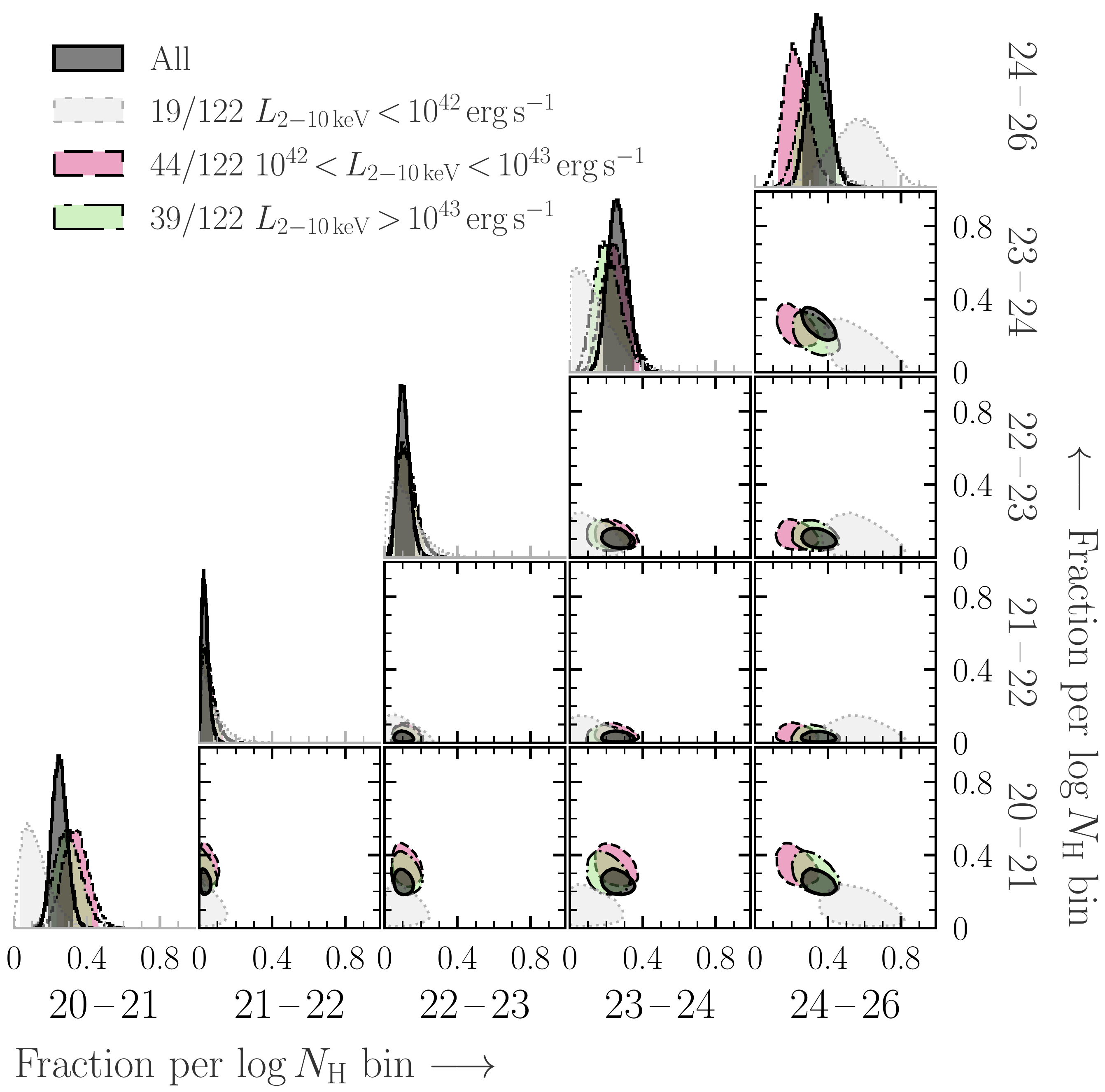}
\caption{\label{fig:nhdist_logLX}$N_{\rm H}$ distribution hierarchical model corner plot, binned by predicted intrinsic 2--10\,keV luminosity from the highest Bayes Factor model fit per source. The binning was performed on the mode of the intrinsic luminosity posterior. Thus for $L_{2-10\,{\rm keV}}$\,$<$\,10$^{42}$\,erg\,s$^{-1}$, the majority are upper limits on intrinsic luminosity caused by poorer quality spectral constraints.
}
\end{figure*}

To investigate trends between intrinsic luminosity and line-of-sight column density, we consider the luminosity posteriors for the highest Bayes Factor models per source only and also the mode of each posterior. We then split the NuLANDS $N_{\rm H}$ distribution corner plot into three sub-groups of intrinsic 2\,--\,10\,keV luminosity, namely $L_{2-10\,{\rm keV}}$\,$<$\,10$^{42}$\,erg\,s$^{-1}$, 10$^{42}$\,erg\,s$^{-1}$\,$<$\,$L_{2-10\,{\rm keV}}$\,$<$\,10$^{43}$\,erg\,s$^{-1}$ and $L_{2-10\,{\rm keV}}$\,$>$\,10$^{43}$\,erg\,s$^{-1}$, resulting in Figure~\ref{fig:nhdist_logLX}. Although a single model is chosen for placing in a given intrinsic luminosity bin, the same Monte Carlo-derived column density distribution which considers all acceptable models per source is used to generate the $N_{\rm H}$ distribution. The process could thus result in some intrinsic luminosity posteriors that may not apply to some column density posteriors, but we expect such scenarios to be infrequent. Furthermore, using the same Monte Carlo technique as the other $N_{\rm H}$ distribution corner plots presented in this section enables a direct comparison since the same technique is used to model the parent line-of-sight column density distribution.

Of the full sample of 102 sources with X-ray data, we find 19 in which the highest Bayes Factor model gives an intrinsic luminosity posterior mode below 10$^{42}$\,erg\,s$^{-1}$. On manual inspection, a number of these sources are either low signal-to-noise, meaning that the intrinsic luminosity posterior is better explained by an upper limit or multi-modal with a lower portion of the intrinsic luminosity posterior mass above the 10$^{42}$\,erg\,s$^{-1}$ threshold. For this reason, these targets are marked with faint shading in Figure~\ref{fig:nhdist_logLX}. Of the sources with higher predicted intrinsic luminosities, the two luminosity bins we consider give fully consistent column density distribution predictions. However, there are only four sources in our sample with $L_{2-10\,{\rm keV}}$\,$>$\,10$^{44}$\,erg\,s$^{-1}$; two type~1 sources (3C\,120 and Mrk\,509) and two Compton-thick type~2 sources (2MASX\,J15504152--0353175 and Mrk\,573), implying a Compton-thick fraction that is still consistent with the entire NuLANDS column density distribution at the highest luminosities in the sample.

Previous works have found evidence for an effect between the fraction of obscured sources and the intrinsic X-ray luminosity (see e.g., \citealt[Extended Data Figure~3]{Ricci17_feedback}). However, considering the relatively small fraction of higher-luminosity sources with e.g., $L_{2-10\,{\rm keV}}$\,$>$\,10$^{44}$\,erg\,s$^{-1}$, one would expect a reduced effect between covering factor and luminosity at these lower luminosities. For example, \citet{Brightman11b} found the fraction of sources with $N_{\rm H}$\,$>$\,10$^{22}$\,cm$^{-2}$ in the 12\,$\mu$ Galaxy Sample to be broadly consistent with constant for intermediate intrinsic luminosities $L_{2-10\,{\rm keV}}$\,$\sim$\,10$^{40}$\,--\,10$^{44}$\,erg\,s$^{-1}$. Thus we may not see a strong effect between obscuration and luminosity because of the overall lower intrinsic luminosities of the sources in NuLANDS as compared to e.g., the \textit{Swift}/BAT sample which contains a higher fraction of sources with $L_{2-10\,{\rm keV}}$\,$>$\,10$^{44}$\,erg\,s$^{-1}$. Though a stronger driver for feedback on the circum-nuclear obscurer of AGN seems to be the Eddington ratio (e.g., \citealt{Fabian08,Ricci17_feedback,Ananna20,Ananna22,Ricci22_ledd}), due to the incompleteness of black hole mass estimates currently in the NuLANDS sample we defer such analyses to future work.

\subsection{Comparison with Other Local AGN Samples Observed in X-rays}\label{subsec:lognhcomp}
Here we compare and contrast the results (primarily $N_{\rm H}$ distributions) with other local AGN samples in the literature.

\subsubsection{Swift/BAT}\label{subsubsec:bat}
A number of works have focused on careful $N_{\rm H}$ constraints for AGN detected by the all-sky \textit{Swift}/BAT monitor. \citet{Ricci17_bassV} combined all 70-month \textit{Swift}/BAT spectra between 14\,--\,195\,keV for the 838 detected AGN with soft X-ray spectra from a number of complementary facilities to constrain line-of-sight $N_{\rm H}$ with a wide array of spectral models. The authors select a sample of 55 Compton-thick AGN candidates from their analysis (see \citealt{Ricci15} for more details of the targets), representing an observed Compton-thick fraction of 7.6$^{+1.1}_{-2.1}$\% out of all non-blazar AGN in the 70-month sample. After considering a number of luminosity and obscuration geometry-dependent bias corrections, \citeauthor{Ricci15} predict a bias-corrected Compton-thick fraction of 27$\pm$4\% in the local Universe (normalised to unity in the $N_{\rm H}$\,=\,10$^{20}$\,--\,10$^{25}$\,cm$^{-2}$ range), which is consistent with the Compton-thick fraction we find for the entire NuLANDS sample. The overall shape of the bias-corrected \nh distribution from \citeauthor{Ricci17_bassV} is also in general agreement with the NuLANDS distribution reported in this paper, suggesting the NuLANDS selection is complimentary to hard X-ray flux-limited selection by identifying Compton-thick AGN in a representative manner.

There have been numerous \textit{NuSTAR} follow-up campaigns of \textit{Swift}/BAT-detected Compton-thick AGN candidates. The largest \textit{NuSTAR} legacy sample dedicated to observing BAT-detected obscured sources is detailed in \citet{Balokovic17b,Balokovic20} who consider type~1.8, 1.9 and 2 AGN from the 70-month \textit{Swift}/BAT compilation of \citet{Baumgartner13}. Sources were selected based on publicly-available \textit{NuSTAR} data within a volume of $z$\,$<$\,0.1 and with observed BAT fluxes $F_{\rm 14\,-\,195\,keV}$\,$>$\,10$^{-11}$\,erg\,s$^{-1}$\,cm$^{-2}$. \citeauthor{Balokovic17b} uses an alternative approach to \citet{Ricci15} to bias-correct the observed Compton-thick fraction in the sample. Different assumptions on the Compton hump strength (parameterized by \texttt{rel\_refl} in the \texttt{pexrav} model) were assumed for the sample to reverse engineer the effective X-ray sensitivity with obscuration level. The resulting bias-corrected Compton-thick fraction (relative to the entire sample of unobscured and obscured AGN with the same selection steps as their sample) was $\gtrsim$\,27\%, in very good agreement with the bias-corrected value from \citet{Ricci15} as well as the observed fraction from NuLANDS.

Similar constraints have been acquired with \textit{NuSTAR} follow-up of the deeper \textit{Swift}/BAT Palermo 100-month catalog \citep{Marchesi18_clemsonI,Marchesi19_clemsonIII,Traina21,TorresAlba21_clemsonVI}. The most recent observed Compton-thick fraction from the sample is $\sim$\,8\% when compared to all AGN selected within a volume of $z$\,$\leq$\,0.05 \citep{Sengupta23}, in agreement with \citet{Ricci15} and \citet{Balokovic17b}. \citet{Zhao21_clemson_bias_correction} then presents a Monte Carlo-based bias-correction using the best-fit geometrical parameters derived from physical torus modelling of the sample. At $\sim$\,37\%, the predicted bias-corrected Compton-thick fraction with this method is higher than previous estimates using the 70-month BAT catalog (though it is still consistent with the lower limit from \citealt{Balokovic17b}), and is in very good agreement with the observed fraction we find for NuLANDS.

More recently, \citet{Tanimoto22} used the \texttt{XCLUMPY} model to analyze 52/55 of the original 70-month \textit{Swift}/BAT-detected Compton-thick candidates from \citet{Ricci15} with publicly-available \textit{NuSTAR} data. Notably, the authors find that after incorporating \textit{NuSTAR} data, 24 of the objects no longer have line-of-sight column densities consistent with the Compton-thick regime within 90\% confidence. The reduction of Compton-thick AGN corresponds to a reduced observed Compton-thick fraction of $\sim$\,3.9\% for the 70-month \textit{Swift}/BAT non-blazar sample. Similar results highlighting the importance of \textit{NuSTAR} data in disentangling parametric degeneracies and constraining line-of-sight column densities are reported in \citet{Marchesi18_clemsonI,Marchesi19_clemsonIII,Marchesi19_clemsonV}. Such results highlight the benefit of comprehensive \textit{NuSTAR} follow-up in deriving robust line-of-sight column density estimations for AGN in combination with sufficiently-sensitive spectral constraints in the soft band simultaneously (e.g., \citealt{Molina24}).

\subsubsection{CfA Seyferts}\label{subsubsec:cfasys}
The CfA Seyfert sample was derived from the parent CfA Redshift Survey \citep{Huchra83}, 2399 galaxies with optical spectroscopy that is complete down to a limiting galaxy magnitude of $m_{\rm Zw}$\,$\leq$\,14.5\,mag.\footnote{$m_{\rm Zw}$ is approximately equivalent to a visual B-band magnitude in the photographic magnitude system.} Of the galaxies in the CfA Redshift Survey, \citet{Huchra92} selected a complete sub-sample of 27 Seyfert~1s and 21 Seyfert~2s (48 Seyferts total\footnote{Note that NGC\,3227 and Mrk\,993 were originally classified as Seyfert~2s, but follow-up spectroscopy has identified broad permitted lines in their spectra \citep{Salamanca94,Corral05}.}) within the magnitude limit of the CfA Redshift Survey. \citeauthor{Huchra92} additionally estimated the Seyfert~2 to Seyfert~1 ratio from the sample to be 2.3$\pm$0.7, since for a given galaxy optical brightness, an intrinsically powerful AGN would always be preferentially detected if a type~1 as opposed to a type~2.

Of the 21 Seyfert~2s from the CfA Seyfert sample, 12 were observed by \textit{NuSTAR} as part of the \textit{Swift}/BAT sample follow-up, NuLANDS or other targeted observations. The remaining 9/21 Seyfert~2s from the CfA Seyfert sample that had not been previously observed were selected for a \textit{NuSTAR} Legacy Survey (PI: J.~Miller; \citealt{Kammoun20}). Of the nine Seyferts observed by \textit{NuSTAR}, \citeauthor{Kammoun20} ruled out two sources as Seyfert~2s (NGC\,5256 and Mrk\,461) based on follow-up optical spectroscopy that placed the targets in the composite star-forming $+$ Seyfert region of the [N\,\textsc{ii}]/H$\alpha$ vs. [O\,\textsc{iii}]/H$\beta$ BPT diagram, leaving 19 Seyfert~2s in total.\footnote{NGC\,5256 has since been studied by \citet{Iwasawa20_mrk266}, finding the South-West component of the merging system to be a Compton-thick AGN.}

\citeauthor{Kammoun20} fit phenomenological (featuring \texttt{pexmon}) and physical (featuring coupled and decoupled variations of \texttt{MYtorus}) models to the 19 targets in the sample, finding between 6\,--\,10 of those Seyfert~2s to be Compton-thick depending on the choice of model and archival results. The resulting observed Compton-thick fraction out of the full 19 Seyfert~2s\,$+$\,27 Seyfert~1s\,=\,46 CfA Seyfert sample (albeit neglecting $N_{\rm H}$ measurement uncertainty) is then between 14$^{+10}_{-7}$\,--\,23$^{+11}_{-9}$\% which is below but consistent with the NuLANDS 90\% confidence range within uncertainties. However, if we correct the Seyfert~2 to Seyfert~1 ratio of the CfA Seyfert sample based on the predicted ratio from \citet{Huchra92} of 2.3\,$\pm$\,0.7, assuming the same fraction of missing Seyfert~2 sources to be Compton-thick as found by \citeauthor{Kammoun20}, we calculate a bias-corrected prediction for the Compton-thick fraction to lie in the range of $\sim$\,22$^{+8}_{-7}$\,--\,37$^{+9}_{-8}$\% for the CfA Seyfert sample. Such a range is in good agreement with the observed value from NuLANDS. As described earlier, Seyfert~2s are expected to be preferentially missed in optical spectroscopic classifications relative to Seyfert~1s at a given optical flux level, since the continuum is typically more suppressed in the former relative to the latter.

\subsubsection{The Complete 15\,Mpc Sample}
As discussed in Section~\ref{sec:isotropic_selection}, near-to-mid infrared lines produced in the narrow line region do not suffer considerably from line-of-sight extinction. The Complete 15\,Mpc Sample (Annuar et al., in prep.) is one such local universe selection, originating from the [Ne\,\textsc{v}] mid-infrared line selection of \citet{Goulding09}. The parent sample is the Revised Bright Galaxy Sample from \textit{IRAS} \citep{Sanders03}, which selects the brightest sources detected by \textit{IRAS} with 60$\mu$m flux densities $f_{60\mu{\rm m}}$\,$>$\,5.24\,Jy. A very local volume cut was imposed on the sample of 15\,Mpc, from which 19 galaxies\footnote{Including two known AGN added to the sample with archival \textit{NuSTAR} data.} were selected as AGN based on significant [Ne\,\textsc{v}] emission in their \textit{Spitzer Space Telescope} \citep{Werner04} high resolution infrared spectra. The [Ne\,\textsc{v}] line has a high excitation potential, making its production unlikely from pure stellar systems. Of the 20 sources, eight were selected as part of a \textit{NuSTAR} program, though X-ray data is available for all, including an additional 17 with \textit{NuSTAR} data. To date, \textit{NuSTAR} has helped robustly confirm two of the sample as Compton-thick AGN: NGC\,5643 \citep{Annuar15} and NGC\,1448 \citep{Annuar17}, as well as NGC\,660 with line-of-sight column density solutions both below and above the Compton-thick threshold \citep{Annuar20}. The sample has additionally identified a number of genuine low-luminosity AGN with 2\,--\,10\,keV luminosities $L_{2-10\,{\rm keV}}$\,$\lesssim$\,10$^{41}$\,erg\,s$^{-1}$, providing further evidence that [Ne\,\textsc{v}] is an extremely effective indicator of AGN activity with little contamination from stellar processes.

Other than NGC\,1068, there is no overlap in sources between the [Ne\,\textsc{v}] sample and NuLANDS. Firstly, NGC\,1068 is the only source in our sample at a distance $<$\,15\,Mpc. But this and the very small overlap with NuLANDS is likely caused by the difficulty of producing warm 25-to-60\,$\mu$m continuum shapes (as required by the NuLANDS selection) when the AGN component is a small fraction of the overall bolometric luminosity of the host galaxy. Low AGN-to-host bolometric fractions are found for the [Ne\,\textsc{v}] sample with many systems having observed X-ray luminosities $L_{2-10\,{\rm keV}}$\,$<$\,10$^{42}$\,erg\,s$^{-1}$ (Annuar et al., in prep.).

\subsubsection{The 12 Micron Galaxy Sample (12MGS)}
The 12MGS \citep{Spinoglio89} was derived from the \textit{IRAS} PSCv2 with (co-added) 12$\mu$m flux densities $f_{12\mu{\rm m}}$\,$>$\,0.3\,Jy. The authors show that typical AGN spectra are broadly isotropic in the mid-infrared, with the 12$\mu$m flux being approximately one-fifth of the bolometric value for all Seyfert types. The extended 12MGS \citep{Rush93} then used the \textit{IRAS} Faint Source Catalog Version 2 to derive an alternative selection of 893 galaxies with a lower flux limit of 0.22\,Jy at 12$\mu$m.

The most comprehensive X-ray follow-up of the 12MGS was reported by \citet{Brightman11a,Brightman11b}, who analysed all publicly-available \textit{XMM-Newton} data of the sample as of 2008-December (126 sources with meaningful spectra). \citet{Brightman11b} find a Compton-thick fraction of 20$\pm$4\% in the X-ray luminous ($L_{2-10\,{\rm keV}}$\,$>$\,10$^{42}$\,erg\,s$^{-1}$) sub-sample, which included optically-classified non-AGN. Though below that of NuLANDS, the 12MGS Compton-thick fraction would likely increase with increased hard X-ray coverage, so this Compton-thick fraction is likely a lower limit.

A number of AGN selected in the 12MGS have been observed by \textit{NuSTAR} to date. \citet{LaCaria19} consider three 12MGS Seyfert galaxies with observed differences between the infrared and X-ray bolometric luminosities of up to three orders of magnitude, finding all targets to be heavily obscured and two to be Compton-thick. \citet{Saade22} alternatively selected a sample of nine Seyfert~2 AGN from the 12MGS with observed 2--10\,keV luminosities significantly below that of their observed [O\,\textsc{iii}] luminosities to investigate the possibility of X-ray obscuration or faded AGN. Using \textit{NuSTAR} data, three galaxies were confirmed to be Compton-thick, with four of the remaining sources being heavily obscured.

\subsubsection{The Great Observatories All-sky LIRG Survey (GOALS)}\label{subsubsec:goals}
Similar to the Complete 15\,Mpc sample of \citet{Goulding09}, GOALS \citep{Armus09}\footnote{\url{http://goals.ipac.caltech.edu}} is fundamentally derived from the Revised Bright Galaxy Sample \citep{Sanders03}, but instead selects all Luminous InfraRed Galaxies (LIRGs, 181 sources) and Ultra-Luminous InfraRed Galaxies (ULIRGS, 21 sources), giving a total sample size of 202 sources at a median distance of 94.8\,Mpc and $z$\,$\leq$\,0.088. Owing to the bright infrared selection of the sample, a large fraction of the sources are confirmed interacting systems \citep{Armus09}.

To date there has been extensive X-ray coverage of the GOALS sample. \citet{Iwasawa11} analysed the \textit{Chandra} data for a complete sub-sample of 44 bright GOALS sources with log\,$L_{8-1000\,\mu{\rm m}}$\,$>$\,11.73\,L$_{\odot}$, finding X-ray detections for all but one target. Considering all sources with hard X-ray colors, a detected 6.4\,keV iron line and a confirmed mid-infrared [Ne\,\textsc{v}] line, the total detected AGN fraction was 48\%. \citet{Koss13} presented a targeted hard X-ray survey of local GOALS-selected LIRGs with \textit{Swift}/BAT, finding 40\,$\pm$\,9\% of the sample to have 14\,--\,195\,keV\,/\,2\,--\,10\,keV band ratios consistent with high or Compton-thick line-of-sight column densities predicted from the \texttt{MYtorus} model. \citet{TorresAlba18} then investigated the \textit{Chandra} data for a lower luminosity sub-sample of 63 GOALS sources, finding a consistent fraction of X-ray-confirmed AGN to the higher luminosity sources analysed in \citet{Iwasawa11}.

To constrain the line-of-sight column density, \citet{Ricci17_mergers} considered all GOALS sources with publicly-available \textit{NuSTAR} data as of 2016-March that were confirmed to be interacting, as well as three systems detected by \textit{Swift}/BAT but not observed by \textit{NuSTAR}. All 30 systems in the sample were found to be obscured with $N_{\rm H}$\,$>$\,10$^{23}$\,cm$^{-2}$, implying a large covering factor for all sources. After additionally binning the sample by observed merger stage, the authors find early-stage mergers to have a Compton-thick fraction of 35$^{+13}_{-12}$\%, consistent with the bias-corrected value of \citet{Ricci15} as well as the value we report for NuLANDS here. In contrast, for the late-stage mergers in the sample a higher Compton-thick fraction of 65$^{+12}_{-13}$\% is observed, which is significantly higher than we find for NuLANDS.

\citet{Ricci21_goals} consider an extended sample of 60 GOALS systems observed by \textit{NuSTAR}, fitting the confirmed AGN with the \texttt{RXtorus} X-ray spectral model \citep{Paltani17,Ricci23_reflex_dust}. The authors find a similarly enhanced Compton-thick fraction in late-stage mergers of 74$^{+14}_{-19}$\%. Complementary X-ray spectral fitting of 57 GOALS sources with publicly-available \textit{NuSTAR} data or 105-month \textit{Swift}/BAT data was performed by \citet{Yamada20,Yamada21}. For the 30 sources detected in the hard X-ray band, the authors fit with \texttt{XCLUMPY} and find Compton-thick fractions of 24$^{+12}_{-10}$\% and 64$^{+14}_{-15}$\% in early and late-stage mergers, respectively, fully consistent with the findings of \citet{Ricci17_mergers,Ricci21_goals}.

\subsection{Comparison with Population Synthesis Models}\label{subsec:popsynmods}
As described throughout this paper, the fraction of Compton-thick AGN amongst the AGN population is currently highly uncertain and a source of significant systematic uncertainty in population synthesis models. Representative samples such as NuLANDS should provide excellent benchmarks for model evaluation. In this sub-section, we compare the observed Compton-thick fraction from NuLANDS to the values predicted across a number of different population synthesis models in the literature as well as other hard X-ray-based analyses of local AGN samples discussed earlier in this Section.

\begin{figure*}
\centering
\includegraphics[width=0.99\textwidth]{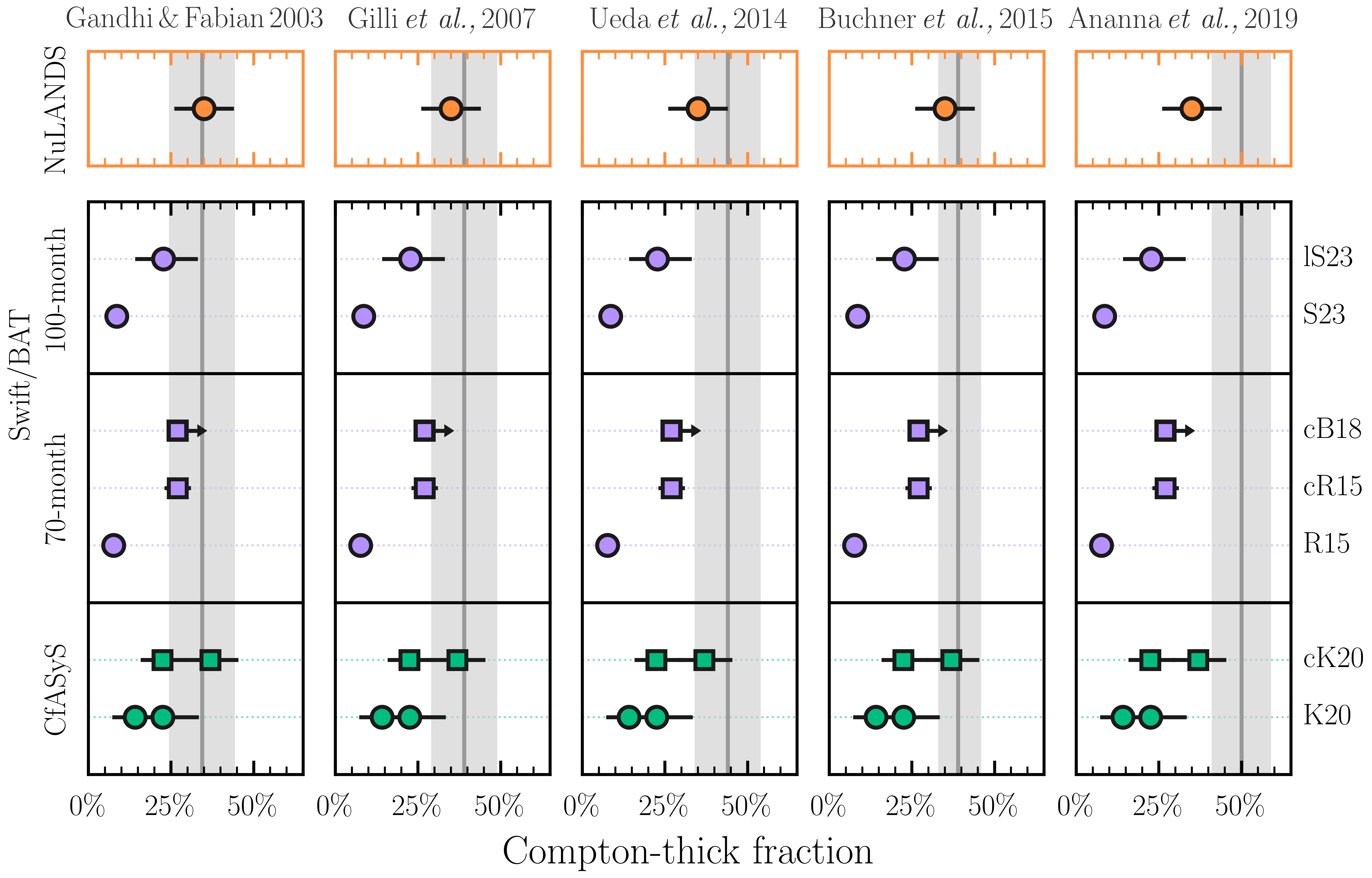}
\caption{\label{fig:ctfracs} A compilation of Compton-thick AGN fractions derived in hard X-ray local AGN sample analyses compared to numerous population synthesis models from the literature. From bottom to top, the Compton-thick fractions are: K20; \citet{Kammoun20}, cK20; the completeness-corrected value derived in Section~\ref{subsubsec:cfasys}, R15; the observed fraction from \citet{Ricci15}, cR15; the bias-corrected value from \citet{Ricci15}, cB18; the bias-corrected value from \citet{Balokovic17b}, and S23/lS23; the observed fraction within $z$\,$\leq$\,0.05/z\,$\leq$\,0.01 from the latest 100-month Palermo BAT sample of \citet{Sengupta23}, respectively. From left to right, the population synthesis models considered are from \citet{Gandhi03}, \citet{Gilli07}, \citet{Ueda14}, \citet{Buchner15} and \citet{Ananna19}.}
\end{figure*}

In Figure~\ref{fig:ctfracs}, we collate the predicted Compton-thick fractions from five population synthesis models in the literature that each use different methodologies and AGN selection functions: \citet{Gandhi03}, \citet{Gilli07}, \citet{Ueda14}, \citet{Buchner15} and \citet{Ananna19}. The model by \citet{Gandhi03} was based on a {\em mid-infrared} selection approach for tackling obscuration selection bias of type~2 AGN, and results therein were one of the original motivations behind the NuLANDS selection. The remaining four models are all popular population synthesis studies in the literature that each use X-ray selected samples, and we defer the reader to the individual papers for specific information regarding each model. For models that do not report uncertainties on the Compton-thick fraction, we assume a default uncertainty of 10\%\ which is typical of the median luminosity function uncertainties for the samples used. But it should be kept in mind that systematic uncertainties related to model assumptions could be higher; these are non-trivial to compare in a self-consistent manner, but some first insights are possible from the scatter across the model predictions.

For any models that report an evolution of the Compton-thick fraction with luminosity we make sure to report the fraction relevant for intrinsic luminosities $L_{2-10\,{\rm keV}}$\,$\lesssim$\,10$^{43}$\,erg\,s$^{-1}$ where possible to match the approximate expected intrinsic luminosities of the NuLANDS sample (see Section~\ref{subsec:nh_v_l210int} and Figure~\ref{fig:nhdist_logLX}). We then compare each model prediction for the Compton-thick fraction to the measured (and bias-corrected where available) Compton-thick fractions in qualitatively similar luminosity ranges from the following hard X-ray local AGN analyses: the \textit{NuSTAR}-based analysis of \citet{Kammoun20} (including the completeness-corrected range derived in Section~\ref{subsubsec:cfasys}), the 70-month BAT analyses of \citet{Ricci15} and \citet{Balokovic17b} and the latest estimates from the ongoing 100-month Palermo BAT (\textit{NuSTAR}-based) analysis of \citet{Sengupta23}.

Figure~\ref{fig:ctfracs} clearly shows that from the five population synthesis models considered, only NuLANDS finds a directly-observed Compton-thick fraction that is consistent with all population synthesis model-predicted values within uncertainties. When considering the bias-corrected values from other surveys, the completeness-corrected fraction for the CfA Seyfert sample \citep{Kammoun20} is consistent with all models at the upper range of possible Compton-thick fractions. The general agreement between the predicted completeness-corrected Compton-thick fraction in the CfA Seyfert sample and the directly-observed NuLANDS sample provides additional support that the NuLANDS selection is representative of type~1 and type~2 AGN.

In terms of \textit{Swift}/BAT-selected samples, the observed Compton-thick fractions from \citet{Ricci15} and \citet{Sengupta23} out to further distances than NuLANDS are inconsistent with all the models considered. The highest observed fraction from BAT-selection that is plotted in Figure~\ref{fig:ctfracs} is from the lowest redshift bin of z\,$<$\,0.01 ($D_{L}$\,$\lesssim$\,45\,Mpc) in the \textit{NuSTAR}-based follow-up of the 100-month Palermo BAT sample \citep{TorresAlba21_clemsonVI,Sengupta23}. Similar results are reported by \citet{Ricci15} for the 70-month sample, finding observed Compton-thick fractions within $\sim$\,50\,Mpc that are broadly consistent with NuLANDS within errors but inconsistent with the model predictions of \citet{Ueda14}, \citet{Buchner15} and \citet{Ananna19}. In terms of bias-corrected values, the lower limit prediction derived by \citet{Balokovic17b} from an analysis of 70-month BAT-selected type~2 AGN is consistent with all models.

The overall consistency between NuLANDS and previous population synthesis models highlights NuLANDS as an optimized sample for future AGN surveys. We note that the largest discrepancy for NuLANDS is with the latest model of \citet{Ananna19}, offering tantalizing evidence that the NuLANDS obscured and Compton-thick fractions could still be a lower limit, in agreement with the sample bias considerations discussed in Section~\ref{sec:representative}.

\subsection{Variability}\label{subsec:var}
The soft X-ray observations analysed in this work were selected to be as quasi-simultaneous with \textit{NuSTAR} as possible per source (see Section~\ref{sec:data} for a breakdown). However, it is not unexpected for variability to affect the log\,$N_{\rm H}$ distribution results presented in this work to some degree (see \citealt{Ricci23_clagn} for a recent review). Several studies have found obscuration variability in both type~1 and~2 AGN (e.g., \citealt{Malzia97,Risaliti02,Kara21}), as well as between Compton-thin and thick obscuration levels which could affect our understanding of the Compton-thick fraction (e.g., \citealt{Risaliti05,Rivers15,Ricci16,Pizzetti22,Marchesi22_clemsonVIII,TorresAlba23,Lefkir23,Pizzetti24}).

\begin{figure*}
\centering
\includegraphics[width=0.99\textwidth]{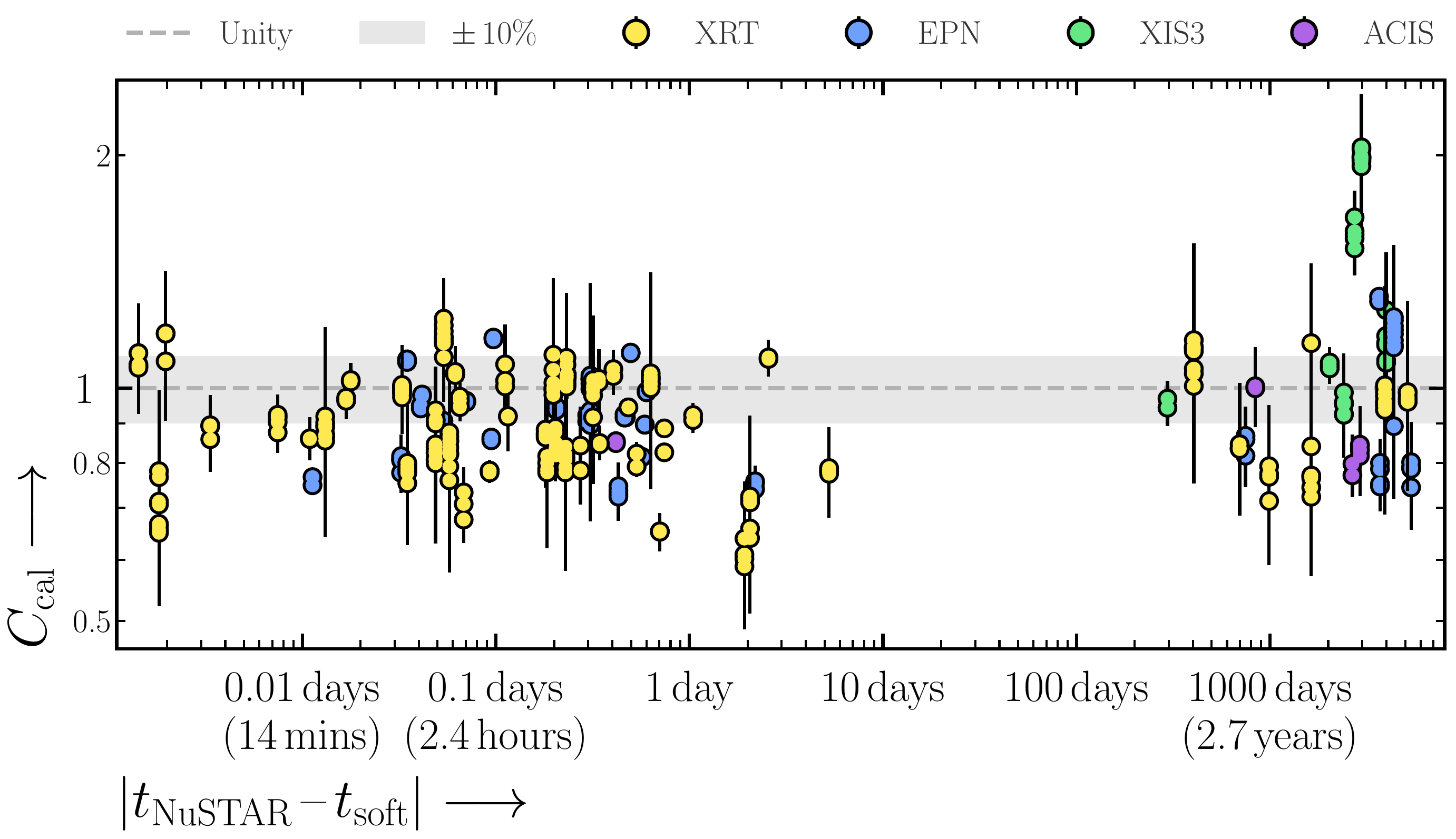}
\caption{\label{fig:nuvar}The difference between the start time of each \textit{NuSTAR} and soft X-ray observation per source analysed in this sample vs. the cross calibration found in the spectral fitting for each model that is selected per source. No large evidence for variability in the sample is observed, with general agreement for each source with unity. The two offsets at large times are both type~2 AGN, and are discussed in Section~\ref{subsec:var}.
}
\end{figure*}

To make a preliminary assessment of any variability effects present in the observations analysed in this work, we use the cross-calibration constant for each model fit. Although cross-calibrations are supposed to account purely for instrumental effects when multi-instrument fits are performed, the posterior for the cross-calibration constant between FPMA and the soft X-ray instruments can be used to indicate the possible presence of variability to zeroth order. Figure~\ref{fig:nuvar} presents the median and 68th percentile range cross-calibration constants for all model fits selected for sources with joint \textit{NuSTAR} and soft X-ray constraints as a function of the time difference of their respective observations. The median cross-calibrations are distributed as log\,$\mathcal{C}$\,=\,-0.03$\pm$0.07, fully consistent with unity and also the cross-calibration constants determined by \citet{Madsen15}. The two sources with the highest cross-calibrations are KUG\,0135--131 and IC\,3639, though both still have values $\lesssim$\,2. There are also a number of \textit{XMM-Newton}/EPN-based fits with cross-calibration values of $\sim$80\% relative to \textit{NuSTAR}/FPMA. As stated in the \textit{XMM-Newton} calibration technical note\footnote{\url{https://xmmweb.esac.esa.int/docs/documents/CAL-TN-0230-1-3.pdf}}, such large cross-calibration differences are not unexpected.

\subsection{Testing Sample Biases on the Column Density Distribution}

\subsubsection{Powerful AGN Missed by NuLANDS}\label{subsec:cool_bat}
NuLANDS is not designed to be a complete AGN population down to a fixed intrinsic luminosity, such that a number of bolometrically powerful AGN that are detected by \textit{IRAS} are also missed by the warm \textit{IRAS} mid-to-far infrared color classification. Two famous examples are NGC\,4051 and NGC\,6240, which both have $\alpha_{25,60}$ values indicative of cooler infrared spectra than we select. In this sub-section we check for any possible bias imposed on the $N_{\rm H}$ distribution (and Compton-thick fraction) by comparing a subset of the NuLANDS warm AGN to correspondingly cool mid-to-far infrared AGN. Our comparison sample is \textit{Swift}/BAT since this is an efficient selector of bolometrically-luminous AGN ($L_{\rm bol}$\,$\sim$\,10$^{44}$\,--\,10$^{46}$\,erg\,s$^{-1}$) in the local universe provided the line-of-sight column density $N_{\rm H}$\,$\lesssim$\,10$^{24}$\,cm$^{-2}$.

To ensure as relevant a comparison for NuLANDS as possible, we match the 70-month \textit{Swift}/BAT catalogue\footnote{Available from \url{https://swift.gsfc.nasa.gov/results/bs70mon/}.} to the \textit{IRAS} Point Source Catalogue v2.1 with a 1\,arcmin matching radius, giving 331 matches. The choice of matching radius may introduce some mis-matches, but chance coincidence with contaminants is unlikely. In addition, since we are interested in population demographics, so long as a suitably large number of sources is considered in any single case, any such mis-matches should not introduce a systematic bias in a given column density bin over another. We then performed the same selection method as NuLANDS, namely removing low Galactic latitudes, the Magellanic Clouds and any sources with upper limits from \textit{IRAS} at 25\,$\mu$m or 60\,$\mu$m. Finally we performed the same warm \textit{IRAS} color selection between 25 and 60\,$\mu$m to classify warm and correspondingly cool sources, giving 60 warm and 53 cool sources.\footnote{The vast majority of \textit{IRAS}-detected 70-month BAT AGN were either warm or cool based on their 25\,--\,60\,$\mu$m color. There are a small subset of sources with 25\,--\,60\,$\mu$m colors hotter than our warm classification, though we include these in the cool sample for simplicity.}

\begin{figure*}
\centering
\includegraphics[width=0.8\textwidth]{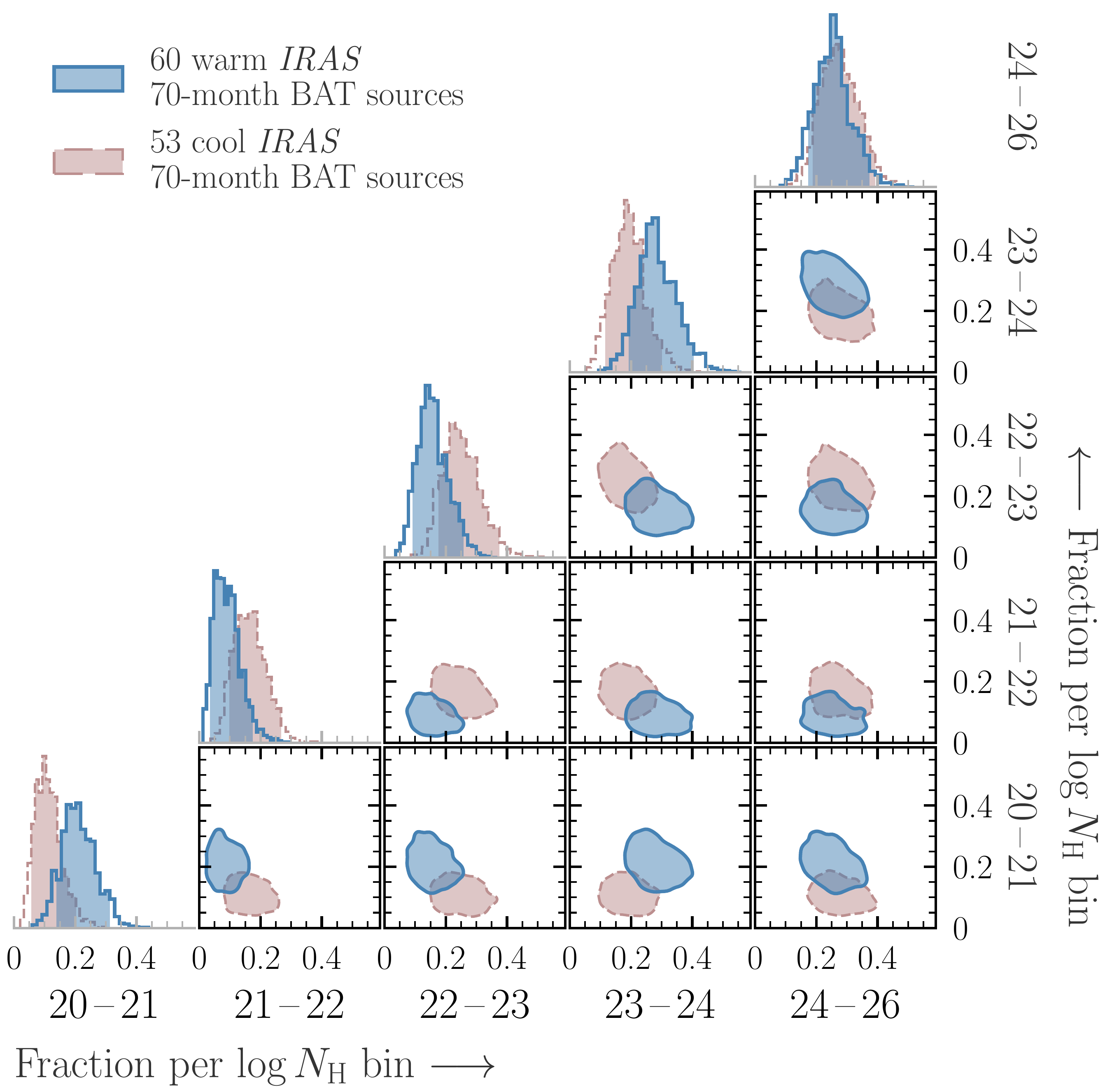}
\caption{\label{fig:warm_cold_bat}The log\,$N_{\rm H}$ distribution for the 70-month BAT sample, after identifying warm \textit{IRAS} sources in the sample, selected and classified as AGN in the same way as for NuLANDS. \lq Cool\rq\ \textit{IRAS} sources are those which are not selected as warm based on their 25--60\,$\mu$m spectral shape. All contours are consistent within 90\%, with the strongest correspondence arising for the Compton-thick fraction.
}
\end{figure*}

For consistency, the X-ray derived $N_{\rm H}$ values for each of the \textit{Swift}/BAT warm and cool \textit{IRAS} sources used the values derived in \citet{Ricci17_bassV}. Using the $N_{\rm H}$ values from our analysis for the warm sources would likely increase the $N_{\rm H}$ distribution uncertainties for the warm sample (as opposed to the cool sample) since our analysis incorporates multiple model solutions per source. For each source, we use the torus model-derived $N_{\rm H}$ estimate from \citet{Ricci17_bassV} if available or the standard \texttt{pexrav}-derived value if not. We then convert the 90\% uncertainties on $N_{\rm H}$ to 68\% assuming a standard Gaussian distribution conversion before approximating the $N_{\rm H}$ parameter posteriors per source by a two-piece Gaussian distribution to incorporate asymmetric errorbars. Finally, we use the same PosteriorStacker method as with the main NuLANDS $N_{\rm H}$ distributions to construct a parent histogram distribution for both the warm and cool \textit{Swift}/BAT sources. The corresponding corner plot for the parent distributions are shown in Figure~\ref{fig:warm_cold_bat}.

We find remarkable agreement in the Compton-thick fraction between warm and cool \textit{Swift}/BAT sources, with values of 26$^{+10}_{-9}$\% and 28$\pm$10\% for the warm and cool sources, respectively. All remaining $N_{\rm H}$ fractions below the Compton-thick limit are consistent between warm and cool sources within 90\% confidence, though with some offsets. The observed offsets do not follow an obvious trend with higher fractions for warm sources in the $N_{\rm H}$\,=\,10$^{20}$\,--\,10$^{21}$\,cm$^{-2}$ and 10$^{23}$\,--\,10$^{24}$\,cm$^{-2}$ bins vs higher fractions for cool sources in the 10$^{21}$\,--\,10$^{22}$\,cm$^{-2}$ and 10$^{22}$\,--\,10$^{23}$\,cm$^{-2}$ bins. Such a lack of trend is indicative of stochastic effects dominating the $N_{\rm H}$ distribution fractions as opposed to some systematic bias in the selection process itself, though quantifying any such effects is outside the scope of this paper. 

Finally, we note there are a number of \textit{Swift}/BAT AGN not detected by \textit{IRAS}, which we have not considered here. Quantifying the effect of AGN not included in NuLANDS due to non-detections from \textit{IRAS} cannot be easily tested in the same manner (e.g., by investigating the $N_{\rm H}$ distribution of AGN detected by \textit{Swift}/BAT but not \textit{IRAS}) since both instruments have their own unique selection functions.

\subsubsection{Elusive AGN Missed by NuLANDS}\label{subsec:s1s2frac}
\begin{figure*}
\centering
\includegraphics[width=0.99\textwidth]{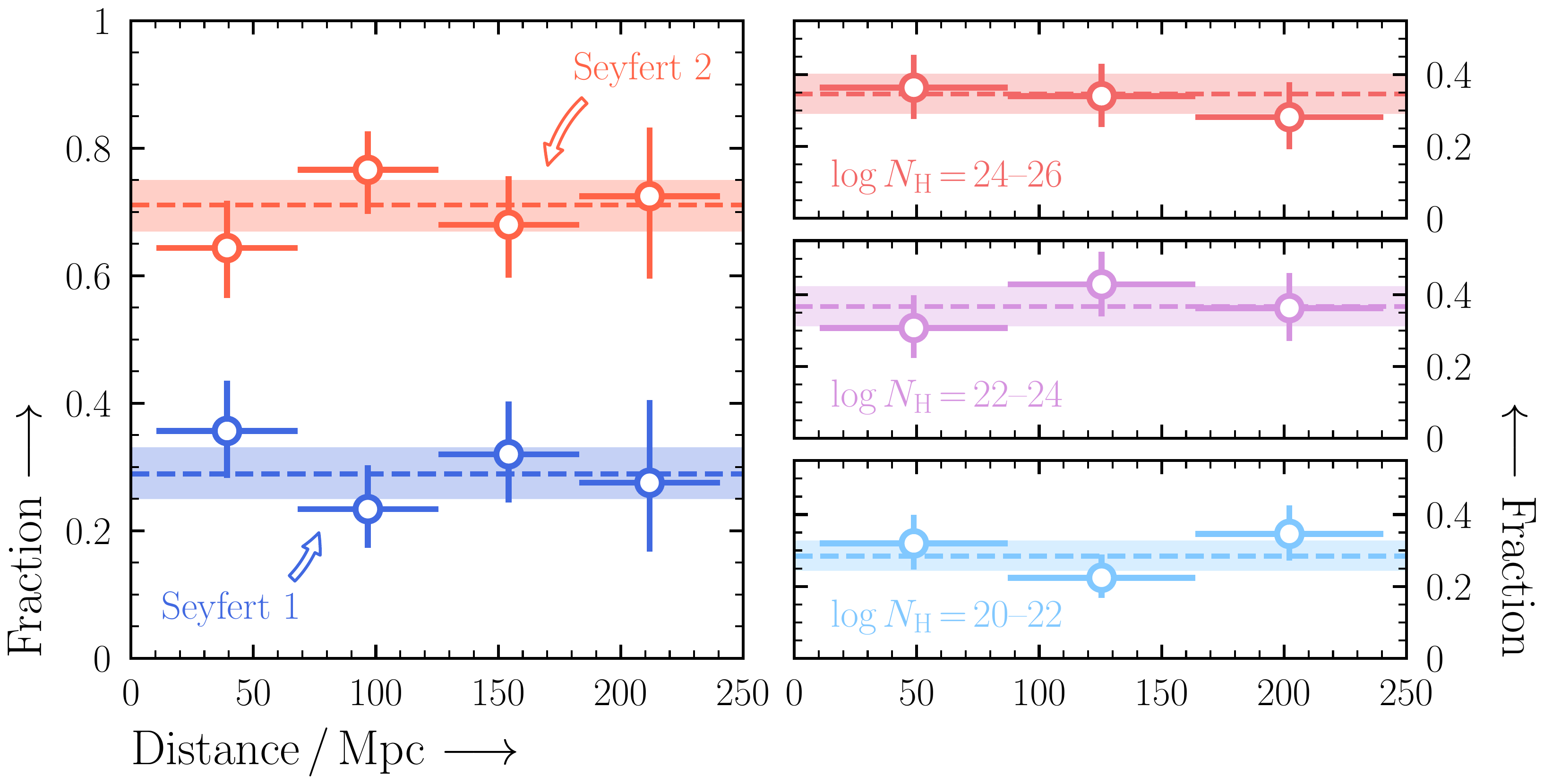}
\caption{\label{fig:S1S2frac} \textit{Left:} The type~1 and type~2 fractions in NuLANDS as a function of distance. For both types, the fraction in each bin of distance is consistent within uncertainties of the average for the whole sample, shown with dashed lines and 68th percentile shading. \textit{Right:} The fraction of sources in particular log\,$N_{\rm H}$ bins from the $N_{\rm H}$ distribution as a function of distance. All fractions are consistent with the values for the whole sample (shown with dashed lines and 68th percentile shading), as expected for an isotropically selected sample.
}
\end{figure*}

As discussed earlier, the derivation of the NuLANDS sample includes optical spectroscopic classifications which can be affected by significant large-scale dust reddening (e.g., \citealt{Goulding09,Greenwell21,Greenwell22,Greenwell24}). If missed, such elusive AGN would be predominantly associated with type~2 AGN due to the overall fainter continuum in such sources. From Figure~\ref{fig:nhdist_seyfert_type}, if a substantial number of such sources were missed in NuLANDS, a reduced number of Type~2 AGN would potentially increase the overall obscured fraction we report. To search for a possible dearth of Seyfert~2 AGN in our sample, we plot the type~1 and type~2 ratios as a function of distance in the left panel of Figure~\ref{fig:S1S2frac}. A reduced overall number of type~2 AGN could be identified by a drop in the overall type~2 fraction with distance, or alternatively an increase in the type~1 ratio with distance. Neither effect is observed, with each bin being fully consistent with the total type~1 and type~2 fractions from the entire sample.

An additional test is shown in the right panel of Figure~\ref{fig:S1S2frac}, in which the fractions of sources in incremental two dex bins of X-ray-derived $N_{\rm H}$ from Figure~\ref{fig:nhdist_DMpc} are plotted in the same format as the left panel versus distance. A similar trend is observed, in which all fractions are found to be consistent with the values found for the entire sample.

Nevertheless the possibility of elusive AGN missed by the optical spectral classifications still remains, meaning that the obscured fraction for NuLANDS can be conservatively considered a lower limit. It is not implausible for a significant number of the HII-classified galaxies to contain Compton-thick AGN. In the subset of sources studied by \citet{Moran02}, 11/18 targets were found to display normal galaxy optical spectra, four of which ($\sim$36\%) have since been confirmed as Compton-thick with \textit{NuSTAR}-based analyses (NGC\,1358; \citealt{Marchesi22_clemsonVIII}, NGC\,2273; \citealt{Masini16a}, NGC\,3982; \citealt{Saade22} and NGC\,5347; \citealt{Kammoun19}). In principle one could test the hypothesis of a substantial fraction of obscured elusive AGN by comparing the host galaxy inclination distribution to the full underlying galaxy population distribution. Such comparisons would require careful consideration of a variety of possible biases, which is outside the scope of this work.

\subsubsection{AGN Types Preferentially Selected by NuLANDS}

Owing to the requirement for Point Source Catalogue v2.1 detections at 60\,$\mu$m, the NuLANDS selection is flux-limited to $F_{60\,\mu{\rm m}}$\,$\gtrsim$\,0.5\,Jy. Other far-infrared flux-limited surveys (with albeit higher flux thresholds) often find large fractions of Ultra/Luminous InfraRed Galaxies (U/LIRGs) (e.g., \citealt{Kewley01_warmIR,Sanders03}) which are found to be far more obscured on average relative to e.g., the \textit{Swift}/BAT AGN sample (e.g., \citealt{Koss13,Ricci17_mergers,Ricci21_goals}). Thus a preferential selection of U/LIRGs in NuLANDS would likely lead to an enhanced obscured and Compton-thick fraction relative to the underlying AGN population.

As outlined in Section~\ref{sec:representative} and Figure~\ref{fig:representativeness}, the flux ratios between [O\,\textsc{iii}] and the 25 and 60\,$\mu$m fluxes for type~1s and type~2s indicate that a strong preference for dusty systems is not present in the sample, possibly partly due to the requirement for detected AGN-dominated [O\,\textsc{iii}] emission in the first place. A sample with considerable contamination from U/LIRGs would likely include a large contribution from large-scale host galaxy dust reddening in the optical (e.g., \citealt{Veilleux95,Veilleux99}).

For a more quantitative test for the presence of U/LIRGs in NuLANDS, we compare with the GOALS sample. As detailed in Section~\ref{subsubsec:goals}, the GOALS sample contains a complete LIRG subset of the Revised \textit{IRAS} Bright Galaxy Sample within a volume of $z$\,$<$\,0.088 (approximately twice the volume of NuLANDS). We determine how many GOALS sources have \textit{warm} 25\,--\,60\,$\mu$m spectra using the same classification as NuLANDS. Out of a total 202 GOALS sources, 7 have mid-to-far infrared spectral slopes consistent with NuLANDS -- the remainder have much colder spectral slopes. Of the 7 matches, one is excluded since it has flux upper limits at 25~and~60\,$\mu$m (VV\,414) and a further two (IRAS\,05223+1908 and NGC\,1275) are excluded since the Galactic latitudes are outside that of the NuLANDS cut (see Section~{\ref{sec:sample}}). Of the remaining four sources, one (the late-stage merger ESO\,350-IG\,038) is classified as an optical HII galaxy and thus not included in this work. \citet{Ricci21_goals} recently reported the 22.6\,ks \textit{NuSTAR} non-detection of this source, finding the \textit{Chandra} spectrum to be described well with a pure star-forming component and no hard X-ray component. The final three sources are in the NuLANDS sample and included in this paper, consisting of two confirmed Compton-thick AGN -- NGC\,1068 \citep{Bauer15}, NGC\,7674 \citep{Gandhi17} -- and a Compton-thin AGN -- MCG\,-03-34-064 \citep{Ricci17_bassV}.

\begin{figure*}
\centering
\includegraphics[width=1.\textwidth]{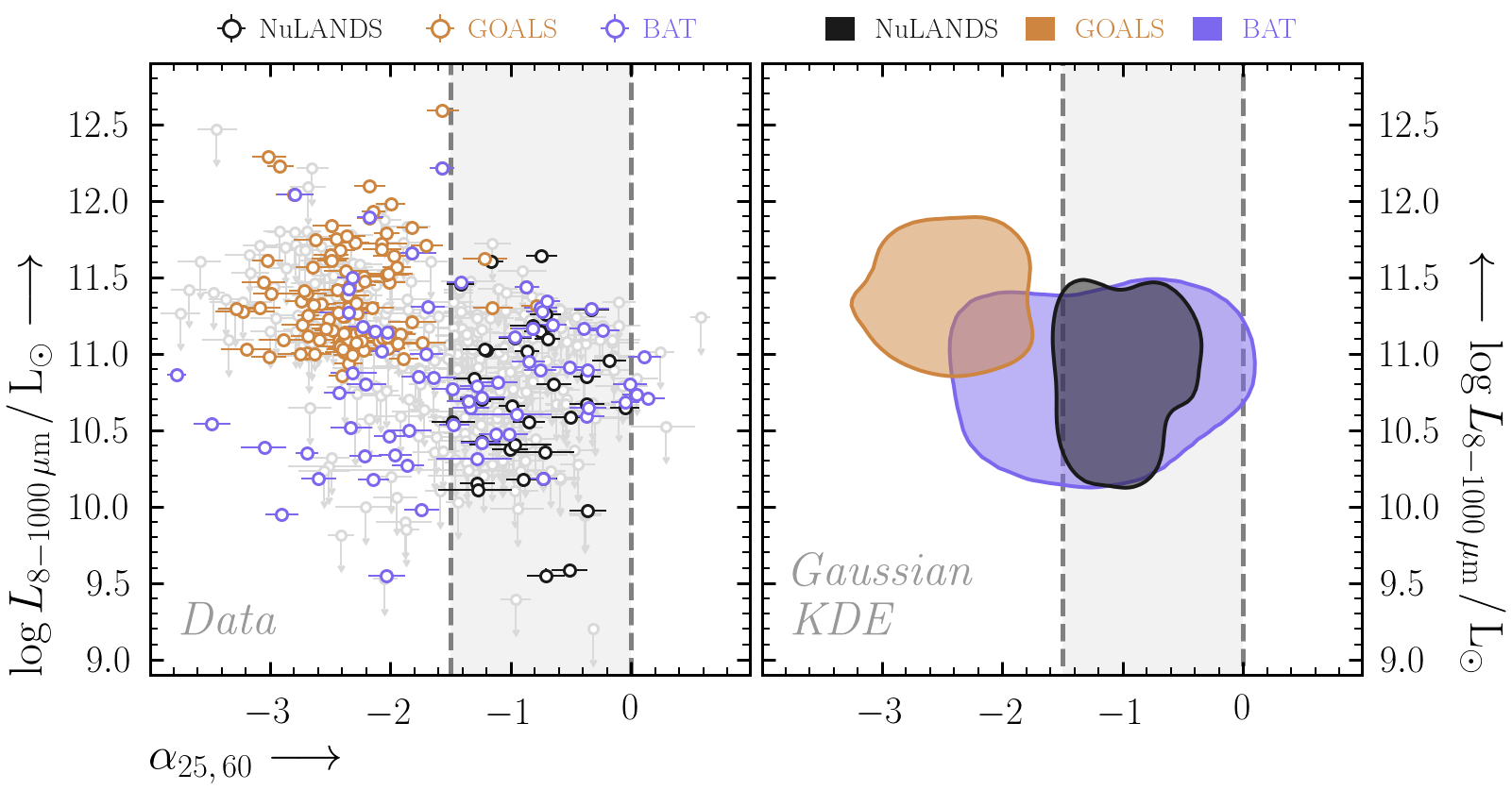}
\caption{\label{fig:goals_comp} Mid-to-far infrared spectral slope (parameterised with $\alpha_{25, 60}$) vs. total infrared luminosity, $L_{8-1000\,\mu{\rm m}}$ for the NuLANDS, GOALS and BAT samples. NuLANDS occupies a distinct area of the parameter space to that of GOALS, likely due to a reduced number of host galaxy-dominated mid-to-far infrared systems. Interestingly NuLANDS occupies a similar parameter space to BAT, indicating that NuLANDS does not over sample infrared-bright sources on average. The NuLANDS warm mid-to-far infrared spectral color is shown with a vertical shaded region bounded by dashed lines. Left and right panels show the same data, displayed in different ways for clarity. Left shows the detections with colored errorbars, and the upper limits are shown in grey in the background. Right shows a Gaussian Kernel Density Estimation containing 68\% of each dataset, including upper limits as detections.}
\end{figure*}

Figure~\ref{fig:goals_comp} plots the NuLANDS, \textit{Swift}/BAT and GOALS samples in the plane of 25-to-60\,$\mu$m color vs. total infrared 8\,--\,1000\,$\mu$m luminosity derived using the relation from \citet{Sanders96}. The 25\,--\,60\,$\mu$m color range used to classify warm \textit{IRAS} sources is shown with a vertical shaded region. For a fair comparison with NuLANDS, we used fluxes reported in the \textit{IRAS} point source catalog for all sources. Since the NuLANDS selection did not exclude flux upper limits at 12 or 100\,$\mu$m and the point source catalog contained a number of flux upper limits in one or more \textit{IRAS} bands for other sources, a number of the values plotted in Figure~\ref{fig:goals_comp} are upper limits. We note that the total infrared luminosity is limited by the requirement for \textit{IRAS} detections, such that the \textit{Swift}/BAT data could potentially extend to lower luminosities than is plotted. However, good agreement is found between \textit{Swift}/BAT and NuLANDS indicating that the warm \textit{IRAS} color criterion is capable of identifying AGN efficiently. The lack of cross-over with GOALS is somewhat more revealing in this plot however, since the GOALS sources are found to be at systematically higher luminosities and significantly cooler mid-to-far infrared colors than NuLANDS. Such cooler colors are associated with mid-to-far infrared spectral energy distributions peaking at colder temperatures than a typical AGN, which indicates a dominant contribution from the host galaxy in the 25\,--\,60\,$\mu$m spectra for the majority of GOALS sources. Indeed AGN are typically very difficult to identify in the GOALS sample owing to the strong nuclear obscuration and contamination from the host galaxy \citep{Iwasawa11,TorresAlba18,Vardoulaki15,DiazSantos17,Falstad21}. Nevertheless Figure~\ref{fig:goals_comp} indicates that U/LIRGs typically occupy a different region of the mid-to-far infrared color vs. total infrared luminosity plane than NuLANDS, which suggests that the Compton-thick and obscured AGN fractions are not significantly boosted by a strong presence of such systems in the sample.

\subsection{Outlook and Prospects for Broadband X-ray Spectroscopy of Local AGN}\label{subsec:hexp}
Robust Compton-thick classifications require robust constraints above 10\,keV by definition to include the underlying reprocessed continuum. The majority of the Compton-thick sources revealed in this paper are too faint for any instrument but \textit{NuSTAR} to do this. However, a subset of the Compton-thick candidates are still faint enough to yield upper limits on the observed X-ray continuum at high energies (see Section~\ref{subsec:nh_v_l210int}). In such cases, BXA is able to yield robust upper limits on intrinsic X-ray luminosity, but the line-of-sight column density posteriors likely allow a wide range of values. The \textit{High Energy X-ray Probe} (\textit{HEX-P}\footnote{\url{https://hexp.org}}) is a probe-class mission concept proposed for launch in 2032 \citep{Madsen24}. The \textit{HEX-P} design includes two telescopes, one Low Energy Telescope (LET) operating between 0.2\,--\,25\,keV and a High Energy Telescope (HET) operating between 2\,--\,80\,keV. Together, the LET and HET provide simultaneous broadband X-ray spectra with dramatically improved spectral sensitivity relative to both \textit{XMM-Newton} and \textit{NuSTAR} combined. \textit{HEX-P} thus holds promise for studying AGN in a wide variety of ways, from black-hole spins of large populations \citep{Piotrowska24}, unveiling the physics of the corona \citep{Kammoun24}, probing the circum-nuclear environment of heavily obscured and Compton-thick AGN \citep{Boorman24}, understanding the demographics of obscured AGN across cosmic time \citep{Civano24} and studying the nature of dual AGN in exquisite detail \citep{Pfeifle24}.

To showcase the importance of \textit{HEX-P} for the NuLANDS sample in modelling the reprocessed continua of faint heavily obscured AGN in our sample, we select five of the faintest heavily obscured NuLANDS targets. To enable a like-for-like comparison with \textit{HEX-P}, load the corresponding NuLANDS model P7 fit per source (based on the \textsc{UXCLUMPY} model) and simulate 200\,ks of operational time using the same background and response files as the real observation. For \textit{HEX-P}, we simulate the same best-fit model with the LET and HET, using the current best estimate response files (v07 - 17-04-2023). However, due to an overall factor of $\sim$\,2 improvement in observing efficiency relative to \textit{NuSTAR} \citep{Madsen24}, 200\,ks of operational time with \textit{HEX-P} is equivalent to 200\,ks of exposure on-source compared to 100\,ks with \textit{NuSTAR}.

\begin{figure*}
\centering
\includegraphics[width=0.99\textwidth]{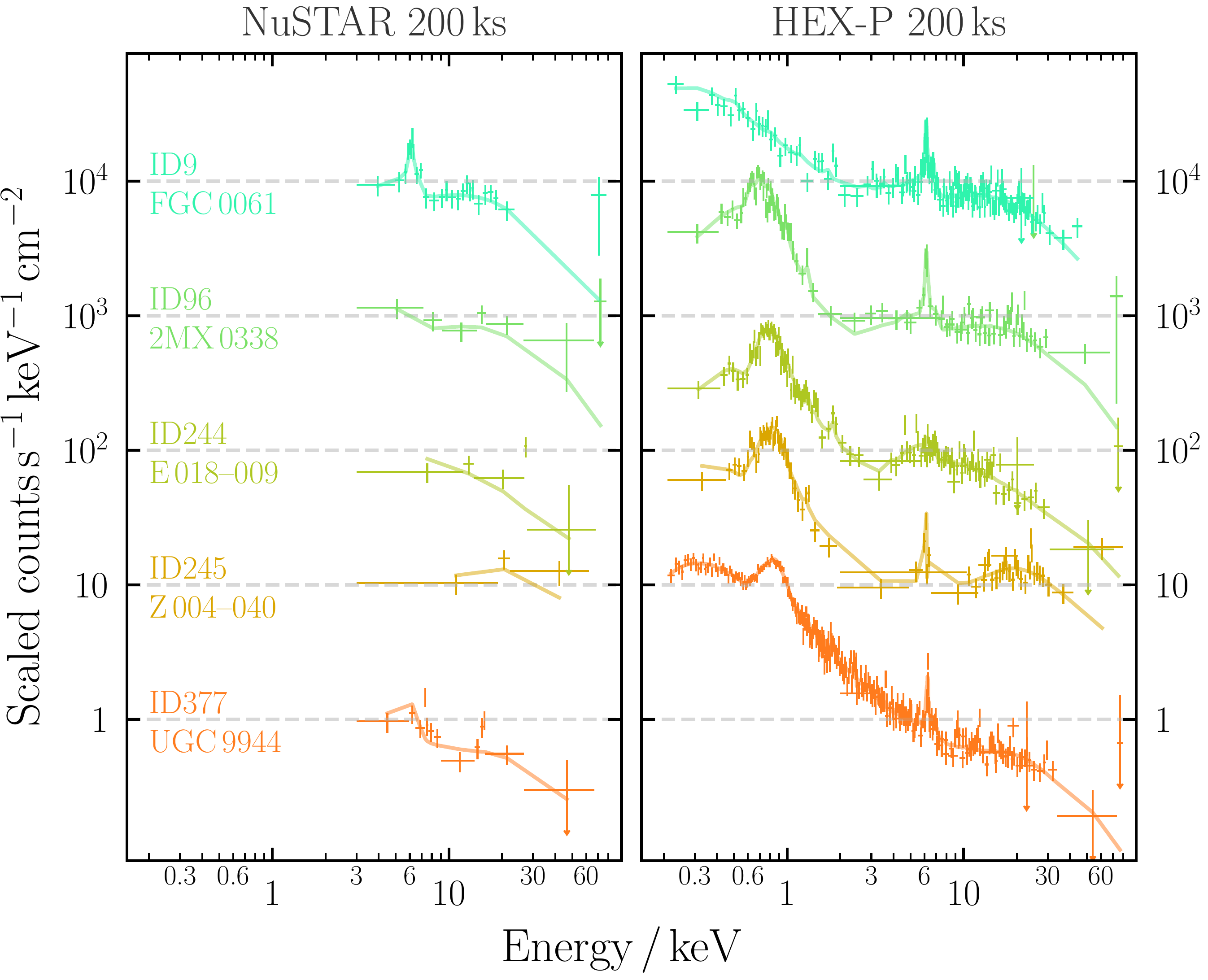}
\caption{\label{fig:hexp}Comparing the current capabilities with \textit{NuSTAR} (left column) to those of the \textit{High Energy X-ray Probe} (\textit{HEX-P}, right column) with five of the faintest heavily obscured candidates in the NuLANDS sample. Each spectrum is scaled logarithmically by their model flux at 5\,keV and plotted in folded units normalised by the effective area. All spectra were simulated for 200\,ks of operational time and then binned visually to have a minimum signal-to-noise ratio of three per bin. We note that 200\,ks of operational time with \textit{NuSTAR} is equivalent to 100\,ks once Earth occultations are accounted for. However, there is no such exposure correction for \textit{HEX-P} due to improved observing efficiencies (see \citealt{Madsen24} for more information). See Section~\ref{subsec:hexp} for more details of the simulations and Figure~\ref{fig:allspec} for definitions of source identifier abbreviations.}
\end{figure*}

Figure~\ref{fig:hexp} presents the corresponding simulated \textit{NuSTAR} (3\,--\,78\,keV) and \textit{HEX-P} (0.2\,--\,80\,keV) spectra, plotted in folded units normalised by the effective area and scaled logarithmically. over  for \textit{NuSTAR} and 0.2\,--\,80\,keV bandpass provided by the LET and HET combined. \textit{HEX-P} is able to reproduce the underlying reprocessed continuum from each AGN and thermal component in a single observation, enabling true broadband spectral modelling without any issues arising from non-simultaneous variability effects. In addition, the LET and HET provide sensitive overlapping spectra over the spectral range associated with the inflection point of the Compton hump between $\sim$\,8\,--\,24\,keV, which has been shown to be critical in deciphering the nature of the circum-nuclear obscurer in detail \citep{Buchner19}.

\section{Summary and Conclusions} \label{sec:summary}
In this work, we present the first paper from the \textit{NuSTAR} Local AGN $N_{\rm H}$ Distribution Survey (NuLANDS). NuLANDS is an X-ray legacy survey of a mid-to-far infrared selected sample of AGN in the nearby universe ($z$\,$<$\,0.044). NuLANDS was constructed to sample optically-classified type~1 AGN with approximately equal efficacy as optically-classified type~2 AGN, with the ultimate goal of providing a sample of AGN that is selected isotropically in terms of line-of-sight column density.

We fit a large library of 23 individual models to each source and check the column density distributions arising from each (Section~\ref{subsec:modelcomp}). Following a Bayesian framework, our fitting process is automated with the \texttt{PyMultiNest} nested sampling implementation in BXA \citep{Buchner14}. Our key findings are as follows:

\begin{enumerate}
    \item \textit{Isotropic selection:} We demonstrate that NuLANDS is isotropically selected based on indistinguishable flux ratios of [O\,\textsc{iii}] to the mid-to-far infrared 25 and 60\,$\mu$m continuum emission for the optically-classified type~1 and~2 AGN (Section~\ref{sec:representative} and Figure~\ref{fig:representativeness}).
    This means that derived quantities, like the distribution of line-of-sight obscuring column density, are closer to the intrinsic distribution than what is seen in X-ray-selected samples. 

    \item \textit{Significant model dependencies:} We show that the choice of spectral model can have significant effects on the parent line-of-sight column density distribution derived for a given sample of sources with X-ray spectral fitting (Section~\ref{subsec:nhdistallmods}). For sources not selected in the 70-month BAT catalog, we find that the line-of-sight column density can vary on average by a factor of $\sim$\,1.4 orders of magnitude between different models, reaching $>$\,2 orders of magnitude in extreme cases (Section~\ref{subsec:modelcomp}, Tables~\ref{tab:logNHlos_Umods} and~\ref{tab:logNHlos_Umods}). To overcome such issues, we develop a Monte Carlo-based Hierarchical Bayesian Model that conservatively propagates the systematic uncertainties associated with individual models into the parent column density distribution for the entire sample (Section~\ref{subsec:nhdist}). 

    \item \textit{The column density distribution:} We find a Compton-thick fraction of 35\,$\pm$\,9\% to 90\% confidence (35\,$\pm$\,6\% to 68\% confidence), which is consistent with the latest estimates from the population synthesis model of \citet{Ananna19} as well as many previous estimates (e.g., \citealt{Ueda14,Buchner15}). We discuss sample selection and classification biases, suggesting our obscured fraction could still be a lower limit.

    \item \textit{Constant selection with distance:} We find no significant systematic trends in any column density distribution bins with distance. Notably, the Compton-thick fraction does not significantly diminish with increased distance out to $\sim$\,200\,Mpc.

    \item \textit{Luminosity dependence:} No significant effect is found for the Compton-thick fraction as a function of total infrared luminosity in the 8\,--\,1000\,$\mu$m wavelength range. Using the \citet{Kennicutt98} relation, we correspondingly find no trend between the Compton-thick fraction and star formation rate though note that AGN contamination may affect our inference of star formation rates derived from infrared fluxes. We additionally find no increase in the Compton-thick fraction with intrinsic luminosities 10$^{42}$\,$<$\,$L_{2-10\,{\rm keV}}$\,$<$\,10$^{43}$\,erg\,s$^{-1}$ and $L_{2-10\,{\rm keV}}$\,$>$\,10$^{43}$\,erg\,s$^{-1}$ within 90\% confidence.
    
    \item \textit{Tests for sample biases:} We find no significant difference in the column density distribution between the warm and cool \textit{IRAS} AGN selected from the \textit{Swift}/BAT sample, indicating that any bolometrically-luminous AGN missed by NuLANDS are not missed in a manner biased against column density measurements (Section~\ref{subsec:cool_bat}). We additionally compare the NuLANDS selection to that of GOALS, finding a significant difference in the infrared properties used to select either sample (Figure~\ref{fig:goals_comp}). Our findings suggest that NuLANDS does not select a disproportionate amount of U/LIRGs that could enhance the obscured and/or Compton-thick fractions of the sample (Figure~\ref{fig:goals_comp} and Section~\ref{subsubsec:goals}).
    
\end{enumerate}

The relatively high Compton-thick fraction reported here is in line with recent estimates that take into account biases against finding the most heavily obscured AGN, and is significantly higher than some older estimates. This implies a much larger fraction of supermassive black hole accretion that is missed by traditional optical and even X-ray surveys. Our work underlines the requirement for pairing multi-wavelength selection and classification techniques with sensitive broadband X-ray spectroscopy in the pursuit of an AGN census.

% \begin{acknowledgments*}
\section*{ACKNOWLEDGEMENTS}
The authors are very grateful to the anonymous referee for their careful reading of the manuscript as well as providing useful and constructive comments that improved the paper. PGB would also like to thank Alberto Masini, Abhijeet Borkar, Emily Moravec, Daniel Kynoch, Ryan Pfeifle, Ari Laor, Giorgio Matt, N\'{u}ria Torres-Alb\`{a}, Lea Marcotulli, Gabriele Matzeu and many others for useful discussions and support throughout the NuLANDS project.

PGB acknowledges financial support from the STFC. PGB additionally thanks the Royal Astronomical Society and Institute of Physics for bursaries awarded in part to support the project, as well as support from the UGC-UKIERI (University Grants Commission - UK-India Education and Research Initiative) Phase 3 Thematic Partnerships. PG (grant reference ST/R000506/1) thanks the STFC for support. MB acknowledges support from the Black Hole Initiative at Harvard University, which is funded by a grant from the John Templeton Foundation. We acknowledge support from ANID-Chile through the Millennium Science Initiative Program ICN12\_009 (FEB), CATA-BASAL FB210003 (CR, FEB, ET), and FONDECYT Regular grants 1200495 (FEB, ET) and  1230345 (CR). IMM acknowledges support from the National Research Foundation of South Africa. JS acknowledges the Czech Science Foundation project No. 22-22643S. AA acknowledges financial support from Universiti Kebangsaan Malaysia through Geran Universiti Penyelidikan code GUP-2023-033. CG acknowledges financial support from the Science and Technology Facilities Council (STFC) through grant codes ST/T000244/1 and ST/X001075/1. MK acknowledges support from NASA through ADAP award 80NSSC22K1126.

This work made use of data from the \textit{NuSTAR} mission, a project led by the California Institute of Technology, managed by the Jet Propulsion Laboratory, and funded by the National Aeronautics and Space Administration. We thank the \textit{NuSTAR} Operations, Software and Calibration teams for support with the execution and analysis of these observations. This research has made use of the \textit{NuSTAR} Data Analysis Software (NuSTARDAS) jointly developed by the ASI Science Data Center (ASDC, Italy) and the California Institute of Technology (USA).

This work made use of data from \textit{XMM-Newton}, an ESA science mission with instruments and contributions directly funded by ESA Member States and NASA.

This work made use of data supplied by the UK \textit{Swift} Science Data Centre at the University of Leicester.

This publication makes use of data products from the \textit{Wide-field Infrared Survey Explorer}, which is a joint project of the University of California, Los Angeles, and the Jet Propulsion Laboratory/California Institute of Technology, funded by the National Aeronautics and Space Administration.

This research has made use of data and/or software provided by the High Energy Astrophysics Science Archive Research Center (HEASARC), which is a service of the Astrophysics Science Division at NASA/GSFC and the High Energy Astrophysics Division of the Smithsonian Astrophysical Observatory.

This research has made use of the NASA/IPAC Extragalactic Database (NED), which is operated by the Jet Propulsion Laboratory, California Institute of Technology, under contract with the National Aeronautics and Space Administration.

This research has made use of NASA's Astrophysics Data System Bibliographic Services.

This research has made use of the \simbad database, operated at CDS, Strasbourg, France.

This work made extensive use of the \texttt{NumPy} \citep{NumPy11}, \texttt{Matplotlib} \citep{Hunter07}, \texttt{SciPy}\citep{Virtanen20}, \texttt{pandas} \citep{Pandas10}, \texttt{Astropy} \citep{Astropy13} {\tt Python} packages.

% \end{acknowledgments*} 

\vspace{5mm}
\facilities{\nustar, \swift, \xmm, \suzaku, \chandra, \iras, \wise}

\appendix
\section{Tables}
\begin{longrotatetable}
\movetabledown=12mm
\startlongtable
\begin{deluxetable*}{ccccccccccccc}
\tablecaption{\label{tab1_src_props} Source properties of all optically-confirmed AGN in the NuLANDS sample, ordered by R.\,A.}
\tabletypesize{\footnotesize}
\tablehead{
\colhead{ID$^{a}$} &
\colhead{NED$^{b}$} &
\colhead{Simbad$^{c}$} &
\colhead{R.\,A.$^{d}$} &
\colhead{Dec.$^{e}$} &
\colhead{Morph.$^{f}$} &
\colhead{Inc.$^{g}$} &
\colhead{Type$^{h}$} &
\colhead{Ref.$^{i}$} &
\colhead{$z^{j}$} &
\colhead{Ref.$^{k}$} &
\colhead{$D^{l}$} &
\colhead{log\,$N_{\rm H,\,Gal.}$\,$^{m}$}
}
\startdata
\multicolumn{13}{c}{\multirow{2}{*}{--\,Type~1 AGN (S1n, S1, S1.2, S1.5, S1.8)\,--}}\\[0.5cm]
\hline
                                 27 &                                 \href{http://ned.ipac.caltech.edu/?q=byname&objname=MRK0359}{MRK\,0359} &                                   \href{http://simbad.u-strasbg.fr/simbad/sim-id?Ident=Mrk359}{MRK\,359} &           21.885 &         19.179 &             S0 &         45.47 &           S1n &  VV10 &  0.017390 &     NED &         78.5 &             20.71 \\
                                 47 &                                 \href{http://ned.ipac.caltech.edu/?q=byname&objname=NGC0931}{NGC\,0931} &                                   \href{http://simbad.u-strasbg.fr/simbad/sim-id?Ident=NGC931}{NGC\,931} &           37.060 &         31.312 &            Sbc &         81.28 &          S1.0 &  VV10 &  0.016650 &     NED &   48.0$^{*}$ &             20.95 \\
                                 54 &                                 \href{http://ned.ipac.caltech.edu/?q=byname&objname=NGC0985}{NGC\,0985} &                                   \href{http://simbad.u-strasbg.fr/simbad/sim-id?Ident=NGC985}{NGC\,985} &           38.658 &         -8.788 &              I &         45.21 &          S1.5 &  VV10 &  0.043140 &     NED &        198.4 &             20.54 \\
                                114 &  \href{http://ned.ipac.caltech.edu/?q=byname&objname=2MASSJ04145265-0755396}{2MASS\,J04145265--0755396} &                                     \href{http://simbad.u-strasbg.fr/simbad/sim-id?Ident=2E955}{2E\,955} &           63.719 &         -7.928 &        \nodata &       \nodata &          S1.5 &  VV10 &  0.038160 &     NED &        174.9 &             20.83 \\
\enddata
% \tablecomments{(1) ID number from \citetalias{deGrijp92}, (2) Identifier from NED, (3) Identifier from Simbad, (4) J2000 right ascension in degrees, (5) J2000 declination ascension in degrees, (6) Host galaxy morphology from HyperLeda, (7) Host galaxy inclination from HyperLeda, (8) Optical AGN spectroscopic classification, (9) Reference for the spectroscopic classification: VV10; \citet{VeronCetty10}, Hale; new spectroscopy from Palomar/Hale, (10) Spectroscopic redshift, (11) Spectroscopic redshift reference. Hale; redshift measured from new spectra with Palomar/Hale, (12) Distance to the source in Mpc. Where available on NED-D, a redshift-independent distance is used, otherwise the luminosity distance is stated, (13) Galactic column density along the line of sight to the source from \citet{Willingale13}.}
\tablenotetext{a}{ ID number from \citetalias{deGrijp92}.}
\tablenotetext{b}{ Identifier from NED.}
\tablenotetext{c}{ Identifier from Simbad.}
\tablenotetext{d}{ J2000 right ascension in degrees.}
\tablenotetext{e}{ J2000 declination in degrees.}
\tablenotetext{f}{ Host galaxy morphology from HyperLeda.}
\tablenotetext{g}{ Host galaxy inclination from HyperLeda in degrees.}
\tablenotetext{h}{ Optical AGN spectroscopic classification.}
\tablenotetext{i}{ Reference for the spectroscopic classification. VV10; \citet{VeronCetty10}, Hale; new spectroscopy from Palomar/Hale.}
\tablenotetext{j}{ Spectroscopic redshift.}
\tablenotetext{k}{ Spectroscopic redshift reference. Hale; redshift measured from new spectra with Palomar/Hale.}
\tablenotetext{l}{ Distance to the source in Mpc. Where available on NED-D, a redshift-independent distance is used and the distance is marked with a *, otherwise the luminosity distance is stated.}
\tablenotetext{m}{ Galactic column density along the line of sight to the source from \citet{Willingale13} in log\,cm$^{-2}$.}
\end{deluxetable*}
\end{longrotatetable}

\begin{longrotatetable}
\startlongtable
\begin{deluxetable*}{cccccccc}
\tablecaption{\label{tab2_multilam_fluxes} Multi-wavelength fluxes used to test the representative nature of the NuLANDS optically-confirmed AGN.}
\tabletypesize{\footnotesize}
\tablehead{
\colhead{ID$^{a}$} &
\colhead{NED\,$^{b}$} &
\colhead{Simbad\,$^{c}$} &
\colhead{$F_{25\,\mu{\rm m}}$\,$^{d}$} &
\colhead{$F_{60\,\mu{\rm m}}$\,$^{e}$} &
\colhead{$\alpha_{25,60}$\,$^{f}$} &
\colhead{log\,$\nu F_{\rm [OIII]}$\,$^{g}$} &
\colhead{log\,$E F_{14-195\,{\rm keV}}$\,$^{h}$}
}
\startdata
% \multicolumn{8}{c}{\multirow{2}{*}{--\,Type~1n AGN\,--}}\\[0.5cm]
% \hline
\multicolumn{8}{c}{\multirow{2}{*}{--\,Type~1 AGN (S1n, S1, S1.2, S1.5, S1.8)\,--}}\\[0.5cm]
\hline
         27 &                                 \href{http://ned.ipac.caltech.edu/?q=byname&objname=MRK0359}{MRK\,0359} &                                   \href{http://simbad.u-strasbg.fr/simbad/sim-id?Ident=Mrk359}{MRK\,359} &            0.48\,$\pm$\,0.07 &            1.26\,$\pm$\,0.14 &  -1.12\,$^{+0.20}_{-0.21}$ &                            -13.06 &             -10.87\,$^{+0.05}_{-0.09}$ \\
         47 &                                 \href{http://ned.ipac.caltech.edu/?q=byname&objname=NGC0931}{NGC\,0931} &                                   \href{http://simbad.u-strasbg.fr/simbad/sim-id?Ident=NGC931}{NGC\,931} &            1.31\,$\pm$\,0.11 &            2.76\,$\pm$\,0.25 &         -0.85\,$\pm$\,0.14 &                            -13.13 &                    -10.21\,$\pm$\,0.02 \\
         54 &                                 \href{http://ned.ipac.caltech.edu/?q=byname&objname=NGC0985}{NGC\,0985} &                                   \href{http://simbad.u-strasbg.fr/simbad/sim-id?Ident=NGC985}{NGC\,985} &            0.55\,$\pm$\,0.06 &            1.44\,$\pm$\,0.13 &         -1.11\,$\pm$\,0.16 &                            -12.65 &             -10.52\,$^{+0.04}_{-0.03}$ \\
        114 &  \href{http://ned.ipac.caltech.edu/?q=byname&objname=2MASSJ04145265-0755396}{2MASS\,J04145265--0755396} &                                     \href{http://simbad.u-strasbg.fr/simbad/sim-id?Ident=2E955}{2E\,955} &            0.54\,$\pm$\,0.05 &            0.66\,$\pm$\,0.06 &         -0.24\,$\pm$\,0.15 &                            -12.20 &             -10.71\,$^{+0.05}_{-0.03}$ \\
\enddata
\tablenotetext{a}{ ID number from \citetalias{deGrijp92}.}
\tablenotetext{b}{ Identifier from NED.}
\tablenotetext{c}{ Identifier from Simbad.}
\tablenotetext{d}{ 25\,$\mu$m flux density in Jy from the \textit{IRAS} v2.1 catalogue.}
\tablenotetext{e}{ 60\,$\mu$m flux density in Jy from the \textit{IRAS} v2.1 catalogue.}
\tablenotetext{f}{ Spectral index between 25--60\,$\mu$m, $\alpha_{25,\,60}$, assuming $F_{\nu}$\,$\propto$\,$\nu^{\alpha}$.}
\tablenotetext{g}{ [OIII] flux from \citetalias{deGrijp92} in log\,erg\,s$^{-1}$\,cm$^{-2}$.}
\tablenotetext{h}{ \textit{Swift}/BAT 105-month 14\,--\,195\,keV logarithmic flux from \citet{Oh18} in erg\,s$^{-1}$\,cm$^{-2}$.}
\end{deluxetable*}
\end{longrotatetable}

\begin{longrotatetable}
\movetabledown=12mm
\startlongtable
\begin{deluxetable*}{ccccccccccccccc}
\tablecaption{\label{tab3_xrayobs} Details of all X-ray data used for the Type~1n AGN in the NuLANDS sample.}
\tabletypesize{\footnotesize}
\tablehead{
\colhead{ID$^{a}$} &
\colhead{\textit{NuSTAR} ID\,$^{b}$} &
\colhead{\textit{NuSTAR} time\,$^{c}$} &
\colhead{FPM\,$^{d}$} &
\colhead{T$^{\rm FPM}$\,$^{e}$} &
\colhead{$N_{\rm src}^{\rm FPM}$\,$^{f}$} &
\colhead{$N_{\rm bkg}^{\rm FPM *}$\,$^{g}$} &
\colhead{S/N$^{\rm FPM}$\,$^{h}$} &
\colhead{Soft Inst.\,$^{i}$} &
\colhead{Soft ID\,$^{j}$} &
\colhead{$\delta$\,/\,days\,$^{k}$} &
\colhead{T$^{\rm soft}$\,$^{l}$} &
\colhead{$N_{\rm src}^{\rm soft}$\,$^{m}$} &
\colhead{$N_{\rm bkg}^{\rm soft *}$\,$^{n}$} &
\colhead{S/N$^{\rm soft}$\,$^{o}$}
}
\startdata
\toprule
\multicolumn{1}{l}{} & \multicolumn{7}{c}{\nustar Observations} & \multicolumn{7}{c}{Soft X-ray Observations} \\ 
\cmidrule(rl){2-8} \cmidrule(rl){9-15} \\
   \multirow{2}{*}{4} &      \multirow{2}{*}{\nodata} &                   \multirow{2}{*}{\nodata} &  \multirow{2}{*}{\nodata} &  \multirow{2}{*}{\nodata} &  \multirow{2}{*}{\nodata} &  \multirow{2}{*}{\nodata} &  \multirow{2}{*}{\nodata} &  \multirow{2}{*}{\nodata} &      \multirow{2}{*}{\nodata} &        \multirow{2}{*}{\nodata} &  \multirow{2}{*}{\nodata} &  \multirow{2}{*}{\nodata} &  \multirow{2}{*}{\nodata} &  \multirow{2}{*}{\nodata} \\
                      &                               &                                            &                           &                           &                           &                           &                           &                           &                               &                                 &                           &                           &                           &                           \\
   \multirow{2}{*}{9} &  \multirow{2}{*}{60361020002} &  \multirow{2}{*}{2019\,Jun\,26,\,23:46:09} &                      FPMA &                      42.8 &                       408 &                     294.3 &                       5.8 &      \multirow{2}{*}{XRT} &  \multirow{2}{*}{00035141001} &       \multirow{2}{*}{-5150.41} &     \multirow{2}{*}{14.7} &       \multirow{2}{*}{10} &      \multirow{2}{*}{1.7} &      \multirow{2}{*}{4.3} \\
                      &                               &                                            &                      FPMB &                      42.5 &                       430 &                     368.2 &                       2.8 &                           &                               &                                 &                           &                           &                           &                           \\
\enddata
\tablenotetext{a}{ ID number from \citetalias{deGrijp92}.}
\tablenotetext{b}{ \textit{NuSTAR} observation ID.}
\tablenotetext{c}{ \textit{NuSTAR} observation UT date and time.}
\tablenotetext{d}{ \textit{NuSTAR} Focal Plane Module.}
\tablenotetext{e}{ \textit{NuSTAR}/FPM filtered exposure time.}
\tablenotetext{f}{ \textit{NuSTAR}/FPM net source counts.}
\tablenotetext{g}{ \textit{NuSTAR}/FPM background counts scaled by to the source region area.}
\tablenotetext{h}{ \textit{NuSTAR}/FPM signal-to-noise ratio in the 3\,--\,79\,keV range.}
\tablenotetext{h}{ Soft instrument, including \textit{Swift}/XRT, \textit{XMM-Newton}/PN, \textit{Chandra}/ACIS and \textit{Suzaku}/XIS3.}
\tablenotetext{j}{ Soft observation ID.}
\tablenotetext{k}{ Number of days after the \textit{NuSTAR} observation. In the event there are no \textit{NuSTAR} data, the UT date is stated.}
\tablenotetext{l}{ Soft filtered exposure time.}
\tablenotetext{m}{ Soft net source counts.}
\tablenotetext{n}{ Soft background counts scaled by to the source region area.}
\tablenotetext{o}{ Soft observation signal-to-noise ratio over instrument full band.}
\end{deluxetable*}

\end{longrotatetable}

\startlongtable
\begin{deluxetable*}{ccccccc}
\tablecaption{\label{tab:logNHlos_Umods} Posterior quantile values for the line-of-sight column density derived from spectral fitting for any type~1 NuLANDS AGN not detected in the 70-month BAT survey. Each row contains the median line-of-sight column densities with associated 90th percentile interquartile ranges found for each model that was selected following the model selection criteria described in Section~\ref{subsec:modelcomp}.}
% \tabletypesize{\footnotesize}
\tablehead{
\colhead{ID$^{a}$} &
\colhead{NED$^{b}$} &
\colhead{U1} &
\colhead{U2} &
\colhead{U3} &
\colhead{U4} &
\colhead{Highest\,Z$^{c}$}
}
\startdata
\multicolumn{7}{c}{\,}\\[-0.2cm]
  260 &      MRK\,1239 &                      -- &                      -- & $21.48_{-0.10}^{+0.08}$ &      $21.49\,\pm\,0.10$ &           U4 \\
  369 &     UGC\,09826 & $22.81_{-0.20}^{+0.18}$ & $22.84_{-0.19}^{+0.16}$ & $22.78_{-0.29}^{+0.17}$ & $22.80_{-0.23}^{+0.18}$ &           U3 \\
  390 &     UGC\,10120 & $20.13_{-0.12}^{+0.28}$ & $20.16_{-0.15}^{+0.34}$ & $20.13_{-0.12}^{+0.24}$ & $20.16_{-0.15}^{+0.27}$ &           U4 \\
  398 &       IC\,1198 &      $22.35\,\pm\,0.10$ & $22.37_{-0.12}^{+0.11}$ & $20.69_{-0.62}^{+1.64}$ & $22.31_{-0.88}^{+0.14}$ &           U3 \\
  531 & ESO\,344--G016 &                      -- &                      -- & $20.34_{-0.29}^{+0.30}$ & $20.33_{-0.29}^{+0.32}$ &           U4 \\
\multicolumn{7}{c}{\,}\\[-0.2cm]
\enddata
\tablenotetext{a}{ ID number from \citetalias{deGrijp92}.}
\tablenotetext{b}{ Identifier from NED.}
\tablenotetext{c}{ The model found to give the highest Bayesian Evidence for a given source.}
\end{deluxetable*}

\begin{longrotatetable}
\movetabledown=12mm
\startlongtable
\begin{deluxetable*}{cccccccccccc}
\tablecaption{\label{tab:logNHlos_Pmods} Posterior quantile values for the line-of-sight column density derived from spectral fitting for any type~2 NuLANDS AGN not detected in the 70-month BAT survey. Each row contains the median line-of-sight column densities with associated 90th percentile interquartile ranges found for each model that was selected following the model selection criteria described in Section~\ref{subsec:modelcomp}.}
\tabletypesize{\footnotesize}
\tablehead{
\colhead{ID$^{a}$} &
\colhead{NED$^{b}$} &
\colhead{P1} &
\colhead{P2} &
\colhead{P3} &
\colhead{P4} &
\colhead{P5A} &
\colhead{P5B} &
\colhead{P6} &
\colhead{P7} &
\colhead{P8} &
\colhead{Highest\,Z$^{c}$}
}
\startdata
\multicolumn{12}{c}{\,}\\[-0.2cm]
    9 &                 FGC\,0061 & $23.76_{-0.26}^{+0.24}$ &                      -- & $24.04_{-0.24}^{+0.67}$ & $24.40_{-0.49}^{+0.98}$ & $24.06_{-1.86}^{+0.85}$ & $23.89_{-1.70}^{+1.00}$ & $23.94_{-0.21}^{+0.15}$ & $25.18_{-1.51}^{+0.72}$ &                      -- &          P5B \\
   26 &                 NGC\,0449 &                      -- &                      -- & $24.69_{-0.55}^{+0.69}$ & $24.60_{-0.48}^{+0.81}$ & $24.07_{-2.01}^{+0.72}$ & $24.05_{-2.00}^{+0.72}$ & $24.15_{-0.17}^{+0.81}$ & $25.29_{-0.81}^{+0.63}$ & $24.58_{-0.45}^{+1.21}$ &           P4 \\
   30 &            KUG\,0135--131 & $20.78_{-0.71}^{+5.13}$ & $21.42_{-0.00}^{+0.37}$ & $25.28_{-0.38}^{+0.20}$ & $24.99_{-0.51}^{+0.45}$ & $24.80_{-0.31}^{+0.18}$ & $24.81_{-0.32}^{+0.17}$ &    $20.00_{-u}^{+6.00}$ & $20.41_{-0.37}^{+0.49}$ & $20.54_{-0.48}^{+0.59}$ &           P2 \\
   33 &                 MRK\,0573 &                      -- &                      -- & $24.25_{-0.05}^{+0.07}$ &                      -- &                      -- &                      -- &                      -- &                      -- & $25.24_{-0.51}^{+0.58}$ &           P8 \\
   37 & 2MASX\,J01500266--0725482 & $21.96_{-0.33}^{+0.29}$ & $22.03_{-0.16}^{+0.17}$ & $22.16_{-0.13}^{+0.19}$ & $22.16_{-0.14}^{+0.16}$ & $22.09_{-0.08}^{+0.16}$ & $22.08_{-0.07}^{+0.13}$ & $22.22_{-0.20}^{+0.21}$ & $21.90_{-0.17}^{+0.16}$ & $22.37_{-0.32}^{+0.15}$ &           P4 \\
   53 &                UGC\,02024 & $24.46_{-4.19}^{+1.39}$ & $23.36_{-1.50}^{+1.32}$ & $24.43_{-2.18}^{+0.96}$ & $24.05_{-1.83}^{+1.27}$ & $24.02_{-1.79}^{+0.89}$ & $23.89_{-1.63}^{+1.01}$ &    $24.22_{-u}^{+1.40}$ & $21.52_{-1.38}^{+4.13}$ & $22.35_{-2.16}^{+3.42}$ &           P2 \\
\enddata
\tablenotetext{a}{ ID number from \citetalias{deGrijp92}.}
\tablenotetext{b}{ Identifier from NED.}
\tablenotetext{c}{ The model found to give the highest Bayesian Evidence for a given source.}
\end{deluxetable*}
\end{longrotatetable}

\bibliography{bibliography}
\bibliographystyle{aasjournal}

\end{document}